\documentclass[aps,pra,superscriptaddress,twocolumn,floatfix,longbibliography,10pt]{revtex4-2}
\usepackage{dcolumn}
\usepackage{graphicx,color}
\usepackage{amsmath,amssymb,bm,amsfonts,dsfont,mathrsfs,amsthm}
\usepackage{braket}
\usepackage{txfonts,comment}
\usepackage[colorlinks=true, allcolors=blue!62!black]{hyperref}

\usepackage{braket}          % Bra-Ket notation
\usepackage{subcaption}
\usepackage{makecell}
\usepackage{orcidlink}       % ORCID iD links
\usepackage{lipsum}          % Dummy text
\usepackage{multirow}
\usepackage{ragged2e}
\usepackage[percent]{overpic}
\usepackage{ulem}
\makeatletter
\long\def\@makecaption#1#2{%
	\vskip\abovecaptionskip
	\sbox\@tempboxa{#1: #2}%
	\ifdim \wd\@tempboxa >\hsize
	{\rightskip=0pt plus 1fil\justifying\noindent #1: #2\par}
	\else
	\global\@minipagefalse
	\hb@xt@\hsize{\hfil\box\@tempboxa\hfil}%
	\fi
	\vskip\belowcaptionskip}
\makeatother



\begin{document}

	\title{ \Large  
		Hybrid boson sampling-neural network architecture for enhanced classification	}
	\author{Mohammad Sharifian\,\orcidlink{0000-0001-9801-239x}
	}
	\email[]{mohammad.sharifian@uestc.edu.cn}
	\affiliation{Institute of Fundamental and Frontier Sciences, University of Electronic Science and Technology of China, Chengdu 611731, China}
	\affiliation{Key Laboratory of Quantum Physics and Photonic Quantum Information, Ministry of Education, University of Electronic Science and Technology of China, Chengdu 611731 , China}
	
	%	\author{...\,\orcidlink{0000-0000-0000-0000}
		%	}
	%	\email[]{...@...}
	%	\affiliation{....}
	% \affiliation{....}
	
	\author{Abolfazl Bayat\,\orcidlink{0000-0003-3852-4558}
	}
	\email[]{abolfazl.bayat@uestc.edu.cn}
	\affiliation{Institute of Fundamental and Frontier Sciences, University of Electronic Science and Technology of China, Chengdu 611731, China}
	\affiliation{Key Laboratory of Quantum Physics and Photonic Quantum Information, Ministry of Education, University of Electronic Science and Technology of China, Chengdu 611731 , China}
	\affiliation{Shimmer Center, Tianfu Jiangxi Laboratory, Chengdu 641419, China}
	%\date{\today}
	
	\begin{abstract}
		\vspace{5pt}
%		\begin{center}
%			\noindent\textbf{ABSTRACT}
%		\end{center}
		\vspace{-3pt}
		
		Demonstration of quantum advantage for classical machine learning tasks remains a central goal for quantum technologies and artificial intelligence. Two major bottlenecks to this goal are the high dimensionality of practical datasets and the limited performance of near-term quantum computers.  Boson sampling is among the few models for which experiments have claimed quantum advantage, yet it has limited practical applications. Here, we propose a hybrid framework that combines the computational power of boson sampling with the adaptability of neural networks to construct quantum kernels that enhance support vector machine classification. The neural network adapts the data features onto a programmable boson sampling circuit, producing quantum states that span a high-dimensional Hilbert space and enable improved classification performance. Using four datasets with various classes, we demonstrate that our model outperforms classical linear and sigmoid kernels. These results highlight the potential of boson sampling-based quantum kernels for practical quantum-enhanced machine learning.

	\end{abstract}

	\maketitle

	\section*{Introduction}

	Quantum computing has emerged as a fundamentally different approach to computation problems through exploiting the principles of quantum mechanics, such as superposition and entanglement~\cite{Nielsen2010quantum}. The superiority of quantum computing is theoretically proven for a number of algorithms that provide exponential or polynomial speedups over their classical counterparts. This includes Shor's algorithm for integer factorization~\cite{Shor1994algorithms}, Grover's algorithm for unstructured search~\cite{Grover1996a} and Harrow-Hassidim-Lloyd algorithm for solving linear equations~\cite{Harrow2017quantum}.
	However, currently available Noisy Intermediate Scale Quantum (NISQ) devices~\cite{Preskill2018quantum,Bharti2022noisy} cannot implement these algorithms on a large scale due to their imperfect operations. To date, sampling problems are the only class of problems for which quantum supremacy has been proposed~\cite{Aaronson2010the,Preskill2012quantum,Hangleiter2023computational} and experimentally reported on NISQ platforms such as superconducting systems~\cite{Arute2019quantum,Wu2021strong,Morvan2024phase,Gao2025establishing} for bit-string sampling of random gates and photonic chips~\cite{Zhong2020quantum,Zhong2021phase,Madsen2022quantum,Liu2025robust,Spring2013boson,Wang2019boson,Zhong2021phase}, for sampling indistinguishable bosons. Nonetheless, in spite of the reported quantum supremacy of boson sampling,  their practical application remains very limited, with exceptions such as  molecular docking~\cite{Banchi2020molecular},  graph problems~\cite{Deng2023solving}, and finding dense subgraphs~\cite{Arrazola2018using}. Therefore, an open question is whether sampling problems can be incorporated into practically useful real-world tasks while benefiting from the computational properties of boson samplers, for which quantum advantage has been experimentally suggested.

	The pursuit of quantum advantage has motivated growing interest in applying quantum systems to machine learning algorithms. Over the past decade, rapid progress in quantum machine learning~\cite{Schuld2021machine,Biamonte2017quantum,Wang2024a,Cerezo2023challenges,Carlesso2023from,Luo2020Yao,Liu2021a,Banchi2021generalization,Banchi2024few,Xiao2023practical,Xiao2024quantum,Li2024experimental,Tangpanitanon2020expressibility,Du2021quantum,Liu2018quantum,Roncallo2025information,Perez2021one} have given rise to the development of quantum kernel methods~\cite{Rebentrost2014quantum,Havlivcek2019supervised}, variational quantum algorithms~\cite{Peruzzo2014aVariational,Cerezo2021Variational,Bharti2021iterative,Jager2023universal,Gong2024enhancing,Li2024ensemble,Acevedo2025variational,Toshio2025practical},  quantum neural networks~\cite{Ren2022experimental,Zheng2023unified,Liu2018differentiable,Li2025quantum,Xiong2023on,Suprano2024photonic,Sannia2024dissipation,Tognini2025solving,Chang2025hybrid,Liu2019solving,Labay2024neural,Ye2025quantum,Liu2022representation,Yu2024expressibility,Zhu2022flexible} and quantum reservoir computing~\cite{Fuji2020quantum,Mujal2021opportunities}. Among these approaches, the quantum Support Vector Machine (SVM) stands as a versatile framework applicable to both regression and classification tasks on classical~\cite{Hoch2025quantum,Ding2019quantum,Yin2024experimental,Montalbano2024quantum,Schuld2019a,Hubregtsen2022training,Yuan2022quantum,Parigi2025supervised} and quantum~\cite{Khosrojerdi2024learning,Sancho2022quantum} data. While an entirely quantum formulation of SVMs has been proposed~\cite{Rebentrost2014quantum}, it relies on quantum linear algebra subroutines and quantum memory access, rendering it unsuitable for current NISQ devices. As a more practical alternative, recent studies employ quantum processors to generate quantum-enhanced kernels, while the core SVM optimization is still performed classically~\cite{Yin2024experimental,Montalbano2024quantum,Schuld2019a,Hubregtsen2022training,Yuan2022quantum,Parigi2025supervised,Khosrojerdi2024learning}. In these NISQ-friendly algorithms, the quantum kernels may capture complex data structures more efficiently than classical kernels, leading to a better classification performance~\cite{Schuld2018quantum}.  Several experimental implementations of the SVM algorithm with quantum-based kernels have been realized~\cite{Havlivcek2019supervised, Peters2021machine,Hubregtsen2022training,Yin2024experimental}. However, most of these experiments rely on artificial datasets with very few features. This is because NISQ computers cannot handle data with large feature sets. In order to achieve quantum advantage with NISQ devices, one has to overcome this limitation. A few questions naturally arise: (i) Is it possible to develop protocols that harness real-life datasets with a large set of features into quantum algorithms such as SVM? and (ii) Can one incorporate boson sampling, whose quantum advantage is experimentally reported on various types of NISQ devices, into the SVM algorithm?

	Here, we propose a protocol that addresses these questions by integrating classical neural networks with quantum boson samplers for classification tasks. The role of the classical neural network is to adapt the large feature sets into a smaller set of parameters, which can be encoded to a boson sampling circuit on a programmable photonic chip.  On the other hand, the boson sampling circuit, whose quantum advantage has been reported on NISQ devices, generates a quantum kernel, which is employed for SVM classification. The integrated architecture of the classical neural network with the quantum boson sampler enhances the accuracy of SVM classification, outperforming both classical linear and non-linear sigmoid kernels as well as neural network-based classifiers. This integration has two immediate advantages. Firstly, it allows us to handle large images without demanding very large quantum circuits. Secondly, it harnesses the computational properties underlying boson sampling, for which quantum advantage has been experimentally suggested. To evaluate the performance of our protocol, we employ different datasets, including Ionosphere~\cite{Sigillito1989ionosphere}, Spambase~\cite{Hopkins1999spambase}, MNIST~\cite{Deng2012the}, and Fashion-MNIST~\cite{Xiao2017fasion}. Our theoretical results show that the enhanced accuracy is achieved by using a sufficiently expressive boson sampling circuit. The expressivity of the boson sampling circuit can be controlled with both the number of modes and the number of injected photons. Our protocol is readily implementable on existing boson sampling photonic chips and can be extended to other supervised learning tasks, such as regression problems.	
	
	%%%%%%%%%%%%%%%%%%%%%%%%%%%%%%%%%%%%%%%%%%%%%%%%%%%%%%%%
		\begin{figure*}[t]
		\centering
		\begin{overpic}[width=0.9\textwidth]{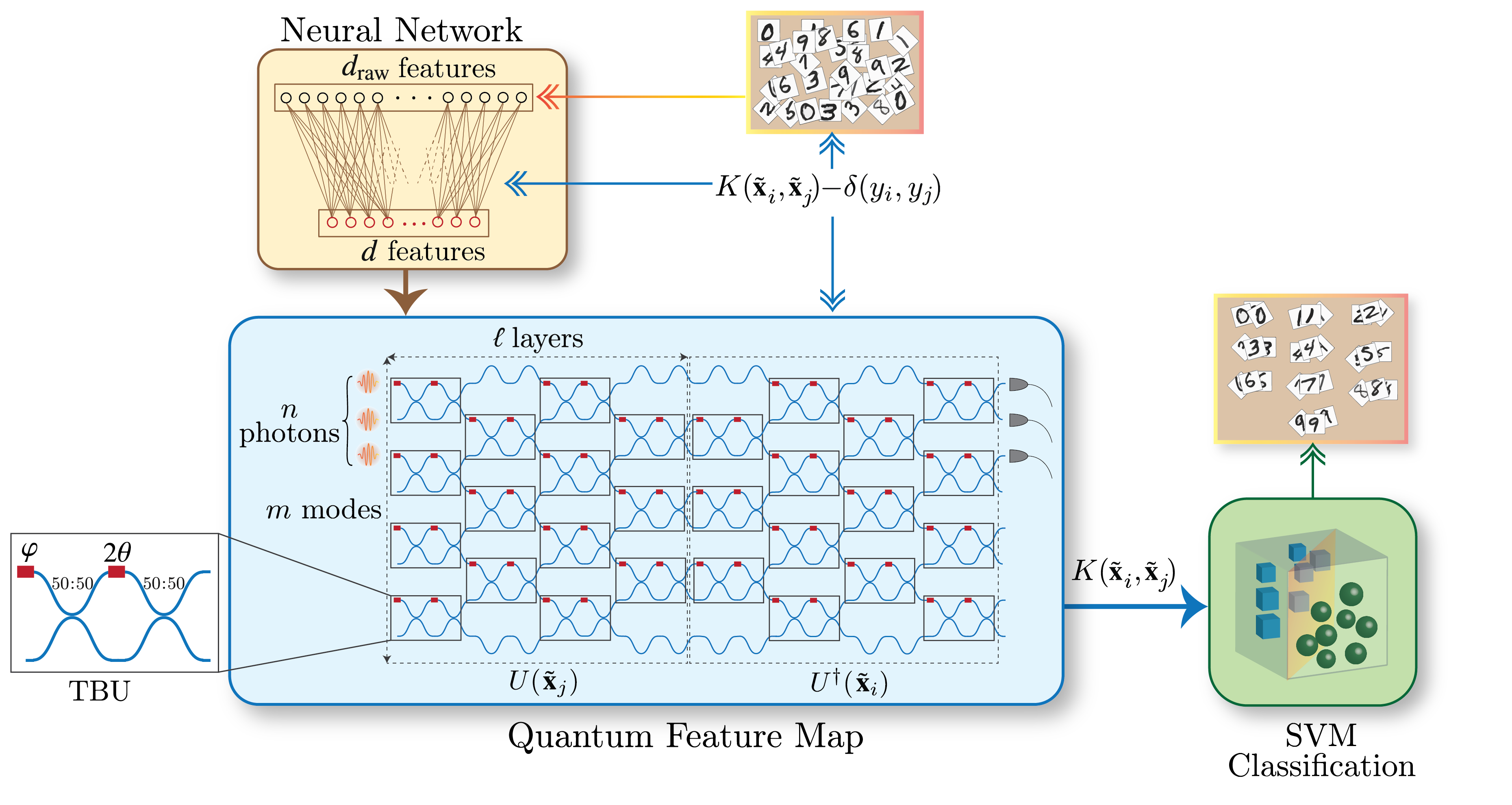}
			% Label for the main image
			\put(0,50){\textbf{(a)}}
			
			% Overlay the second image in the top-right corner
			% Adjust position (x,y) and width as needed
			\put(77,45){%
				\begin{minipage}{0.16\textwidth}
					\begin{overpic}[width=\textwidth]{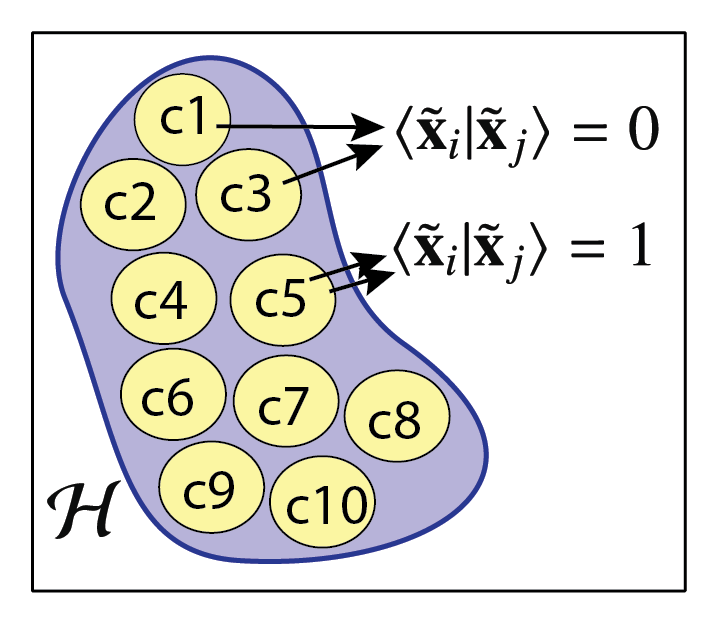}
						\put(-15,70){\textbf{(b)}}
					\end{overpic}
				\end{minipage}
			}
		\end{overpic}
		\vspace{-15pt}
		\caption{Schematic overview of the proposed model. a) The input dataset is first processed by a neural network that transforms its features to match the number of tunable phase shifters (red rectangles) in the boson sampling circuit. Each kernel value for a pair of data points corresponds to the probability of coincidence detection at the circuit output. During training, the mean squared error between the estimated kernel values and the target pairwise label equivalence is used as the loss function to optimize the neural network parameters. After training, the computed kernel matrix is supplied to an SVM classifier to predict the labels of test data. b) Conceptually, the model learns to increase the quantum-state fidelity between samples of the same class while suppressing it for samples from different classes, thereby enhancing class discrimination in the Hilbert space representation.}
		\label{fig:schematic}
	\end{figure*}
	
		\section*{RESULTS} 
	
	\noindent\textbf{Background Review}
	
	\noindent SVMs are powerful supervised machine learning algorithms for both classification and regression~\cite{Cortes1995support}. They benefit from kernel methods that employ mapping data into higher dimensional feature spaces for enhancing their accuracy. On the other hand, boson sampling has been proposed~\cite{Aaronson2010the} and experimentally realized~\cite{Spring2013boson,Wang2019boson} on photonic setups as a demonstration of quantum supremacy. In this paper, we propose a hybrid structure of a classical neural network and a quantum boson sampler, realizable on photonic chips, to define an effective kernel for an SVM classification. The hybridization of classical neural networks and boson samplers will permit dealing with datasets with a large number of features, benefiting from inherent quantum advantage of boson samplers and yet remain NISQ-friendly. Before explaining our protocol, we provide a brief review on the SVM classification algorithm as well as the theory behind boson sampling. The readers who are familiar with these concepts, can skip the following two sections and continue from the next one. 
	\\

	\noindent\textbf{Support vector machines}
	
	\noindent Linear SVM classification searches for a hyperplane that divides two classes of data points with a maximum distance from them. We consider a training set $\{(\mathbf{x}_i,y_i)\}_{i=1}^{N_{\mathrm{train}}}$,
	where $N_{\mathrm{train}}$ is the number of training samples, $\mathbf{x}_i {\in} \mathbb{R}^d$ is the
	$i$-th data point vector, $y_i {\in} \{{+}1{,}{-}1\}$ denotes its class label, and
	$i{=}1,\ldots,N_{\mathrm{train}}$ indexes the training samples.  The hyperplane is parametrized by a vector $\mathbf{w} {\in} \mathbb{R}^d$ and a bias term $b {\in} \mathbb{R}$, such that points $\mathbf{x} {\in} \mathbb{R}^d$ lying on it satisfy $\mathbf{w} {\cdot} \mathbf{x} {+} b {=} 0$. A label is assigned to the new test data point, $\mathbf{x}_j^{\textrm{test}}$, based on which side of the hyperplane this data point is, using the decision hypothesis
	\begin{eqnarray}
		\hat{y}_j^{\textrm{test}}=\operatorname{sign}(\mathbf{w} \cdot \mathbf{x}_j^{\textrm{test}}+b)~.
	\end{eqnarray}
	The perpendicular distance of a data point $\mathbf{x}_i$ from the hyperplane is $D(\mathbf{x}_i) {=} y_i(\mathbf{w} {\cdot} \mathbf{x}_i {+} b)/\|\mathbf{w}\|$, and the training data points with the smallest absolute distance, i.e., lying closest to the decision boundary, are called support vectors. Since the function associated with the hyperplane $(\textbf{w}, b)$ does not change if one rescales the hyperplane to $(\textbf{w}/\|\textbf{w}\|,b/\|\textbf{w}\|)$ there is an inherent degree of freedom in the definition of linear classifiers. To deal with this ambiguity one can impose $ {y_i}{\left(\mathbf{w} {\cdot} \mathbf{x}_i {+} b\right)}{=}1$ for the support vectors. In this manner, the distance of the support vectors from the hyperplane, i.e. $y_i (\textbf{w}{\cdot} \mathbf{x}_i {+} b)/\|\textbf{w}\|$, simplifies into $1/\|\textbf{w}\|$ and the problem reduces to the following quadratic form
	\begin{align} \label{eq:SVMcost1}
		\min  \quad  \frac{1}{2}\|\mathbf{w}\|^2& \\
		\text{subject to} \quad  y_i\left(\mathbf{w} \cdot \mathbf{x}_i + b\right) \geq 1,& \quad \forall i = 1, \ldots, N_\mathrm{train}~. \nonumber
	\end{align}
	Then, a dual formulation is obtained by incorporating the constraints in the Lagrangian
	\begin{eqnarray}\label{eq:primalL}
		\mathcal{L}(\mathbf{w},b, \boldsymbol{\alpha})=\frac{1}{2}\|\mathbf{w}\|^2{-}\sum_{i=1}^{N_\mathrm{train}}\alpha_i \Big( y_i\left(\mathbf{w} \cdot \mathbf{x}_i+b\right)-1 \Big),
	\end{eqnarray}
	with $\alpha_i {\geq} 0$ as Lagrange multipliers chosen to ensure the constraints of Eq.~\eqref{eq:SVMcost1} are satisfied~\cite{Cortes1995support,Cervantes2020a}. Based on Eq.~\eqref{eq:SVMcost1} the Lagrangian should be minimized with respect to $\mathbf{w}$ and $b$ and maximized with respect to $\alpha_i$. It can be readily solved by differentiating $\mathcal{L}$ with respect to $\mathbf{w}$ and $b$ 
	\begin{eqnarray}
		& \frac{\partial \mathcal{L}(\mathbf{w}, b, \boldsymbol{\alpha})}{\partial \mathbf{w}}=\mathbf{w}-\sum_{i=1}^{N_\mathrm{train}} y_i \alpha_i \mathbf{x}_i=\mathbf{0}~, \nonumber \\
		& \frac{\partial \mathcal{L}(\mathbf{w}, b, \boldsymbol{\alpha})}{\partial b}=\sum_{i=1}^{N_\mathrm{train}} y_i \alpha_i=0~.
	\end{eqnarray}
	Resubstituting the above relations into the primal Lagrangian of Eq.~\eqref{eq:primalL} leads to the following optimization problem of the Lagrange multipliers $\boldsymbol{\alpha}=\left\{\alpha_i\right\}_{i=1 \ldots {N_\mathrm{train}}}$
	\begin{eqnarray}\label{eq:LagrangeOpt}
		&\underset{\boldsymbol{\alpha}}{\operatorname{max}~~}& \mathcal{L}(\boldsymbol{\alpha})=\sum_{i=1}^{N_\mathrm{train}} \alpha_i-\frac{1}{2}\sum_{i, j=1}^{N_\mathrm{train}}y_iy_j\alpha_i\alpha_j \mathbf{x}_i \cdot\mathbf{x}_j~,\nonumber\\
		&\text{subject to: }&\sum_i \alpha_i y_i=0~, \nonumber\\
		&&\alpha_i \geq 0~, \quad \forall i=1, \cdots, N_\mathrm{train}~.
	\end{eqnarray}
	The optimization gives $\boldsymbol{\alpha}$ as the coefficients which can be used to assign a new label to a test data point by 
	\begin{equation} 
		\label{eq:decision}
		\hat{y}(\mathbf{x}_j^{\textrm{test}})=\operatorname{sign}\left(\sum_{i=1}^{N_\mathrm{train}}\alpha_i\, y_i \,\mathbf{x}_i \cdot\mathbf{x}_j^{\textrm{test}}+b\right) .
	\end{equation}

	By introducing a feature map, the data can be transferred into a high dimensional Hilbert space $\boldsymbol{\Phi}{:} \mathbb{R}^d {\rightarrow} \mathcal{H}$ named feature space. After choosing an appropriate feature map, the SVM classifier can be applied to the mapped data in $\mathcal{H}$ instead of $\mathbb{R}^d$. The optimization problem now can be solved in a higher dimensional feature space for the vectors $\boldsymbol{\Phi}(\mathbf{x}_i)$. It follows the same formulation as in Eq.~\eqref{eq:LagrangeOpt} and Eq.~\eqref{eq:decision} except that the inner product $\mathbf{x}_i {\cdot}\mathbf{x}_j$ is replaced by the kernel function $K(\mathbf{x}_i, \mathbf{x}_j){=}\boldsymbol{\Phi}(\mathbf{x}_i) {\cdot} \boldsymbol{\Phi}(\mathbf{x}_j)$ which  is the inner product of the mapped data in the higher dimensional space.
	
	Kernel method in SVM is efficient because it avoids directly creating a mapped version of data in the higher dimensional feature space. Instead, SVM exploits the scalar inner products in that space during the whole algorithm. In the classical machine learning, commonly used kernels include the linear kernel $K(\mathbf{x}_i,\mathbf{x}_j){=}\mathbf{x}_i{\cdot}\mathbf{x}_j$ and the sigmoid kernel $K(\mathbf{x}_i,\mathbf{x}_j){=}\tanh\left(\mathbf{x}_i{\cdot}\mathbf{x}_j{+}1\right)$, as a non-linear transformation. We have employed these two kernels as benchmarking with our quantum  kernel constructed by a boson sampler quantum circuit, which will be described later.
	\\

	\noindent\textbf{Boson sampling}
	
	\noindent One of the computational schemes that exhibits a quantum advantage over classical computers is boson sampling. The calculation of boson distribution under a unitary transformation is a \#P-hard problem but it can be readily implemented on the multimode interferometric circuits  thanks to the progress in integrated photonics. Considering an $m$-modes interferometer with input creation operators $a^\dagger_i$ and output ones $b^\dagger_i$ satisfying bosonic commutation relations, the input and output Fock states of the circuit are given by
	\begin{eqnarray}
		\ket{\psi_{\mathrm{in}}}&=&\prod_{i=1}^{m} \frac{(a_i^\dagger)^{s_i}}{\sqrt{s_i!}} \ket{0}~,\quad\mathrm{and}\quad
		\ket{\psi_\mathrm{out}}= \prod_{i=1}^m \frac{(b^\dagger_i)^{t_i}}{\sqrt{t_i!}}\ket{0}~,\nonumber\\
	\end{eqnarray}
	where $s_i$ and $t_i$ denote the photon numbers in the $i$-th input and output mode, respectively.  The total number of bosons in the input state $\ket{\psi_{\rm in}}$ and the output state $\ket{\psi_{\rm out}}$ is then given by $\sum_i s_i$ and $\sum_i t_i$, respectively. 
	The relation between input creation operators and output ones is  $b^\dagger_i{=}\sum_{j=1}^m V_{ij}a^\dagger_j$ in which $V$ is an $m{\times} m$ unitary operator governing the quantum circuit.

	The goal is to find the probability of observing $\ket{\psi_{\rm out}}$ at the output of the quantum circuit assuming that the input is $\ket{\psi_{\rm in}}$. Since $V$ conserves the number of bosons, this probability is nonzero only if the number of bosons in the two quantum states are exactly the same, namely $\sum_i s_i {=} \sum_i t_i{=}n$.  
	To find such probability it is useful to write the output state in terms of input bosonic operators as 
	\begin{eqnarray}
		\ket{\psi_\mathrm{out}}= \prod_{i=1}^m \frac{\left(\sum_{j=1}^m V_{ij}a^\dagger_j\right)^{t_i}}{\sqrt{t_i!}}\ket{0}~.
	\end{eqnarray}
	Then, the probability can be computed as~\cite{Gard2015an,Scheel2004permanents}
	\begin{eqnarray}\label{eq:qprob}
		P=\left|\braket{\psi_{\mathrm{out}}|\psi_{\mathrm{in}}}\right|^2=\frac{\left|\mathrm{Per}(V_{S,T})\right|^2}{s_1!\cdots s_m!\,t_1!\cdots t_m!}~,
	\end{eqnarray}
	where $V_{S,T}$ is a $n{\times} n$ matrix extracted from $V$ and $\mathrm{Per}(\cdots)$ denotes the permanent function whose classical computation is known to be a \#P-hard problem~\cite{Lund2017quantum,Tichy2015sampling,Renema2018efficient}. However, from a quantum mechanics perspective, the above probability $P$ can be efficiently computed just by sampling the output of the quantum circuit. To build the $V_{S,T}$ matrix, two steps are required: (i) selecting rows of $V$ according to the output state by taking $t_i$ copies of the $i$-th row that constructs an $n{\times} m$ intermediate matrix; (ii) from this intermediate matrix extracting $s_j$ copies of $j$-th column to form the final $n{\times} n$ matrix $V_{S,T}$.

				\begin{table*}[t]
		\centering
		\begin{tabular}{lccccccccc}
			\hline
			\textbf{Dataset} & \textbf{Total data} & \textbf{Features} & \textbf{Classes} & \textbf{Used data} & \textbf{Modes} & \textbf{Quantum} & \textbf{Linear} & \textbf{Sigmoid} & \textbf{Neural Network} \\
			\hline
			Ionosphere    & 351    & 34  & 2 (pass, reflect)  & 351   & 6  & $93.24{\pm}2.1\%$ & $86.20{\pm}4.1\%$ & $64.79{\pm}6.4\%$ & $64.79{\pm}6.4\%$ \\
			Spambase      & 4,601  & 57  & 2 (spam, not spam) & 4,601 & 11 & $90.84{\pm}0.9\%$ & $89.49{\pm}0.6\%$ & $79.28{\pm}1.0\%$ & $60.50{\pm}1.7\%$ \\
			MNIST         & 70,000 & 784 & 10 (digits 0--9)   & 5,000 & 12 & $92.94{\pm}0.5\%$ & $91.82{\pm}0.4\%$ & $83.17{\pm}0.8\%$ & $80.31{\pm}1.8\%$ \\
			Fashion-MNIST & 70,000 & 784 & 10 (T-shirt, bag, etc.) & 5,000 & 12 & $82.89{\pm}0.8\%$ & $80.36{\pm}0.8\%$ & $35.77{\pm}3.5\%$ & $60.76{\pm}1.4\%$ \\
			\hline
		\end{tabular}
		\caption{
			Summary of benchmark datasets, photonic circuit configurations, and test accuracies. Quantum accuracies correspond to using five photons in the photonic circuit. Reported values are mean $\pm$ standard deviation over five independent runs.
		}
		\label{tab:merged_datasets}
	\end{table*}

	The possibility for efficient estimation of the inner product $P$ on a quantum computer paves the way for defining a quantum kernel function for developing a quantum enhanced SVM classifier~\cite{Gong2025enhance,Yin2024experimental}. In this situation, we consider a photonic chip with $m$ modes which is injected by an input Fock state $\ket{\phi}$ containing $n$ bosons (with $n{\le} m$), each injected into a different mode.  Therefore, the Hilbert space of such system has a dimension of $\mathcal{D}_\mathrm{boson}{=}\binom{m + n - 1}{n}$, that represents the number of distinct configurations of $n$ indistinguishable photons in $m$ modes. In a parameterized quantum circuit, the unitary operator $U(\mathbf{x}_i)$ is defined such that its parameters represent the features of the input data $\mathbf{x}_i$.
	Note that both the $\mathcal{D}_\mathrm{boson} {\times} \mathcal{D}_\mathrm{boson} $ unitary matrix $U$ and the $m{\times}m$ unitary matrix $V$ describe the same quantum circuit, however, $V$ represents the operation of the circuit on bosonic operators and $U$ denotes the same operation in the Hilbert space.  The quantum state at the output is thus given by $\ket{\mathbf{x}_i}{=} U(\mathbf{x}_i)\ket{\phi}$. The quantum kernel of the two distinct input data $\mathbf{x}_i$ and $\mathbf{x}_j$ is thus given by 
	\begin{equation}\label{eq:Qkernel}
		K(\mathbf{x}_i,\mathbf{x}_j)=|\langle \mathbf{x}_i|\mathbf{x}_j\rangle|^2=\left|\bra{\phi}U^\dagger(\mathbf{x}_i) U(\mathbf{x}_j)\ket{\phi}\right|^2.
	\end{equation}

			\begin{figure*}[t]
		\centering
		
		% =========================
		% Left side: plots (two rows)
		% =========================
		\begin{minipage}[c]{.8\textwidth}
			\centering
			
			% ----- Row 1 -----
			\begin{subfigure}{.39\textwidth}
				\begin{overpic}[width=\textwidth]{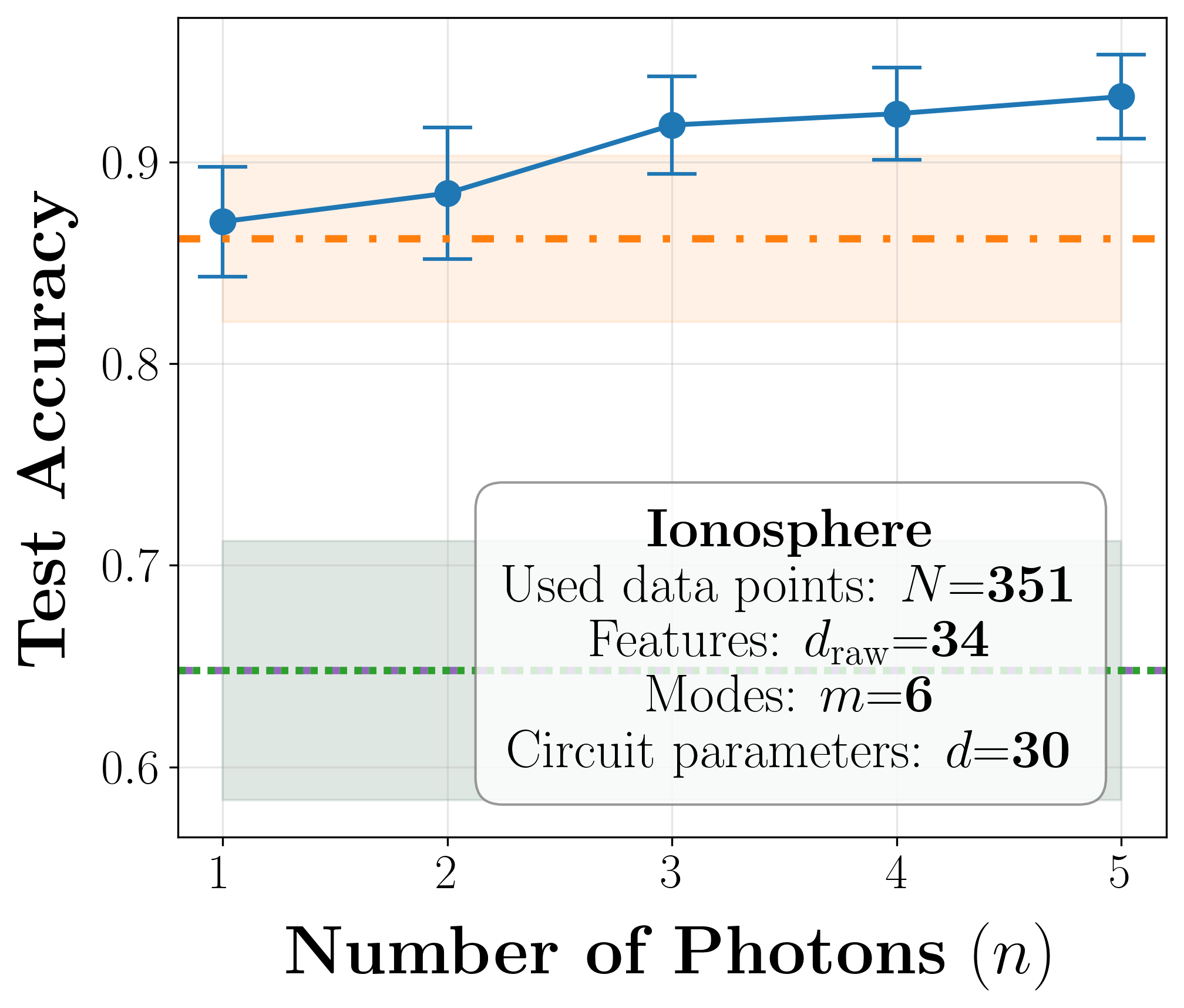}
					\put(-3,83){\textbf{(a)}}
				\end{overpic}
				\label{fig:mult2}
			\end{subfigure}
			\hspace{0.04\textwidth}
			\begin{subfigure}{.39\textwidth}
				\begin{overpic}[width=\textwidth]{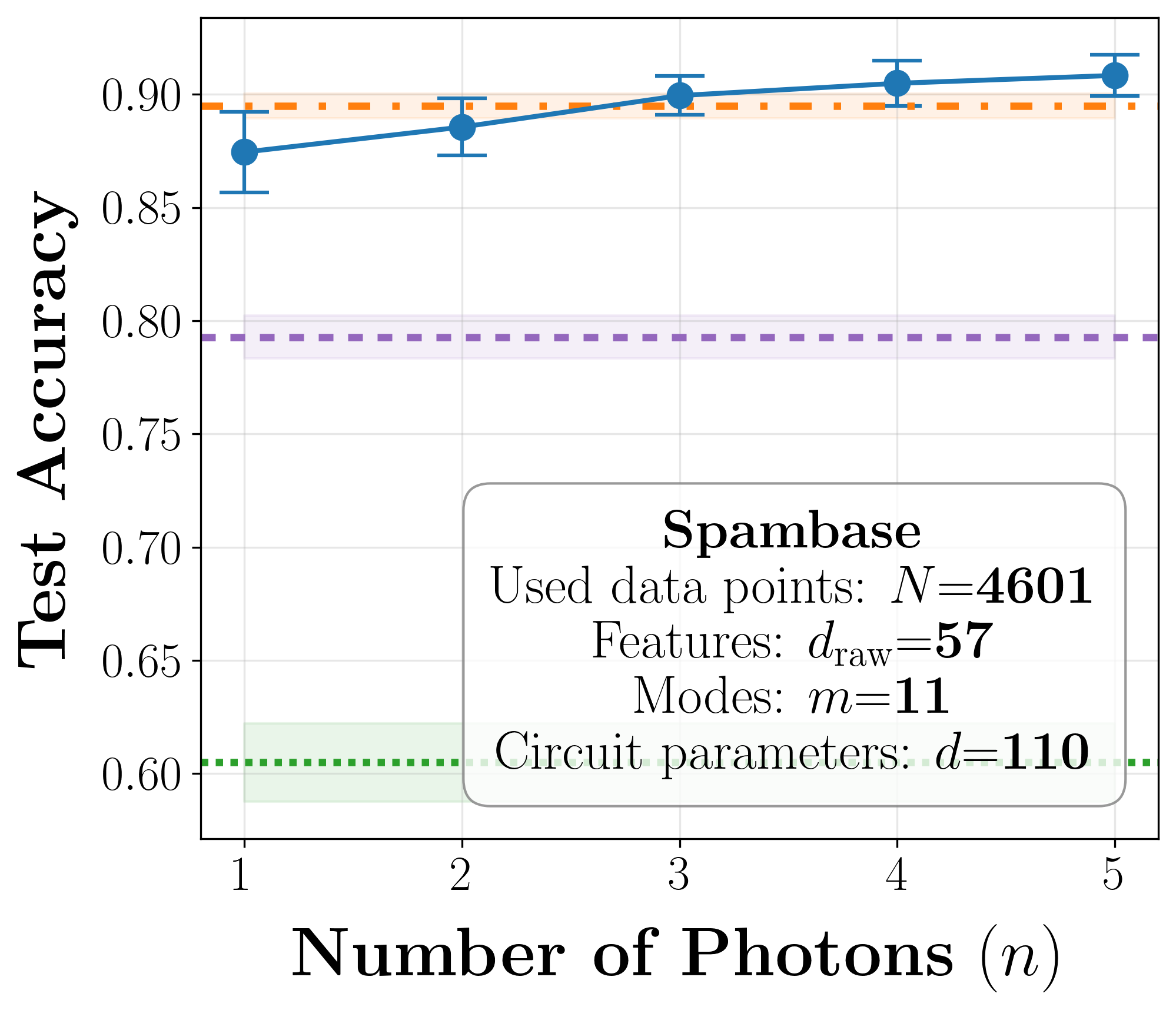}
					\put(-3,83){\textbf{(b)}}
				\end{overpic}
				\label{fig:mult4}
			\end{subfigure}
			
			\vspace{0cm}
			
			% ----- Row 2 -----
			\begin{subfigure}{.39\textwidth}
				\begin{overpic}[width=\textwidth]{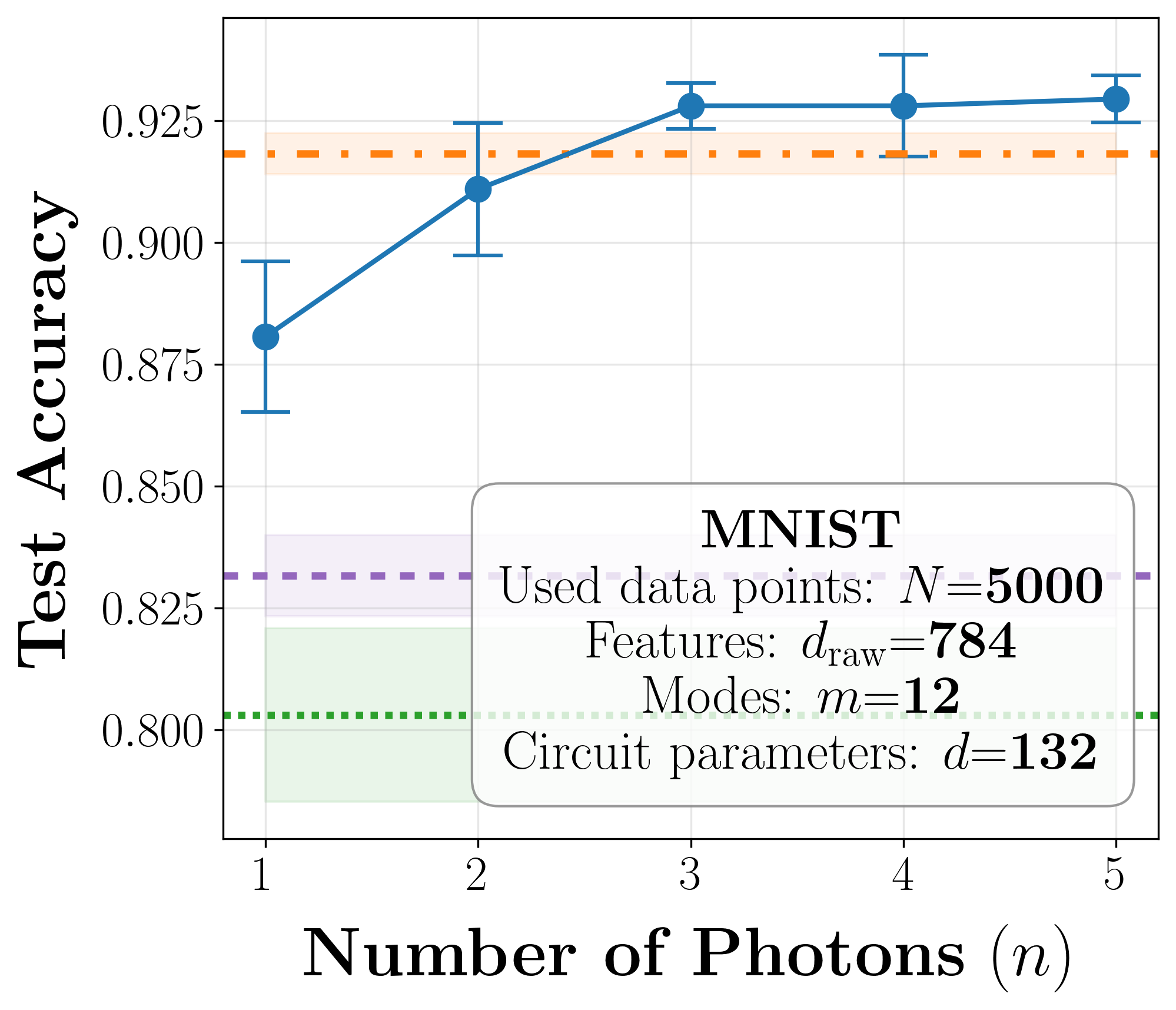}
					\put(-3,83){\textbf{(c)}}
				\end{overpic}
				\label{fig:mult5}
			\end{subfigure}
			\hspace{0.04\textwidth}
			\begin{subfigure}{.39\textwidth}
				\begin{overpic}[width=\textwidth]{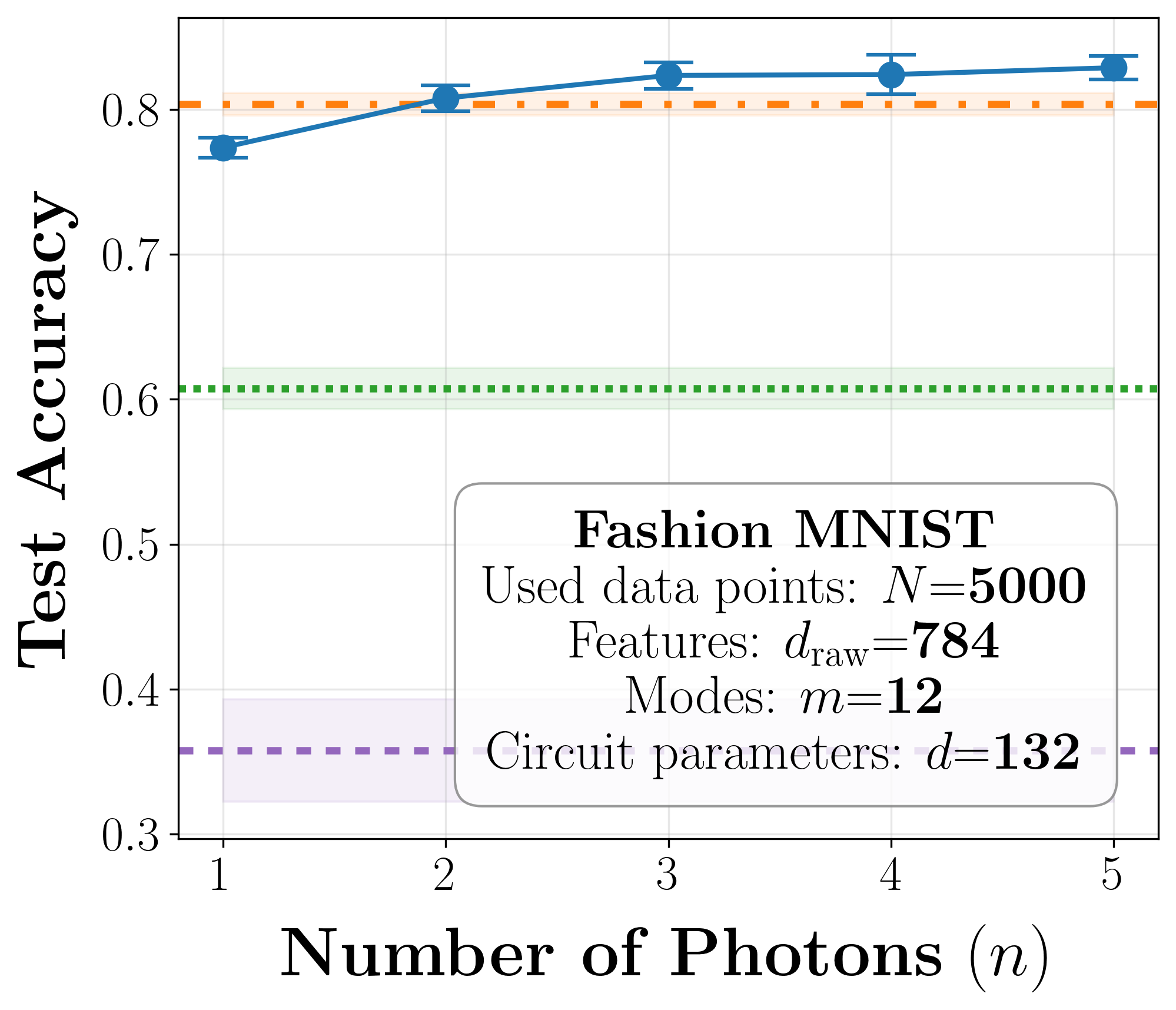}
					\put(-3,83){\textbf{(d)}}
				\end{overpic}
				\label{fig:mult6}
			\end{subfigure}
		\end{minipage}
		\hspace{-15pt}
		% =========================
		% Right side: tall legend
		% =========================
		\begin{minipage}[c]{.13\textwidth}
			\centering
			\includegraphics[width=\textwidth]{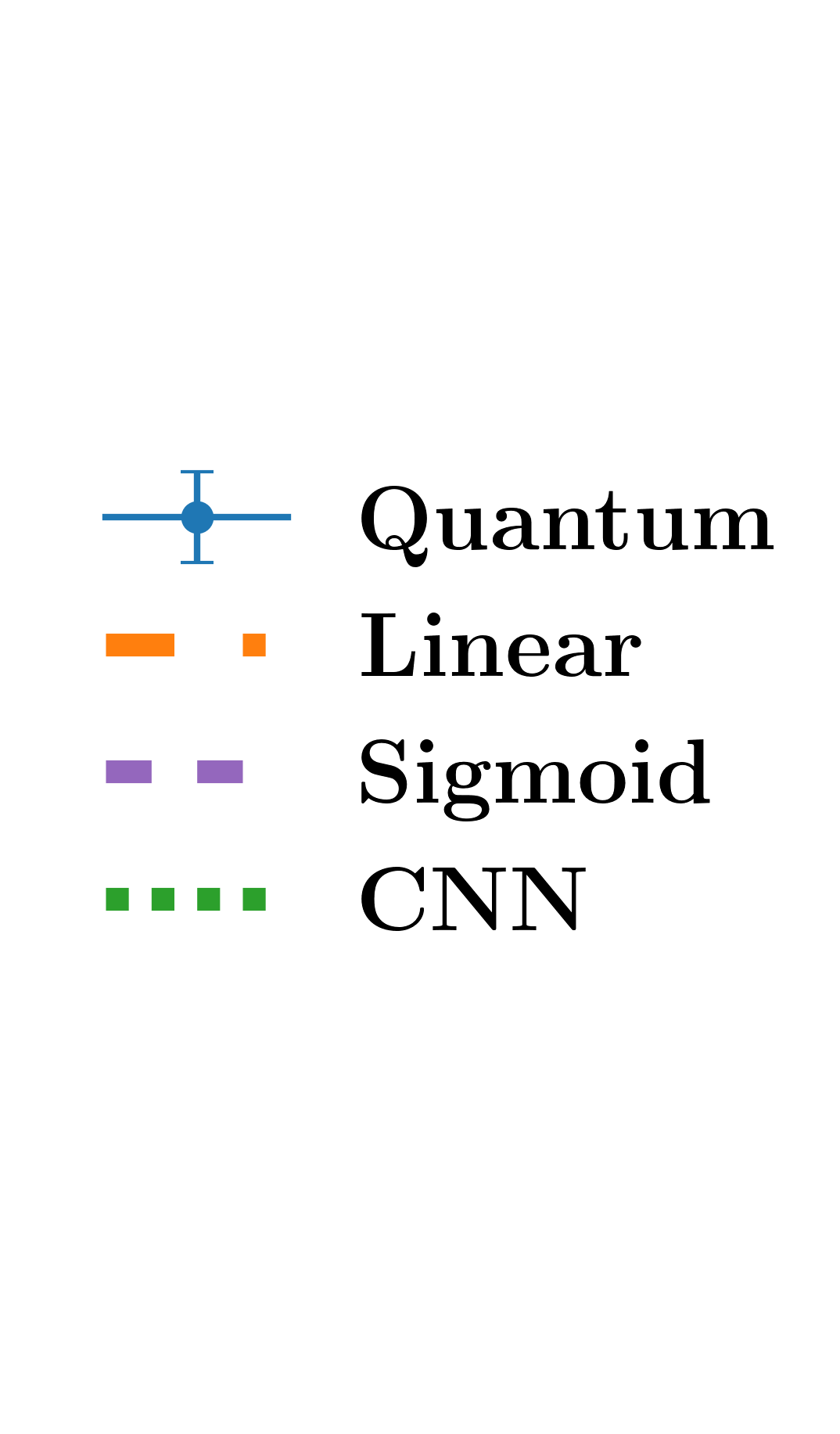}
		\end{minipage}
%		\vspace{-15pt}
		% ===== Caption =====
		\caption{Mean test accuracy of the proposed model as a function of the number of indistinguishable photons. Results are presented for a) Ionosphere, b) Spambase, c) MNIST, and d) Fashion-MNIST datasets. The solid blue line denotes the average classification accuracy obtained using the hybrid quantum kernel for different photon numbers. Dashed lines correspond to the mean accuracy achieved with SVM using classical kernels, whereas the dotted curve shows the performance of the corresponding standalone classical neural network without any kernel integration. All values represent averages over five independent runs, with error bars indicating the standard deviation.}
		\label{fig:acc_photons}
	\end{figure*}

	There are two different approaches to compute this quantum kernel. First, one can use swap test~\cite{Blank2020quantum} for determining the overlap between the two quantum states which requires to access to both $\ket{\mathbf{x}_i}$ and $\ket{\mathbf{x}_j}$ simultaneously. Second, one can compute the above quantum kernel through sampling the same input state at the output of the quantum circuit which realizes the operation of $U^\dagger(\mathbf{x}_i) U(\mathbf{x}_j)$, which has been used in Refs.~\cite{Yin2024experimental,Hoch2025quantum,Gong2025enhance}. While the first approach relies on doubling the number of qubits, for accessing both  $\ket{\mathbf{x}_i}$ and $\ket{\mathbf{x}_j}$, the second approach relies on a quantum circuit with the double depth compared to $U(\mathbf{x}_i)$. In this paper, we take the second approach. 
	
	The boson-sampling kernels also offer superior expressivity over their qubit counterparts. The dimension of the Hilbert space in a boson sampler with $m$ modes and $n$ photons, $\mathcal{D}_\mathrm{boson}{=}\binom{m+n-1}{n}$,  is larger than their qubit counterparts with $m'$ qubits and $n'$ excitation, $\mathcal{D}_{\mathrm qubit}{=}\binom{m'}{n'}$. This extension naturally allows that the encoded quantum states of different classes have larger distances in the Hilbert space. Therefore, with the same number of resources a boson sampler can result in a better accuracy in comparison with a qubit circuit. 
	\\

	\noindent\textbf{Model architecture}
	
	\noindent Our framework consists of three components: (i) a classical neural network; (ii) a boson sampling circuit; and (iii) a support vector machine (SVM) classifier. The first two components are interconnected to form a hybrid quantum–classical kernel generator. This kernel is used by the third component to classify the images. In general, classical images have many features which makes it hard to encode them in NISQ simulators. Thus we employ  a classical neural network whose main task is to map the input data into a representation that matches the number of tunable elements in the boson sampling circuit. The output of the classical neural network are directly fed into the boson sampler. Therefore, the classical neural network is the only component which is trained. Once training is accomplished, the interconnected part generates the kernels between all pairs of data which are then used by an SVM to perform the final classification. These parts are illustrated schematically in Fig.~\ref{fig:schematic}a.

	One of the key obstacles for adopting quantum machine learning for classification tasks is the large size of classical images. In our protocol, a classical neural network maps the original input data 
	$\mathbf{x}_i {\in} \mathbb{R}^{d_{\mathrm{raw}}}$ to an encoded vector 
	$\tilde{\mathbf{x}}_i {\in} \mathbb{R}^{d}$, which matches the boson sampling parameters. To accomplish this, we employ a classical neural network whose first layer consists of $d_{\mathrm{raw}}$ neurons, taking each original data $\mathbf{x}_i$ as its input. The output layer of the neural network  contains $d$ nodes corresponding to the parameters which are used in the boson sampler. In order to simplify the training we use the simplest possible neural network, namely a shallow fully connected network without any hidden layer that uses a sigmoid activation function to ensure each feature remains between 0 and 1. 
	Therefore, every pair of input data $(\mathbf{x}_i,\mathbf{x}_j)$ is mapped to  $(\tilde{\mathbf{x}}_i,\tilde{\mathbf{x}}_j)$ through the neural network. The parameters of the neurons are trained in fifty training epochs according to the following cost function that aligns with kernel based classification
	\begin{eqnarray}\label{eq:nncost}
		\mathcal{C} = \frac{1}{N_\mathrm{train}^2} \sum_{i=1}^{N_\mathrm{train}} \sum_{j=1}^{N_\mathrm{train}} \left( K(\tilde{\mathbf{x}}_i,\tilde{\mathbf{x}}_j) - \delta(y_i,y_j) \right)^2~,
	\end{eqnarray}
	where $K(\tilde{\mathbf{x}}_i,\tilde{\mathbf{x}}_j)$ denotes the kernel value between the outputs of the neural network, namely  $\tilde{\mathbf{x}}_i$ and $\tilde{\mathbf{x}}_j$. The kernel is efficiently measured using  the boson sampling circuit, as will be explained below. The above mean square error cost function encourages kernel values to be maximized for pairs of data points from the same class while being minimized for pairs from different classes. By increasing the fidelity between same class quantum states and suppressing it between different classes, this approach mirrors the concept of contrastive learning~\cite{Khosla2020computer}.

	The second part of the protocol is the boson sampler circuit which can be constructed using tunable basic units (TBUs) as adjustable components designed for encoding data on a circuit~\cite{Capmany2020programmable,Arrazola2021quantum,Tan2023scalable,Zhu2024a}. TBUs operate between a pair of modes and can be implemented using different photonic architectures, such as balanced Mach-Zehnder interferometers. The TBUs play the role of two-mode entangling gates in the circuit and thus make the exploration of the Hilbert space possible. As shown in Fig.~\ref{fig:schematic}a, such a TBU consists of two tunable thermo-optics phase shifters and two 50:50 beam splitters. The operation of such TBU is like a lossless beam splitter and (up to a global phase) is described by the unitary matrix
	\begin{equation}\label{eq:TBU_matrix}
		T(\theta, \varphi) = 
		\begin{bmatrix}
			e^{i\varphi} \cos\theta & -\sin\theta \\
			e^{i\varphi} \sin\theta & \cos\theta
		\end{bmatrix}~.
	\end{equation}
	The parameter $\varphi{\in}[0,2\pi]$ introduces a phase shift and $\theta{\in}[0,\pi/2]$ determines the beam splitter’s reflectivity and transmissivity~\cite{Clements2016an}. Each $\tilde{\textbf{x}}_i$ can be normalized to lie within the valid range of $\theta$ and $\varphi$, and then mapped to the corresponding phase shifters. The circuit is designed with an alternating  arrangement of TBUs so that in the first layer, TBUs couple adjacent modes with odd indexed connections, while in the second layer, they act on even indexed couplings, see  Fig.~\ref{fig:schematic}a. This alternating pattern continues across the circuit layers.  In general, for a quantum circuit of $m$ modes and $\ell$ layers the number of required TBUs is
	\begin{eqnarray}\label{eq:TBU}
		N_{\text{TBU}}(m, \ell) = \left\{
		\begin{array}{ll}
			\frac{\ell}{2}(m - 1)+\frac{1}{2}, & m \text{ is even and } \ell \text{ is odd,}\\
			\frac{\ell}{2}(m - 1), & \text{otherwise.}
		\end{array}
		\right.\nonumber\\
	\end{eqnarray}
	Since for a universal unitary operation one requires $\ell{=}m$ layers~\cite{Clements2016an}, we also  fix $\ell{=}m$. Note that every TBU has two controllable parameters and thus the number of  neurons at the output of the neural network will be $d{=}2N_{\text{TBU}}(m, \ell{=}m)$. The values of the output neurons is a real number between 0 and 1. To adapt the neural network outputs to the TBU parameters, each value is multiplied by $\pi/2$ and $2\pi$ to obtain the phase shifts $\theta$ and $\varphi$, respectively, as defined in Eq.~\eqref{eq:TBU_matrix}.

			\begin{figure*}[t]
		\centering
		
		% =========================
		% Row of figures with overpic captions
		% =========================
		\begin{subfigure}{.4\textwidth}
			\begin{overpic}[width=\textwidth]{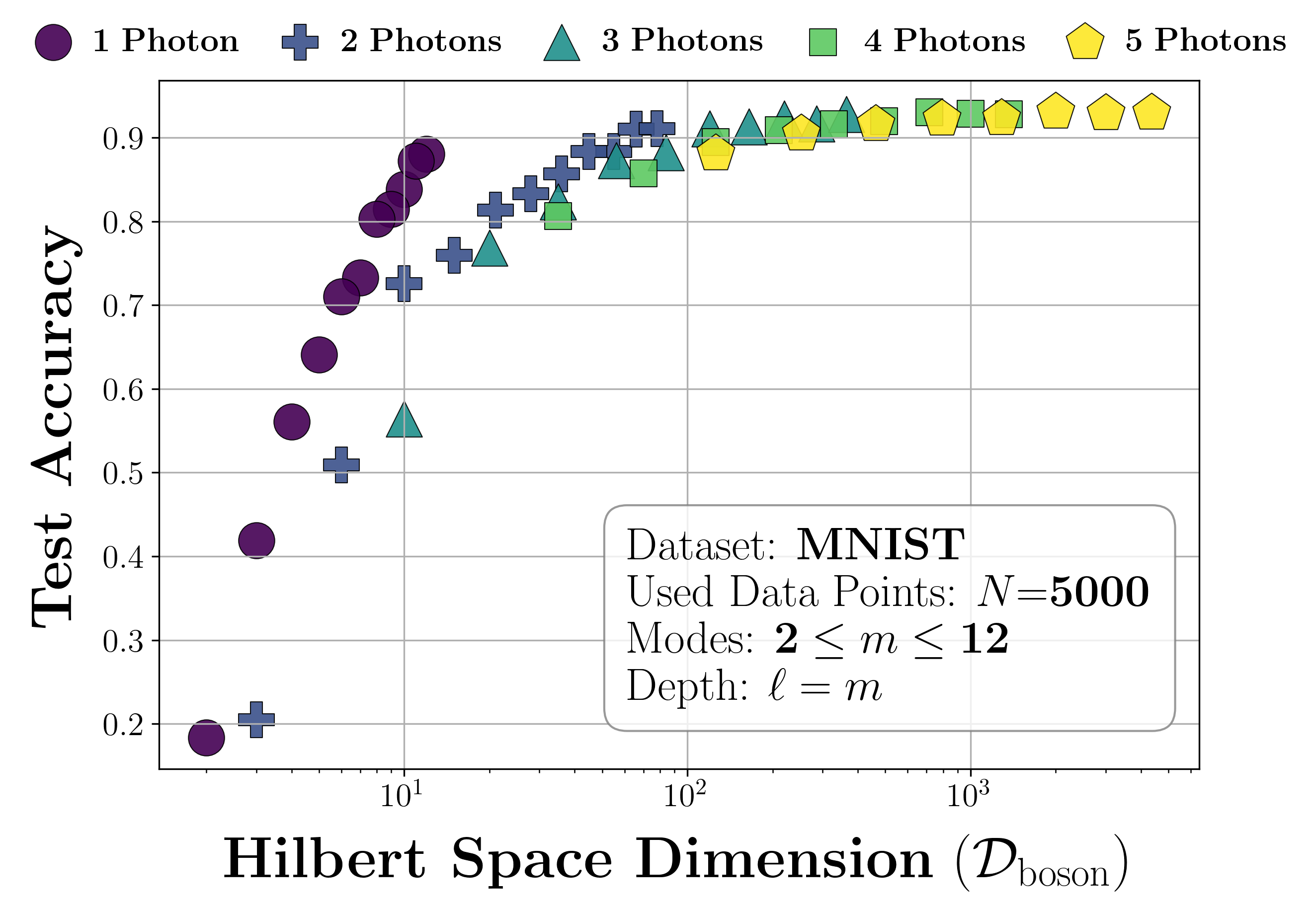}
				\put(-5,66){\textbf{(a)}}
			\end{overpic}
			\label{fig:Hilbert}
		\end{subfigure}
		\quad
		\begin{subfigure}{.4\textwidth}
			\begin{overpic}[width=\textwidth]{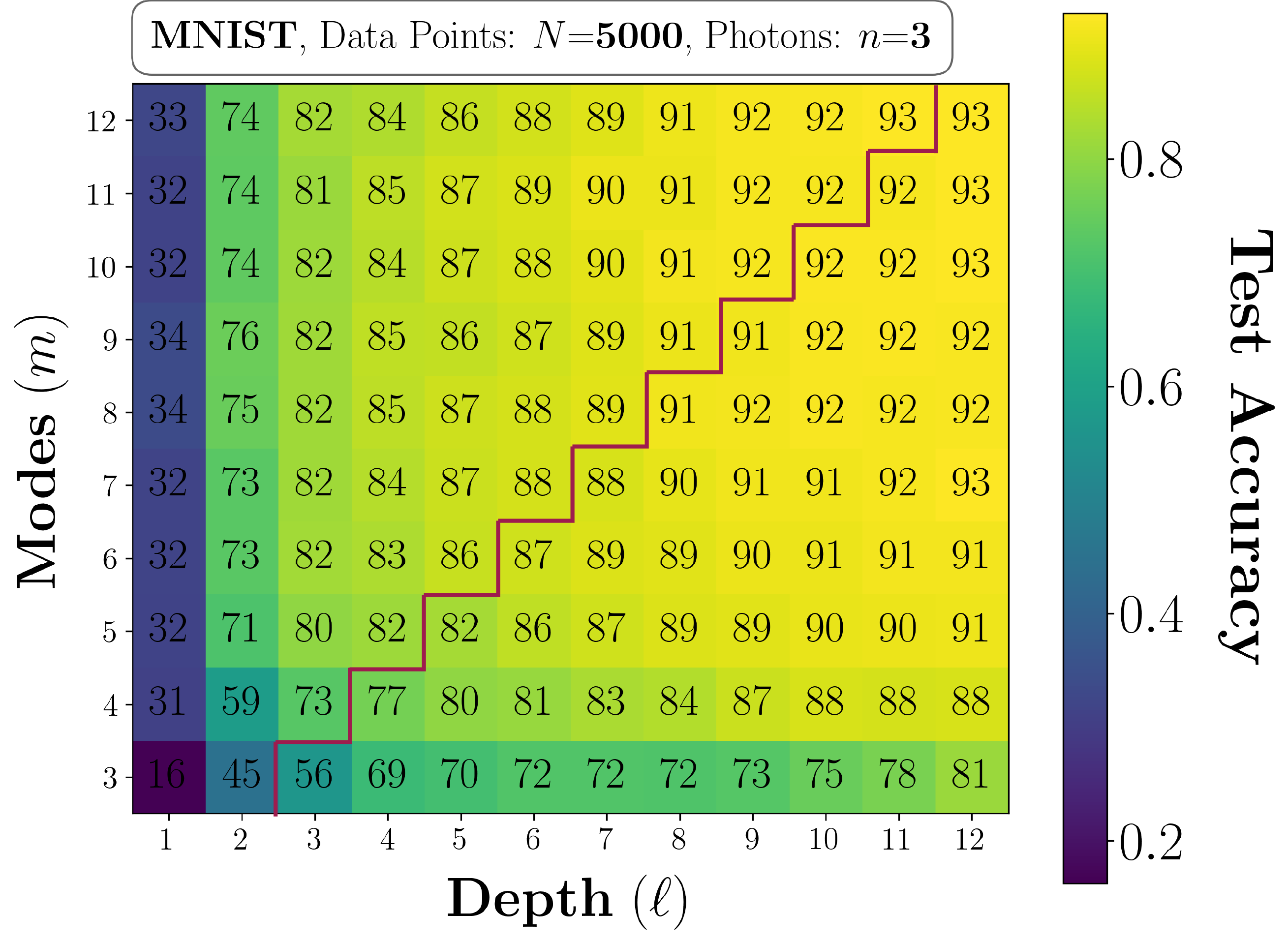}
				\put(-3,68.5){\textbf{(b)}}
			\end{overpic}
			\label{fig:depth_mode}
		\end{subfigure}
		\vspace{-15pt}
		% ===== Caption =====
		\caption{Effect of Hilbert space dimension and circuit size on MNIST classification accuracy. a) shows the test accuracy as a function of Hilbert space dimension, and b) shows the effect of varying the number of modes and layers. In the panel (b), $n{=}3$ photons are used and the red line indicates the minimum number of layers ($\ell{=}m{-}1$) required for full connectivity between the first and last modes. All results are averaged over five independent runs.}
		\label{fig:Hilbert_depth}
	\end{figure*}

	In order to construct a boson sampler that produces kernel values between different pairs of data points 
	$K(\tilde{\mathbf{x}}_i{,}\tilde{\mathbf{x}}_j){=}\left|\bra{\phi}U^\dagger(\tilde{\mathbf{x}}_i) U(\tilde{\mathbf{x}}_j)\ket{\phi}\right|^2$, the photonic circuit should consist of two sequential circuits. The first circuit applies the unitary transformation $U(\tilde{\mathbf{x}}_j)$ and is followed by a second interferometer that implements $U^\dagger(\tilde{\mathbf{x}}_i)$. The second interferometer is constructed by reversing the layer order of the first and replacing each TBU with its Hermitian conjugate, such that the overall circuit performs the operation $U^\dagger(\tilde{\mathbf{x}}_i)U(\tilde{\mathbf{x}}_j)$. We note that all $\varphi$, and $\theta$ parameters between the two unitary operations are different since the vectors $\tilde{\mathbf{x}}_i$ and $\tilde{\mathbf{x}}_j$ denote different data points. Therefore, the overall unitary matrix of the circuit is calculated and then the corresponding submatrix is extracted from it to compute the kernel value using Eq.~\eqref{eq:qprob}.

	Thanks to the above hybrid quantum classical procedure, the classical neural network is trained and the corresponding kernel matrix $K(\tilde{\mathbf{x}}_i,\tilde{\mathbf{x}}_j)$ is obtained. Now one can optimize the SVM classifier of Eq.~(\ref{eq:LagrangeOpt}) using a classical optimization toolbox, such as~\cite{Pedregosa2011scikit}. For new test data points, one has to send each data point $\mathbf{x}_j^{\textrm{test}}$ through the trained neural network to get $\tilde{\mathbf{x}}^{\textrm{test}}_j$. Then the assigned label to this data point in a reduced linear SVM can be obtained as
	\begin{equation} 
		\label{eq:quantum_decision}
		\hat{y}^{\textrm{test}}(\tilde{\mathbf{x}}_j^{\textrm{test}})=\operatorname{sign}\left(\sum_{i=1}^{N_\mathrm{train}} \alpha_i\, y_i \,K(\tilde{\mathbf{x}}_i ,\tilde{\mathbf{x}}_j^{\textrm{test}})+b\right).
	\end{equation}
	Note that the kernel elements $K(\tilde{\mathbf{x}}_i {,}\tilde{\mathbf{x}}_j^{\textrm{test}})$ are extracted from the boson sampling circuit between the given test data $\tilde{\mathbf{x}}_j^{\textrm{test}}$ and all the training data points $\tilde{\mathbf{x}}_i$. As we will see in the following sections, one may significantly reduce this as only a handful number of training data points are enough to obtain the label with good accuracy. Then the accuracy of the model is calculated by 
	\begin{equation}
		\textrm{Test Accuracy} = \frac{1}{N_\mathrm{test}} \sum_{j=1}^{N_\mathrm{test}} \delta\left(\hat{y}_j^{\textrm{test}}, y_j^{\textrm{test}}\right)~,
	\end{equation}
	where $ N_\mathrm{test} $ is the total number of test data points, $\hat{y}_j^{\textrm{test}}$ is the predicted label for the $j$-th test sample, and $ \delta(a, b) $ is the Kronecker delta function. 
	
	\begin{figure*}[t]
		\centering
		\begin{subfigure}{.06\textwidth}
			\includegraphics[width=\textwidth]{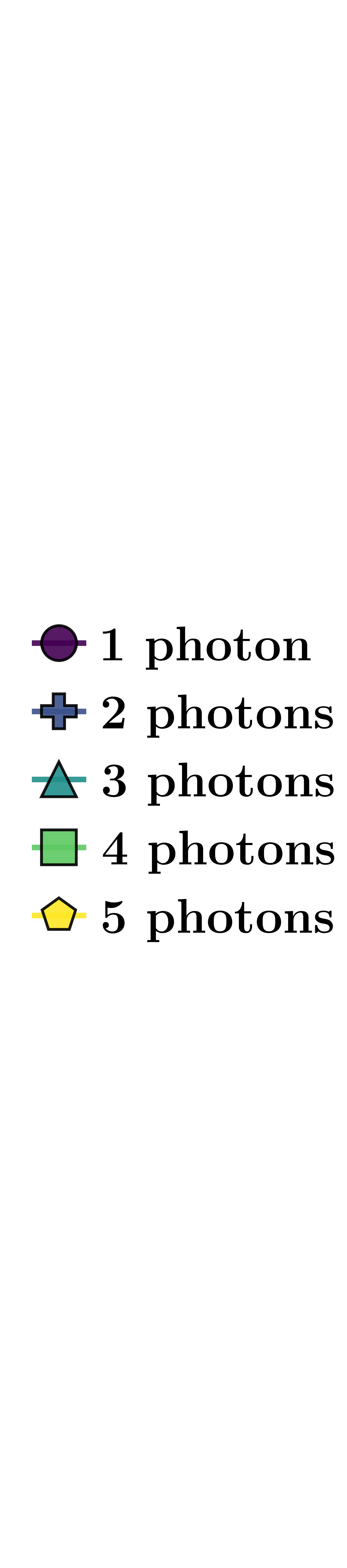}
		\end{subfigure}
		\begin{subfigure}{.23\textwidth}
			\begin{overpic}[width=\textwidth]{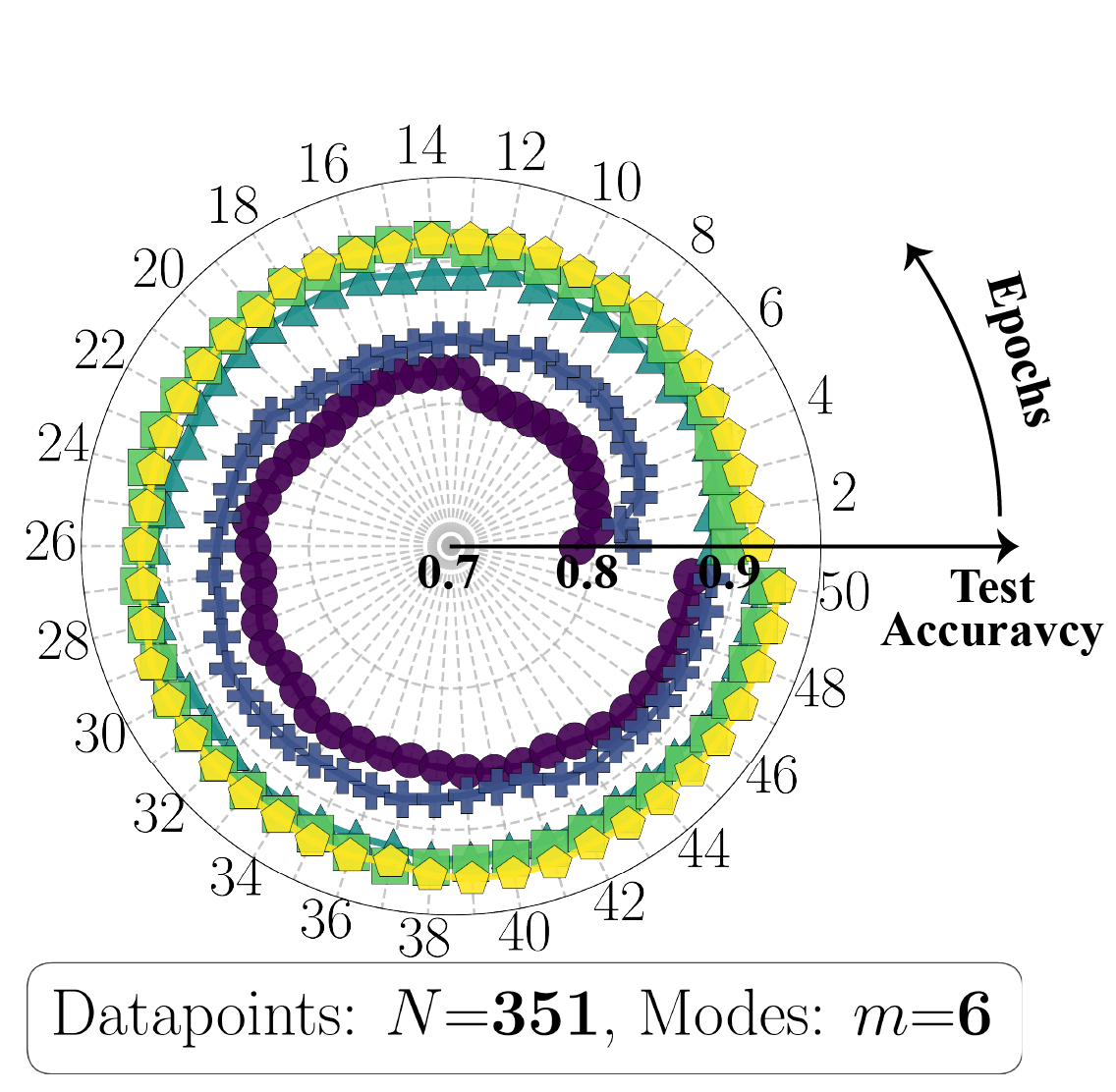}
				\put(4,90){\textbf{(a)}} % <-- x=4%, y=90% from bottom-left
			\end{overpic}
			\label{fig:ionosphere_epochs}
		\end{subfigure}
		\begin{subfigure}{.23\textwidth}
			\begin{overpic}[width=\textwidth]{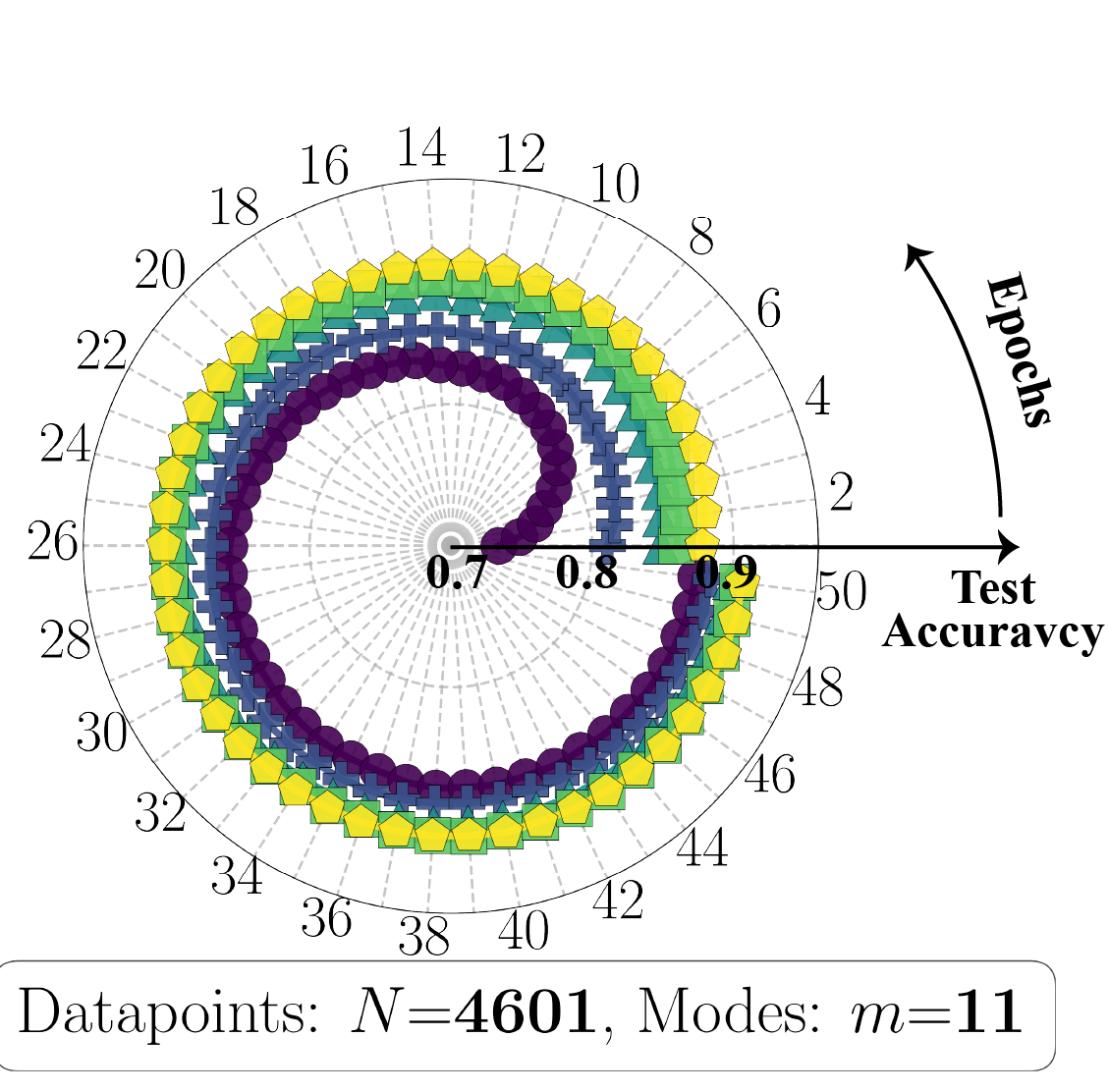}
				\put(4,90){\textbf{(b)}}
			\end{overpic}
			\label{fig:spambase_epochs}
		\end{subfigure}
		\begin{subfigure}{.23\textwidth}
			\begin{overpic}[width=\textwidth]{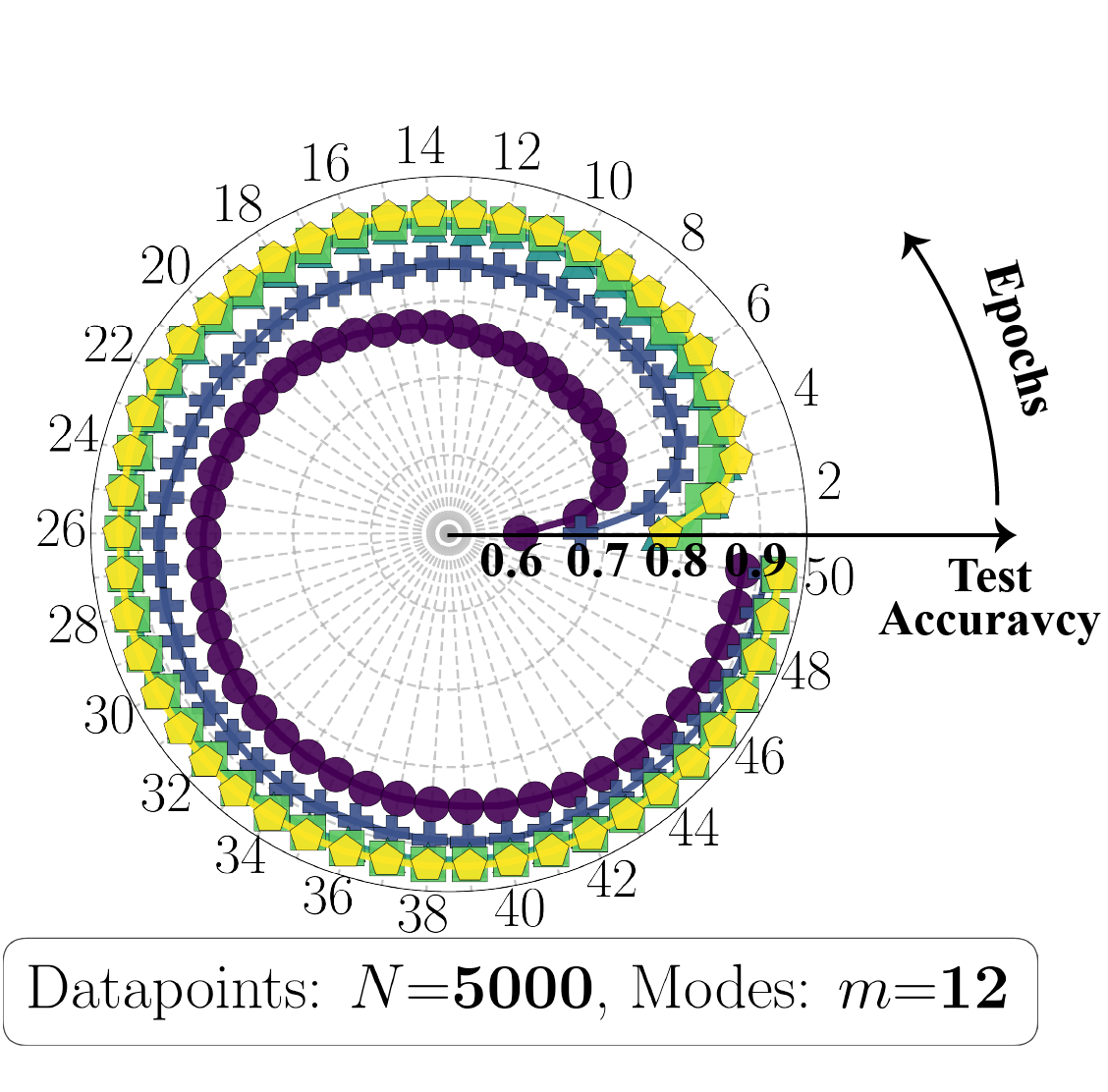}
				\put(4,90){\textbf{(c)}}
			\end{overpic}
			\label{fig:MNIST_epochs}
		\end{subfigure}
		\begin{subfigure}{.23\textwidth}
			\begin{overpic}[width=\textwidth]{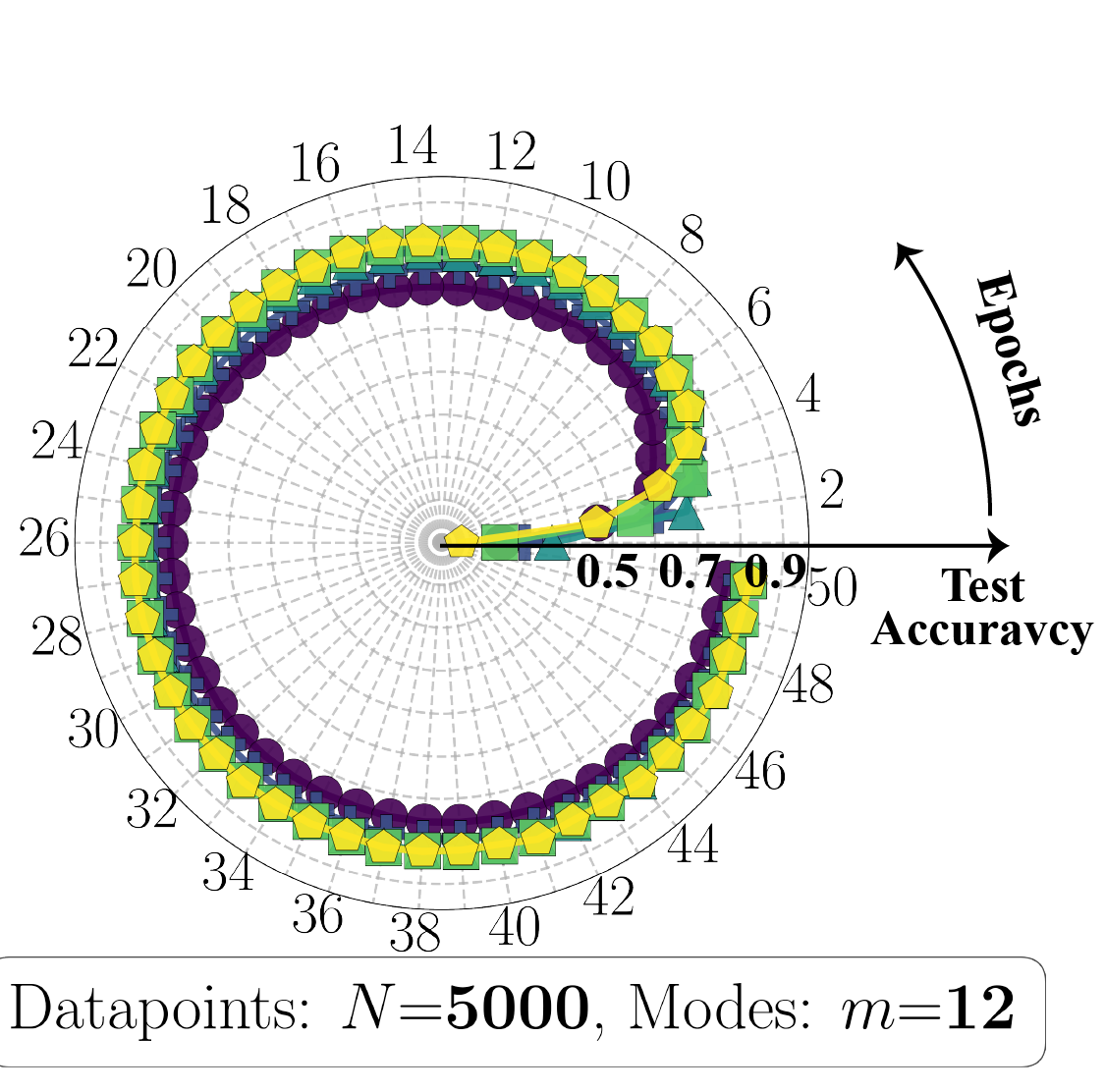}
				\put(2,90){\textbf{(d)}}
			\end{overpic}
			\label{fig:fashionMNIST_epochs}
		\end{subfigure}
		\vspace{-.4cm}
		
		\caption{Evolution of test accuracy during neural network training. The panels show the test accuracy as a function of training epoch for a) Ionosphere, b) Spambase, c) MNIST, and d) Fashion-MNIST datasets. }
		\label{fig:epochs}
	\end{figure*}

	As a benchmark, we can compare the accuracy of the proposed boson sampling based quantum kernel with classical linear and nonlinear sigmoid  kernels which are directly applied to the original data points $\{\mathbf{x}_j\}$, as opposed to the quantum kernel which is used for $\{ \tilde{\mathbf{x}}_j\}$. Applying the classical algorithms to the original data avoids any information loss due to feature mapping and results in a better accuracy. 
	As another benchmark, to ensure that any performance improvement is not solely attributed to the neural network itself, we also evaluate the accuracy of the corresponding classical neural network  without the quantum kernel. The classical neural network is trained directly for label prediction using the cross-entropy loss function. As described earlier, the neural network used in our quantum model has a two linear layer architecture (without any hidden layer), where the input layer neurons are equal to the number of features in the dataset, and the output layer matches the number of tunable parameters in the boson sampling circuit. Accordingly, for the classical benchmark, we employ a fully connected three layer classical neural network, the first layer corresponds to the dataset features, the second layer to the circuit parameters, and the final layer to the number of classes, consistent with the cross entropy objective. This configuration allows the neural network to be used for solving the classification problem while keeps its complexity equivalent to the one which is used for feature mapping in our protocol.  \\

	\noindent\textbf{Interpretability and rationale for high accuracy}
	
	\noindent The  protocol has a clear interpretation at both the algorithmic and physical levels. The classical neural network in our hybrid model functions as a feature adapter that maps raw data $\{ \mathbf{x}_j \}$ into a latent representation $\{ \tilde{\mathbf{x}}_j\}$ compatible with the boson sampling circuit. By encoding the latent information into the boson sampling circuit $U(\tilde{\mathbf{x}}_j)$, the information is encoded in a quantum state $U(\tilde{\mathbf{x}}_j)\ket{\phi}$ whose Hilbert space dimension $\mathcal{D}_\mathrm{boson}$ may even be larger than the original image size. This provides a higher-dimensional embedding that can enhance class separability despite the preceding feature transformation. The choice of the cost function  in Eq.~(\ref{eq:nncost}) implies that  the quantum states $U(\tilde{\mathbf{x}}_i)\ket{\phi}$ and $U(\tilde{\mathbf{x}}_j)\ket{\phi}$ become orthogonal (i.e. $K(\tilde{\mathbf{x}}_i,\tilde{\mathbf{x}}_j){\sim}0$) if the input data points $\tilde{\mathbf{x}}_i$ and $\tilde{\mathbf{x}}_j$ belong to different classes. On the other hand, if the two data points belong to the same class the hybrid classical-quantum procedure forces the two corresponding quantum states have a large overlap (i.e. $K(\tilde{\mathbf{x}}_i,\tilde{\mathbf{x}}_j){\sim}1$). This interpretation is schematically shown in Fig.~\ref{fig:schematic}(b). In fact, the huge dimension of the Hilbert space allows to have multiple islands  in which the quantum states have large overlaps while they remain orthogonal to quantum states from other islands. This allows to reach high accuracy in multi-categories classification problems. In addition, our protocol is naturally trained to be resistive against kernel concentration. This is imposed by the cost function,   Eq.~(\ref{eq:nncost}), in which imposes the kernel to be near 1 (0) for the two images  $\tilde{\mathbf{x}}_i$, and $\tilde{\mathbf{x}}_j$ from the same (different) classes. 
	\\

	% \section*{Numerical Simulations}
	
	\noindent\textbf{Datasets and Initialization}\\
	\noindent We use four different datasets, whose details are given in table~\ref{tab:merged_datasets}, to evaluate the performance of our protocol. The datasets represent both binary and multi-class problems: Ionosphere and spambase are for binary classification, while MNIST and Fashion-MNIST have $10$ classes each.
	For multiclass datasets the one-vs-rest technique is used in SVM. In this approach, a separate binary SVM is done for each class, treating all other classes as a single combined class. During prediction, the test data is assigned to the class whose classifier outputs the highest decision score. All input raw data $\mathbf{x}_i$ were normalized to be in the interval of $[0{,}1]$  using min-max normalization before being fed in the neural network.  The features at the output of the neural network, $\tilde{\textbf{x}}_i$, are also between $[0{,}1]$. Then to be applied on the phase shifters, they were scaled by $\pi/2$ and $2\pi$ and assigned as $\theta$ and $\varphi$, respectively,  (see Eq.~\eqref{eq:TBU_matrix}). For the binary datasets, a $1/4$ test-to-train ratio was used and for MNIST and Fashion-MNIST the standard $1{:}6$ split was applied.  In the photonic circuit, each mode contains at most one photon, and all photons are injected in the first modes at the input. For instance in a setup with $m{=}5$ modes and $n{=}2$ photons, the initial state is $\ket{\phi}{=}\ket{1,1,0,0,0}$. Note that the number of photons detected at the output is equal to the number of photons at the input. All of the results are obtained by averaging over at least five times of running the model.
	\\
		\begin{figure*}[t]
		\centering
		
		% =========================
		% Left side: plots (two rows)
		% =========================
		\begin{minipage}[c]{.8\textwidth}
			\centering
			
			% ----- Row 1 -----
			\begin{subfigure}{.39\textwidth}
				\begin{overpic}[width=\textwidth]{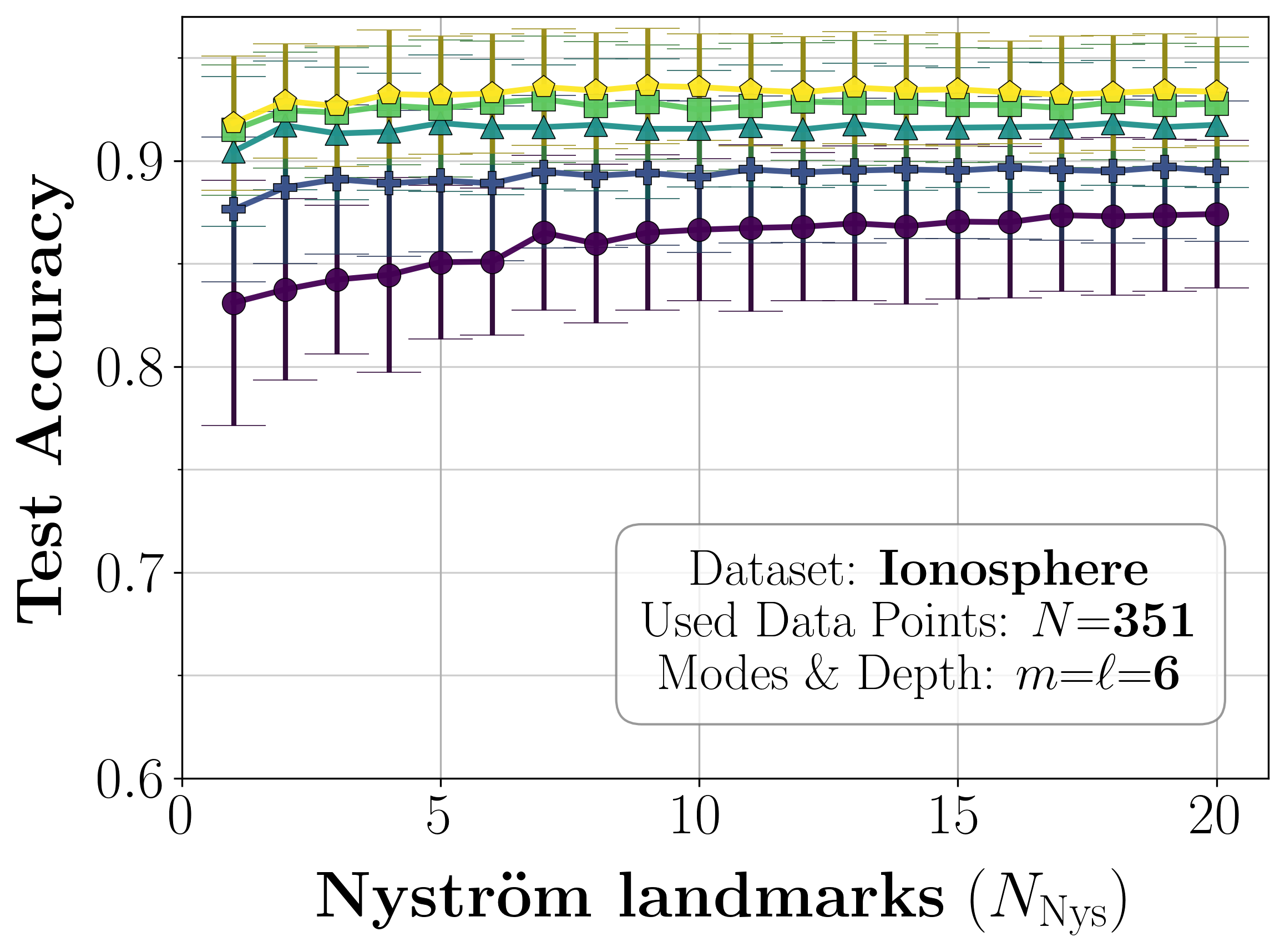}
					\put(-3,72){\textbf{(a)}}
				\end{overpic}
				%	\label{fig:ion_readout}
			\end{subfigure}
			\hspace{0.04\textwidth}
			\begin{subfigure}{.39\textwidth}
				\begin{overpic}[width=\textwidth]{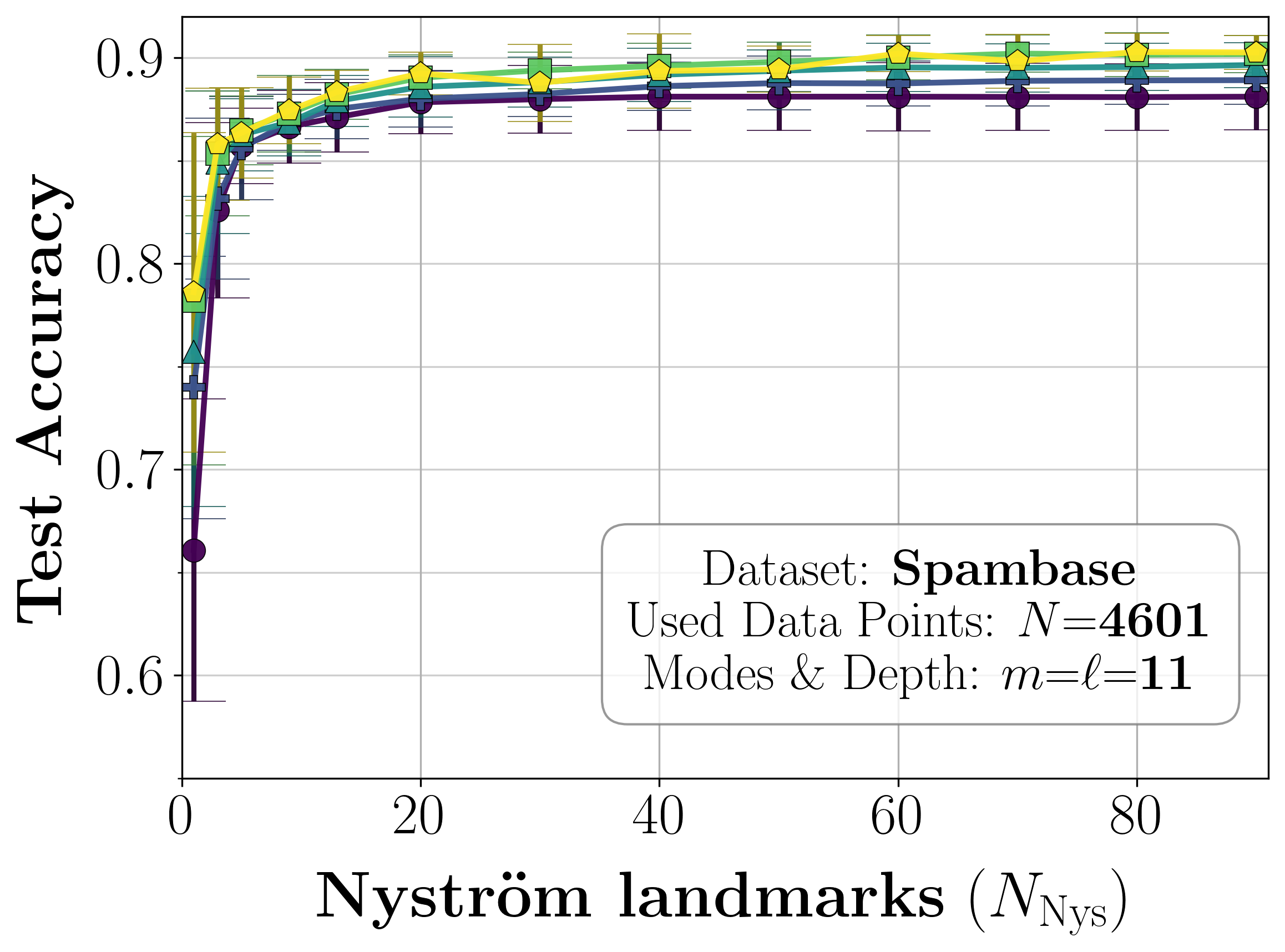}
					\put(-3,72){\textbf{(b)}}
				\end{overpic}
				%	\label{fig:spam_readout}
			\end{subfigure}
			
			\vspace{0.3cm}
			
			% ----- Row 2 -----
			\begin{subfigure}{.39\textwidth}
				\begin{overpic}[width=\textwidth]{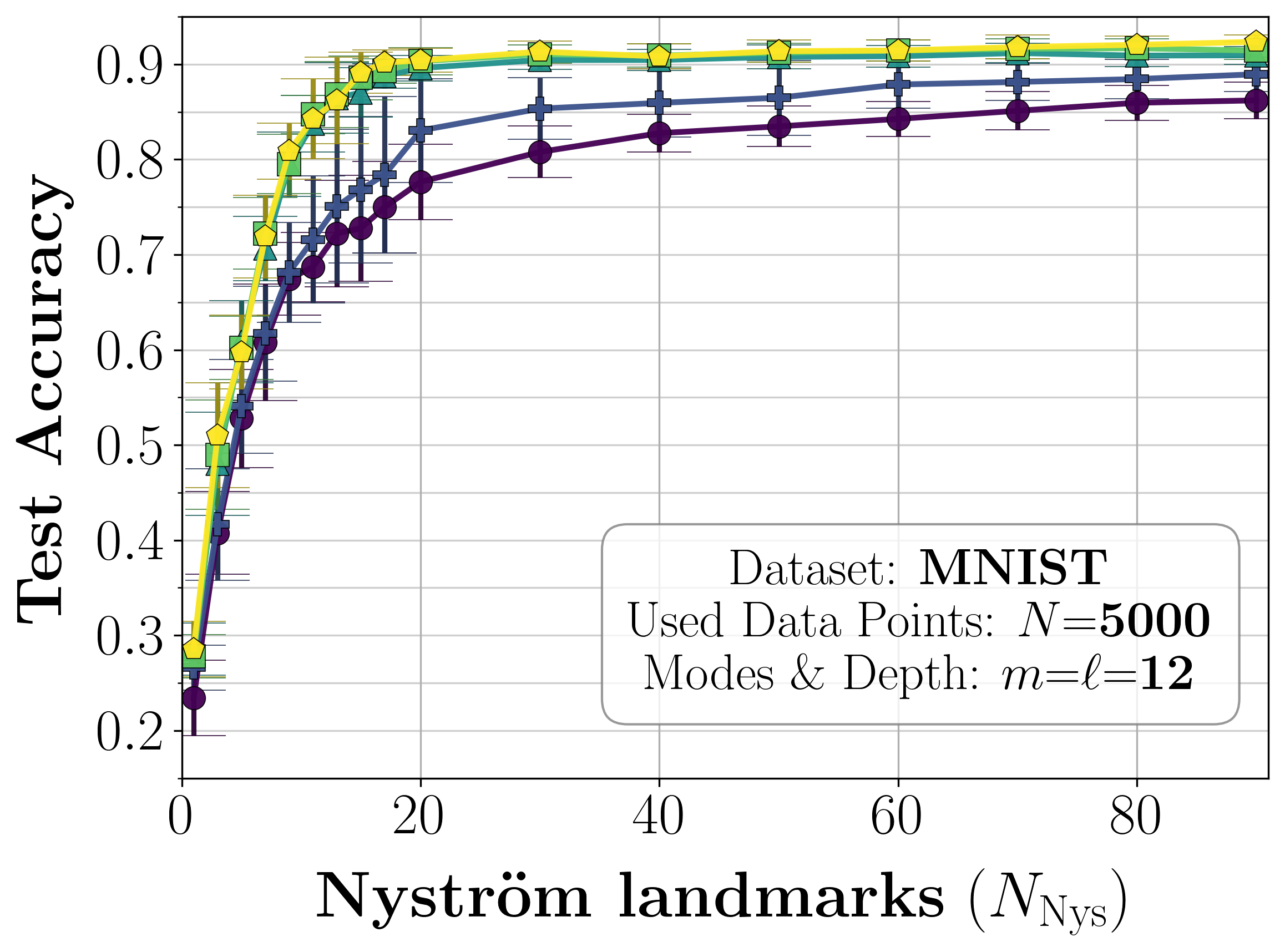}
					\put(-3,72){\textbf{(c)}}
				\end{overpic}
				%	\label{fig:MNIST_readout}
			\end{subfigure}
			\hspace{0.04\textwidth}
			\begin{subfigure}{.39\textwidth}
				\begin{overpic}[width=\textwidth]{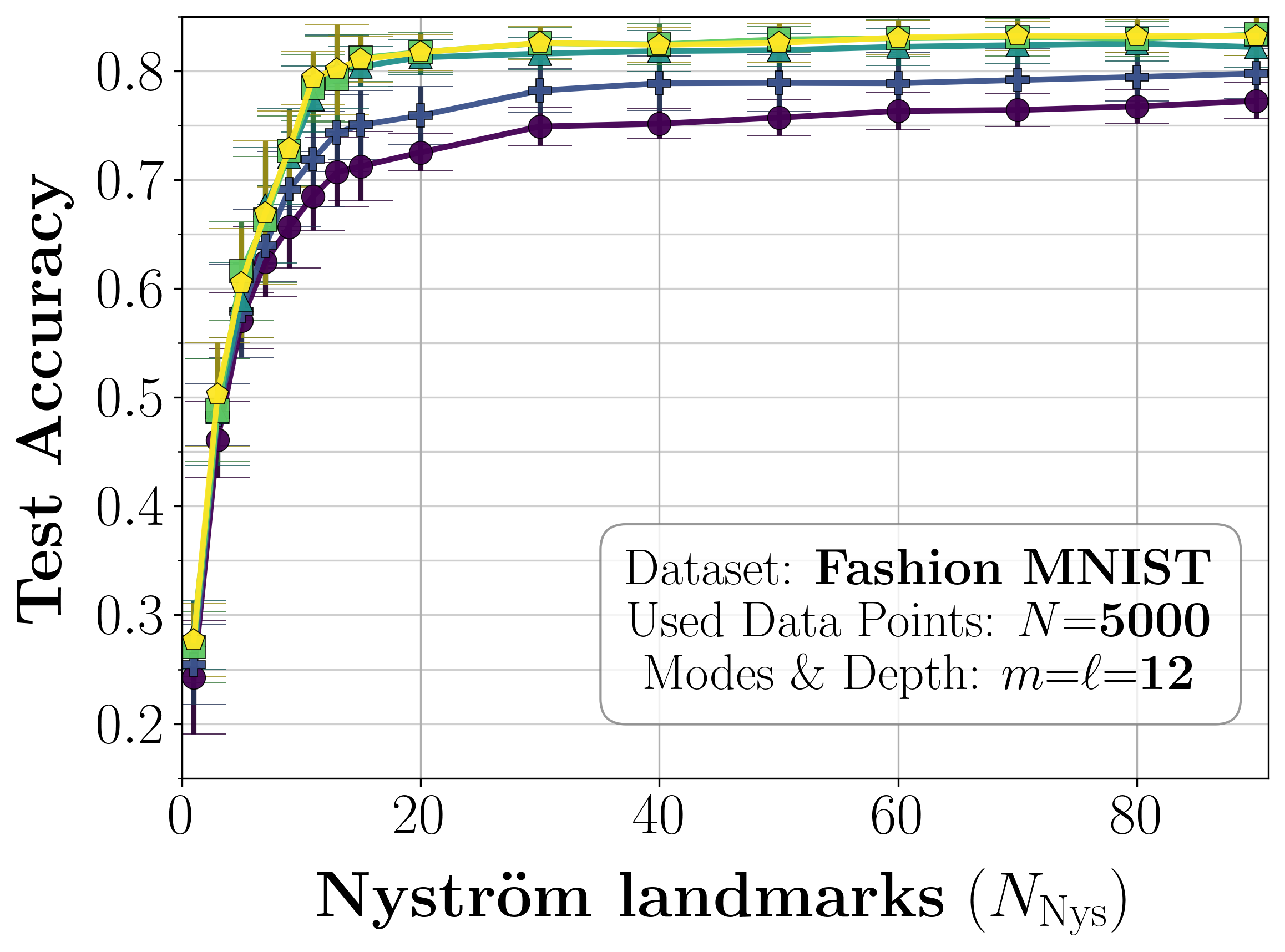}
					\put(-3,72){\textbf{(d)}}
				\end{overpic}
				%	\label{fig:fashion_readout}
			\end{subfigure}
		\end{minipage}
		\hspace{-0.04\textwidth}
		% =========================
		% Right side: tall legend
		% =========================
		\begin{minipage}[c]{.1\textwidth}
			\centering
			\includegraphics[width=\textwidth]{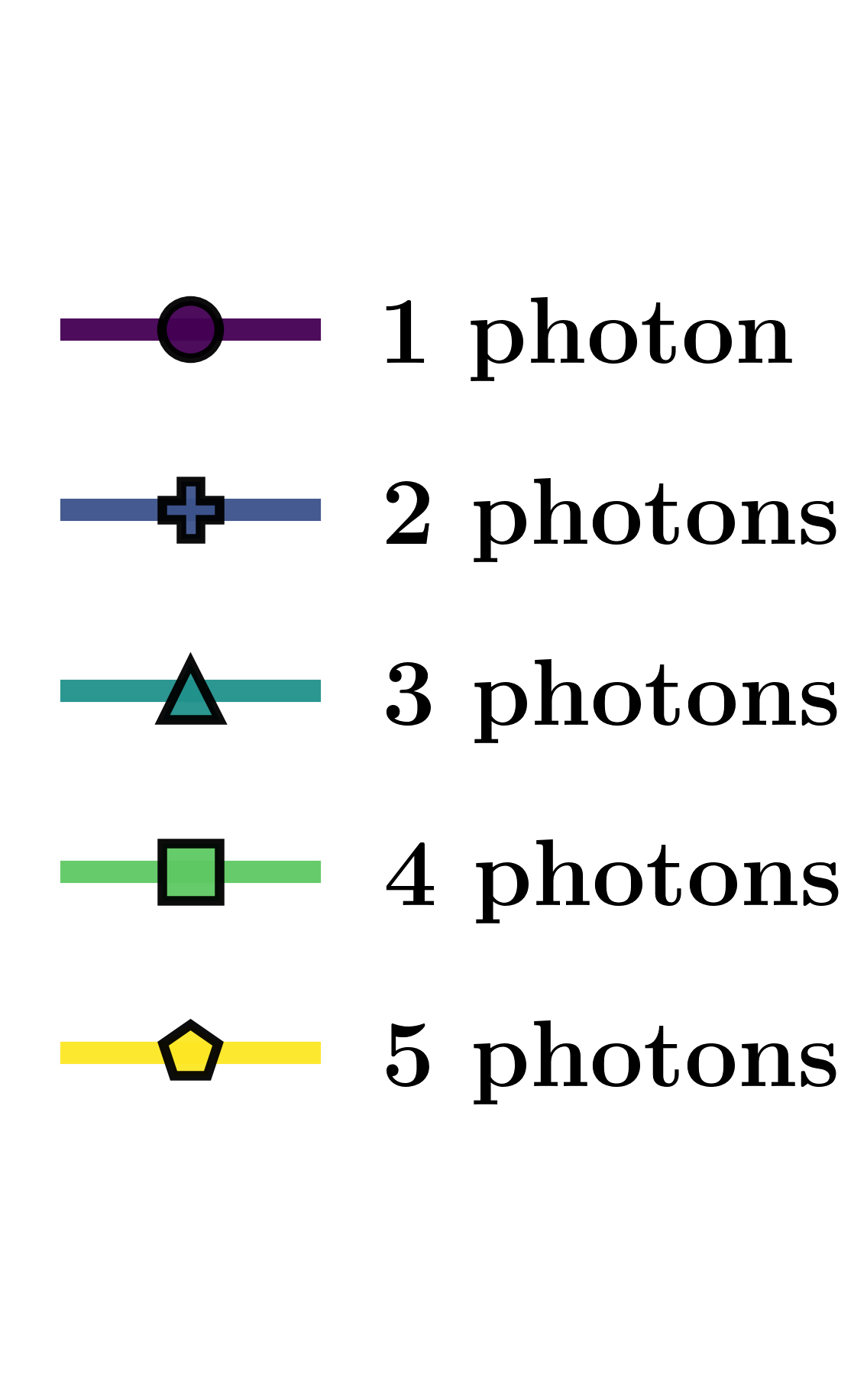}
		\end{minipage}
		\vspace{5pt}
		% ===== Caption =====
		\caption{Test accuracy of the model using fewer landmark points with Nystr\"{o}m approximation. The panels show the results for a) Ionosphere, b) Spambase, c) MNIST, and d) Fashion-MNIST datasets with different numbers of photons.}
		
		\label{fig:Nystrom}
	\end{figure*}

	\noindent\textbf{Effect of photon number on accuracy}
	
	\noindent First, we evaluate the framework by increasing the number of injected indistinguishable photons $n$ into the boson sampling circuit, while keeping the circuit depth equal to the number of modes  (i.e. $\ell{=}m$). Figs.~\ref{fig:acc_photons}(a)-(d) illustrate the effect of increasing the photon number on model accuracy for all the four datasets, respectively. In order to build the kernel matrix $K(\tilde{\mathbf{x}}_i,\tilde{\mathbf{x}}_j)$, we use $N{=}351$ data points for the Ionosphere, $N{=}4601$ data points for the spambase, $N{=}5{\times}10^3$ data points for the MNIST and $N{=}5{\times}10^3$ for the Fashion-MNIST datasets. As evidenced in Fig.~\ref{fig:acc_photons}, by increasing the  number of indistinguishable photons $n$, the accuracy of the quantum model consistently improves. This enhancement ultimately enables the boson-sampling-based quantum kernel to surpass both classical kernels (linear and sigmoid) and the pure classical neural network baseline, demonstrating robust performance on both low- and high-dimensional data. Specifically, using five photons, the quantum model achieves test accuracies of $93.24{\pm}2.1\%$ with $m{=}6$ modes on the full Ionosphere dataset, $90.84{\pm}0.9\%$ with $m{=}11$ modes on the full Spambase dataset, $92.94{\pm}0.5\%$ with $m{=}12$ modes on $N{=}5000$ data points from MNIST, and $82.89{\pm}0.8\%$ with $m{=}12$ modes on $N{=}5000$ data points from Fashion-MNIST.
	Table~\ref{tab:merged_datasets} summarizes the accuracies and corresponding parameters for each dataset and compares them with classical ones. To ensure experimental feasibility of the programmable boson sampling circuit, we confine the system to a maximum of twelve modes and five photons~\cite{Hoch2022reconfigurable,Yin2024experimental,Anguita2025experimental,Taballione2020a,Broome2013photonic,Taballione2023_20mode}. 
	\\

	\noindent\textbf{Impact of circuit depth and Hilbert space dimension}

	\noindent By increasing the number of modes and photons in a bosonic quantum system, the dimension of the corresponding Hilbert space grows rapidly. This expansion in Hilbert space helps to have a richer representation of quantum states, and more complex hyper planes in SVM classification~\cite{Peters2021machine, Jolly2025harnessing}. In Fig.~\ref{fig:Hilbert_depth}a, we plot the accuracy of the MNIST dataset, as the representative of the datasets, as a function of the Hilbert space dimension using $N{=}5000$ data points. In order to vary the dimension of the Hilbert space, we change the number of photons $n$ and the number of modes $m$, while keeping the circuit depth equal to the number of modes $\ell{=}m$. As the figure shows, increasing the dimension of Hilbert space positively influences the test accuracy of the model.

	The effect of Hilbert-space dimensionality can also be examined by varying the number of modes $m$ and the circuit depth $\ell$ independently. To see this effect, in Fig.~\ref{fig:Hilbert_depth}b we present the achievable accuracy of the MNIST dataset, as the representative model, when the  number of photons is fixed to $n{=}3$ and instead the circuit depth $\ell$ and the number of modes $m$ are varied.
	By increasing $m$, the dimension of the Hilbert space increases, and thus the obtainable accuracy is enhanced, which is evident in Fig.~\ref{fig:Hilbert_depth}b. On the other hand, increasing the circuit depth $\ell$ enhances the expressivity of the unitary transformation, enabling stronger mode-mode correlations. In our circuit design (Fig.~\ref{fig:schematic}a), establishing full connectivity between the first and last modes requires a minimum depth of $\ell {\ge} m{-}1$, as indicated by the red line in Fig.~\ref{fig:Hilbert_depth}b. 		
	\\

	\noindent\textbf{Training resource analysis}
	
	\noindent In the training part of the protocol, the  time needed for accomplishing the task is dominantly determined by two major factors,  namely the size of the kernel matrix  as well as the number of epochs required for the training of the neural network. Here, we investigate these factors and potential ways to optimize them.

	The number of epochs that are used for training the classical neural network accounts for the complexity of the optimization in the protocol. We examine the impact of training epochs on classification accuracy. The growth of test accuracy with respect to the training epochs is illustrated in Figs.~\ref{fig:epochs}(a)-(d) for the four datasets. One can use fewer epochs if a specific accuracy threshold is reached in the training. 
	
			\begin{figure*}[t]
		\centering
		
		% =========================
		% Left side: plots (two rows)
		% =========================
		\begin{minipage}[c]{.8\textwidth}
			\centering
			
			% ----- Row 1 -----
			\begin{subfigure}{.39\textwidth}
				\begin{overpic}[width=\textwidth]{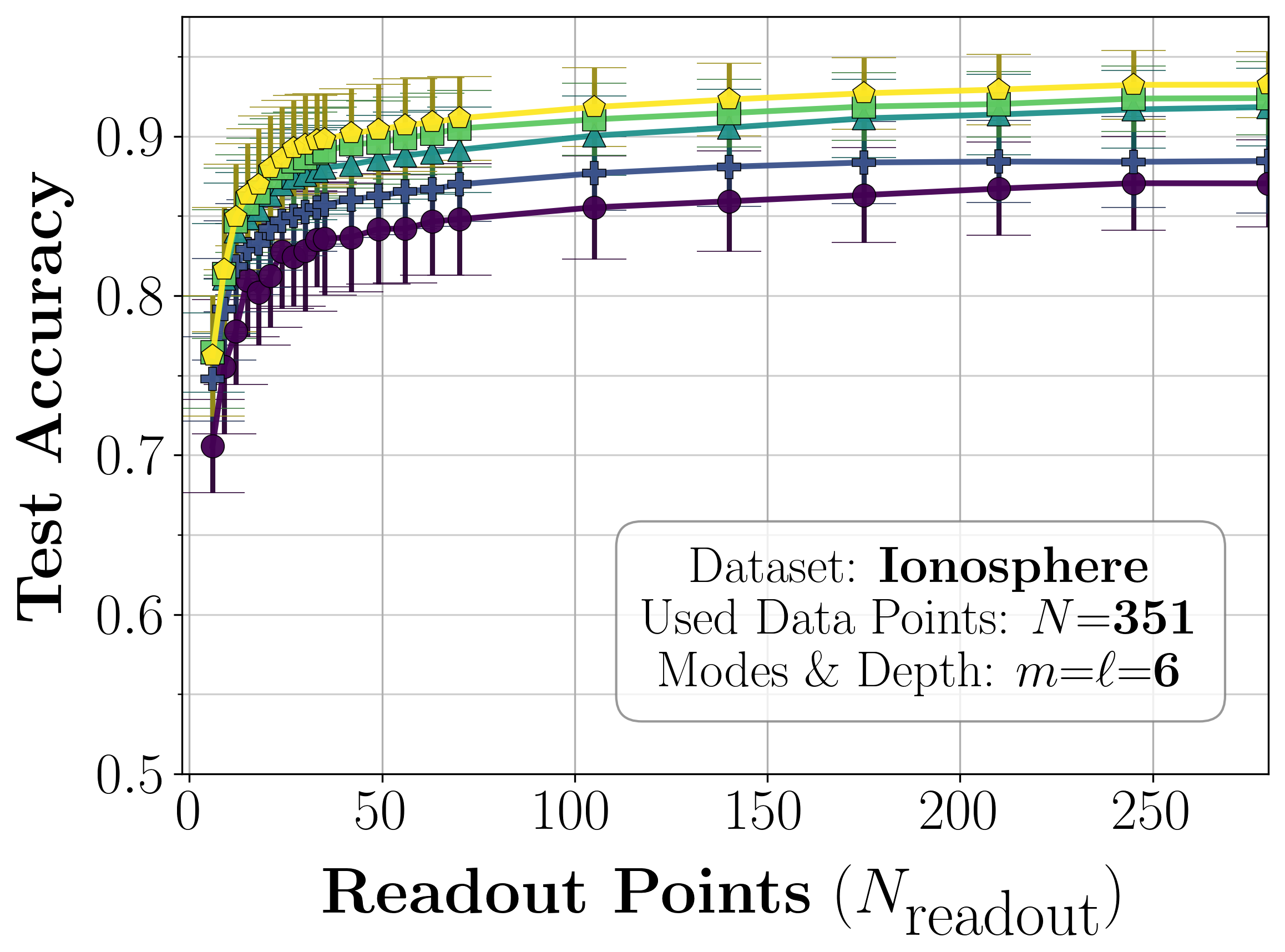}
					\put(-3,72){\textbf{(a)}}
				\end{overpic}
				\label{fig:ion_readout}
			\end{subfigure}
			\hspace{0.04\textwidth}
			\begin{subfigure}{.39\textwidth}
				\begin{overpic}[width=\textwidth]{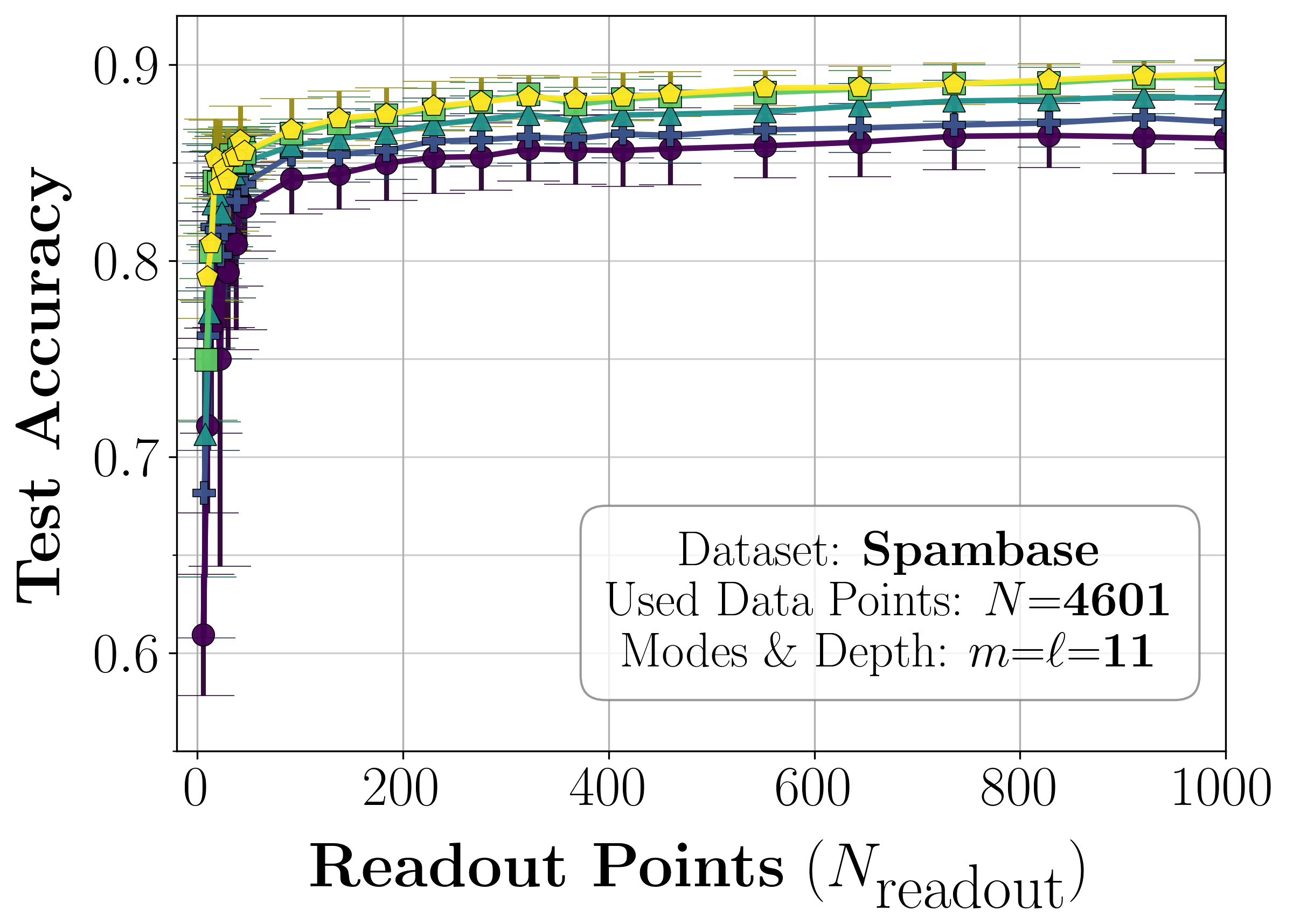}
					\put(-3,72){\textbf{(b)}}
				\end{overpic}
				\label{fig:spam_readout}
			\end{subfigure}
			
			\vspace{0.3cm}
			
			% ----- Row 2 -----
			\begin{subfigure}{.39\textwidth}
				\begin{overpic}[width=\textwidth]{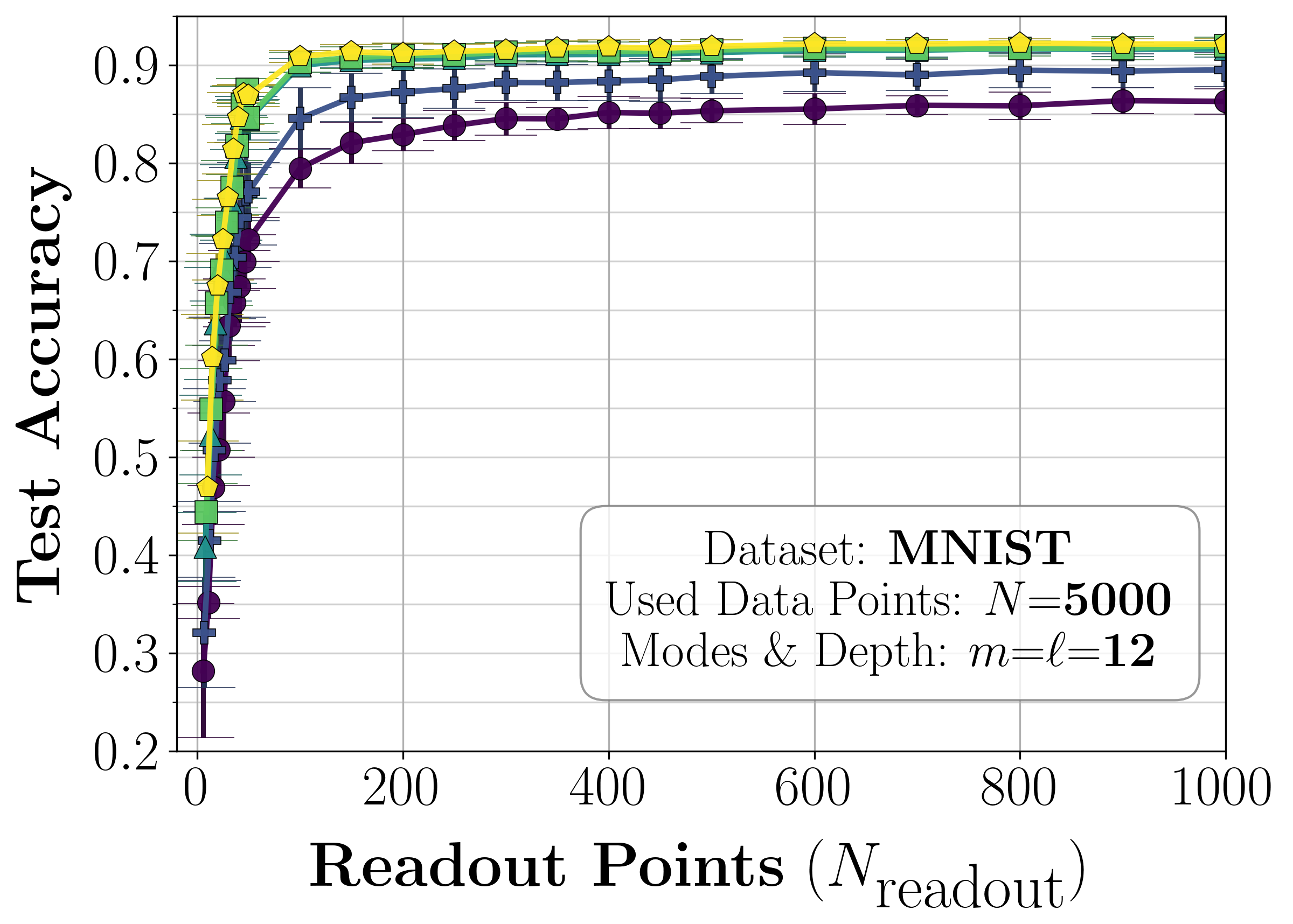}
					\put(-3,72){\textbf{(c)}}
				\end{overpic}
				\label{fig:MNIST_readout}
			\end{subfigure}
			\hspace{0.04\textwidth}
			\begin{subfigure}{.39\textwidth}
				\begin{overpic}[width=\textwidth]{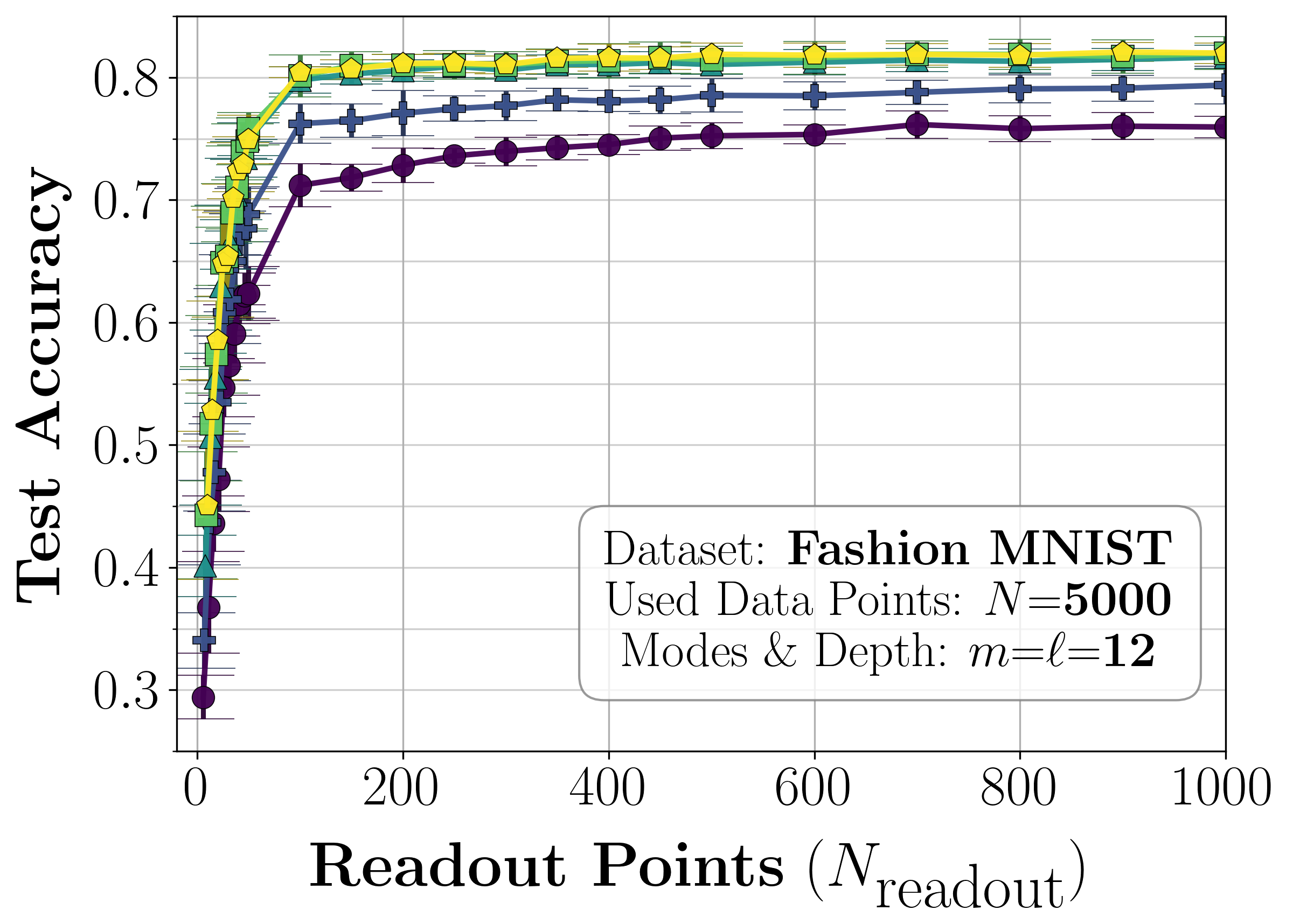}
					\put(-3,72){\textbf{(d)}}
				\end{overpic}
				\label{fig:fashion_readout}
			\end{subfigure}
		\end{minipage}
		\hspace{-0.04\textwidth}
		% =========================
		% Right side: tall legend
		% =========================
		\begin{minipage}[c]{.1\textwidth}
			\centering
			\includegraphics[width=\textwidth]{figures/accuracy_readout_legend.png}
		\end{minipage}
		
		% ===== Caption =====
		\caption{
			Test accuracy of the model as a function of the number of readout points. The panels show the results for a) Ionosphere, b) Spambase, c) MNIST, and d) Fashion-MNIST datasets with different numbers of photons. }
		
		\label{fig:readout}
	\end{figure*}

	\noindent So far, for the training part we build the kernel matrix using a large number of data points $N_\mathrm{train}$. Since the Kernel matrix is a square matrix it requires $N_\mathrm{train}^2$ evaluations. This makes the training very time consuming. Here, we show that one can indeed accomplish the training task by using far fewer evaluations using  Nystr\"{o}m approximation~\cite{Williams2000using}. In this approach, instead of computing the full $N_{\mathrm{train}}{\times} N_{\mathrm{train}}$ kernel matrix, we pick a small set of landmark training points $X_{\mathrm{Nys}}$ with $N_{\mathrm{Nys}}{\ll} N_{\mathrm{train}}$ and only evaluate (i) the landmark--landmark kernel block $W{=}K(X_{\mathrm{Nys}},X_{\mathrm{Nys}})$ and (ii) the training--landmark kernel block $C_{\mathrm{tr}}{=}K(X_{\mathrm{train}},X_{\mathrm{Nys}})$. The training kernel matrix is then approximated by
	\begin{equation}
		\widehat K_{\mathrm{tr}} \;=\; C_{\mathrm{tr}}\,W^{+}\,C_{\mathrm{tr}}^{\top},
	\end{equation}
	where $W^{+}$ denotes the Moore--Penrose pseudoinverse of $W$. This approach reduces the number of kernel evaluations from $N_{\mathrm{train}}^{2}$ to $N_{\mathrm{train}}N_{\mathrm{Nys}}$. We apply this approach to all datasets. Fig.~\ref{fig:Nystrom}(a)-(d) show that, across all datasets, the test accuracy rapidly reaches its maximum as $N_{\mathrm{Nys}}$ increases, indicating that evaluating the full kernel matrix is unnecessary. The Ionosphere dataset has the lowest number of data points so its accuracy saturates faster than other datasets (see Fig.~\ref{fig:Nystrom}(a)). For MNIST and Fashion-MNIST in Figs.~\ref{fig:Nystrom}(c)-(d), using only $N_{\mathrm{Nys}}{=}20$ landmark training points with an $n{=}3$ photon configuration already achieves near-maximal accuracy, while requiring kernel evaluations for only ${\approx} 0.47\%$ of all training pairs ($N_{\mathrm{Nys}} / N_{\mathrm{train}}$). \\

	\noindent\textbf{Classification resource analysis}
	
	\noindent After training, the classification of a new test data point using an SVM classifier requires computing the kernel values between the new test data point and all training data points, (see Eq.~(\ref{eq:quantum_decision}). From a practical perspective, however, this is very costly and resource consuming. Hence, it is very desirable to achieve the label allocation with a  smaller subset of training points. Let's consider the number of readout points that are used for label allocation to be $N_\text{readout}{<}N_\text{train}$. In other words,  in Eq.~(\ref{eq:quantum_decision}), the kernel is only computed between the new test point and $N_\text{readout}$ random data points  in the training set. We implemented this subset-based SVM scheme and plot the achievable accuracy as a function of the number of readout points $N_\text{readout}$ for various photon counts in Figs.~\ref{fig:readout}(a)-(d), for all the four datasets respectively. The results indicate that one can reach a desired accuracy using a small fraction of the training data points instead of all of them. Interestingly, increasing the number of photons $n$ further reduces the required readout points. \\
		\\
		
			\begin{figure*}[t]
			\centering
			
			% ===== Legend =====
			\begin{subfigure}{.35\textwidth}
				\centering
				\includegraphics[width=\textwidth]{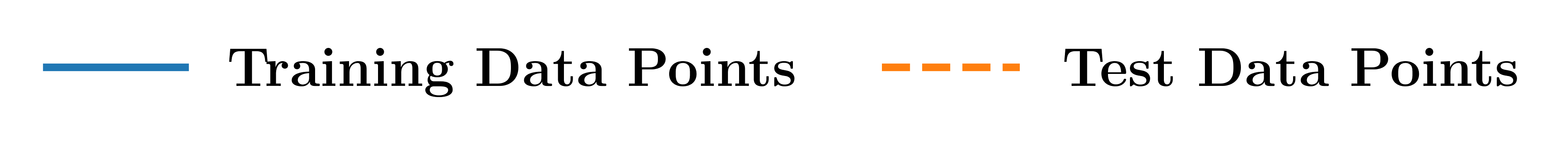}
				\vspace{.1cm}
			\end{subfigure}
			
			% =========================
			% Row 1: SAME-CLASS
			% =========================
			\begin{subfigure}{.23\textwidth}
				\begin{overpic}[width=\textwidth]{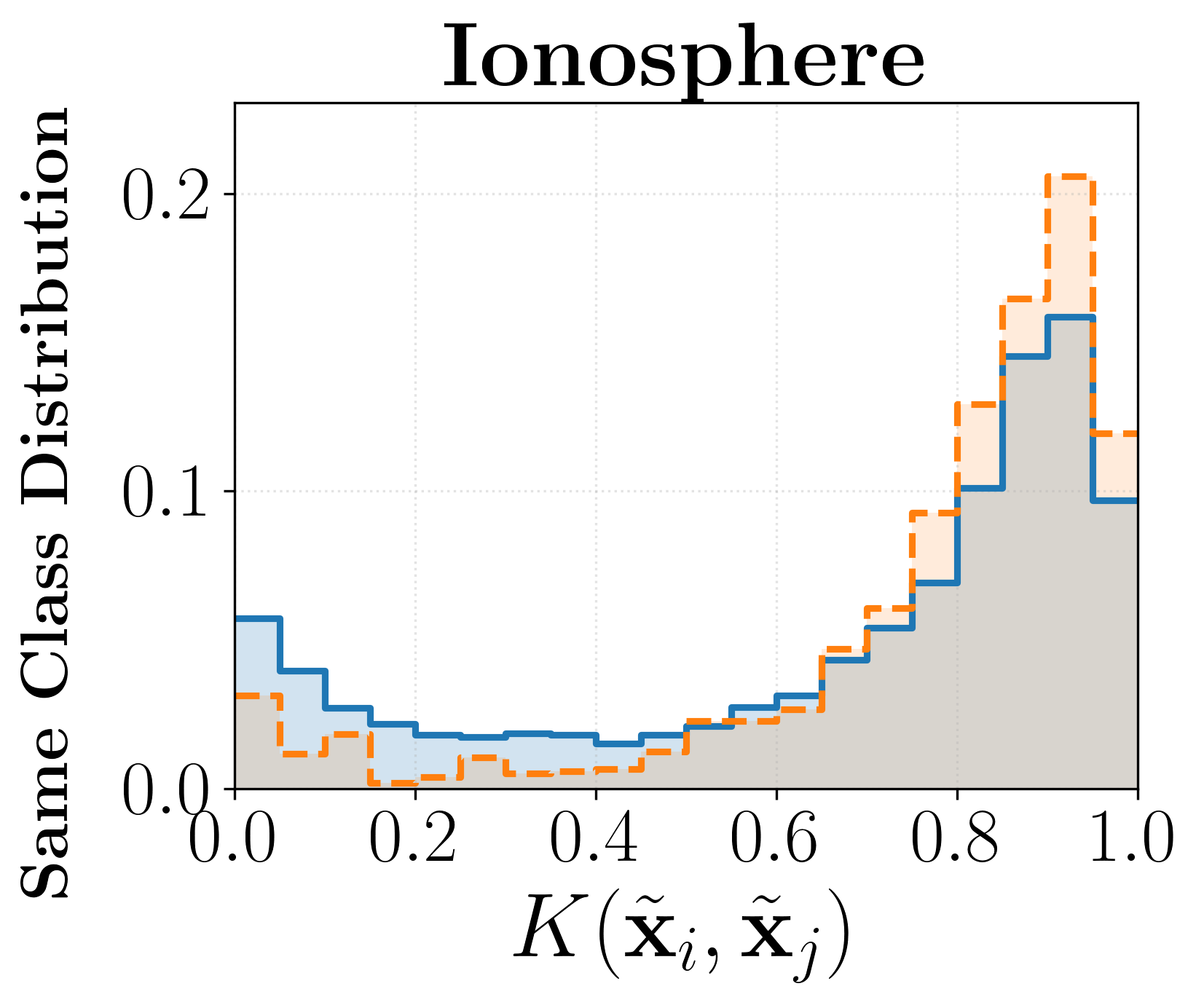}
					\put(4,90){\textbf{(a)}}
				\end{overpic}
				\label{fig:ion_hist_same}
			\end{subfigure}
			\begin{subfigure}{.23\textwidth}
				\begin{overpic}[width=\textwidth]{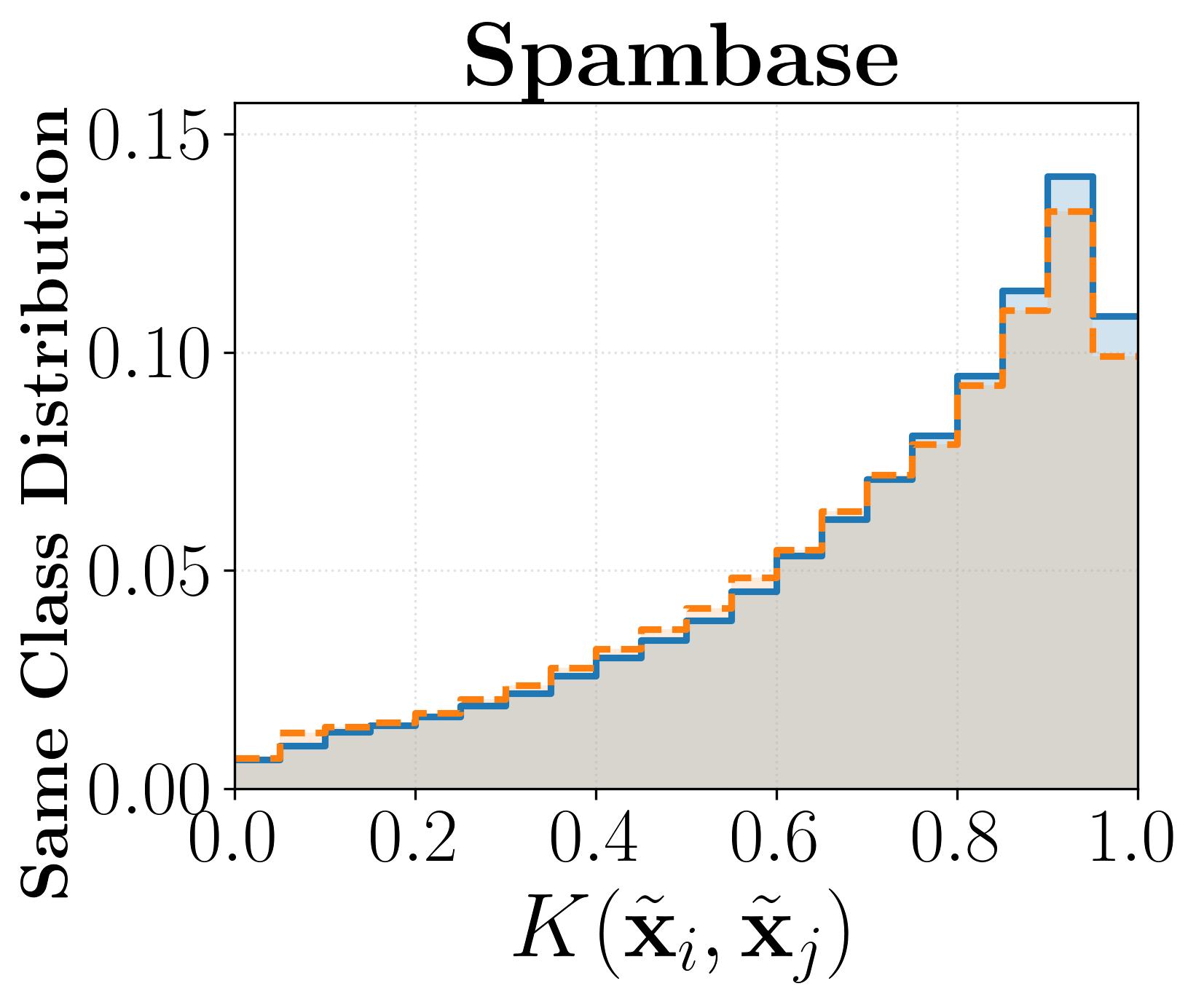}
					\put(4,90){\textbf{(c)}}
				\end{overpic}
				\label{fig:spam_hist_same}
			\end{subfigure}
			\begin{subfigure}{.23\textwidth}
				\begin{overpic}[width=\textwidth]{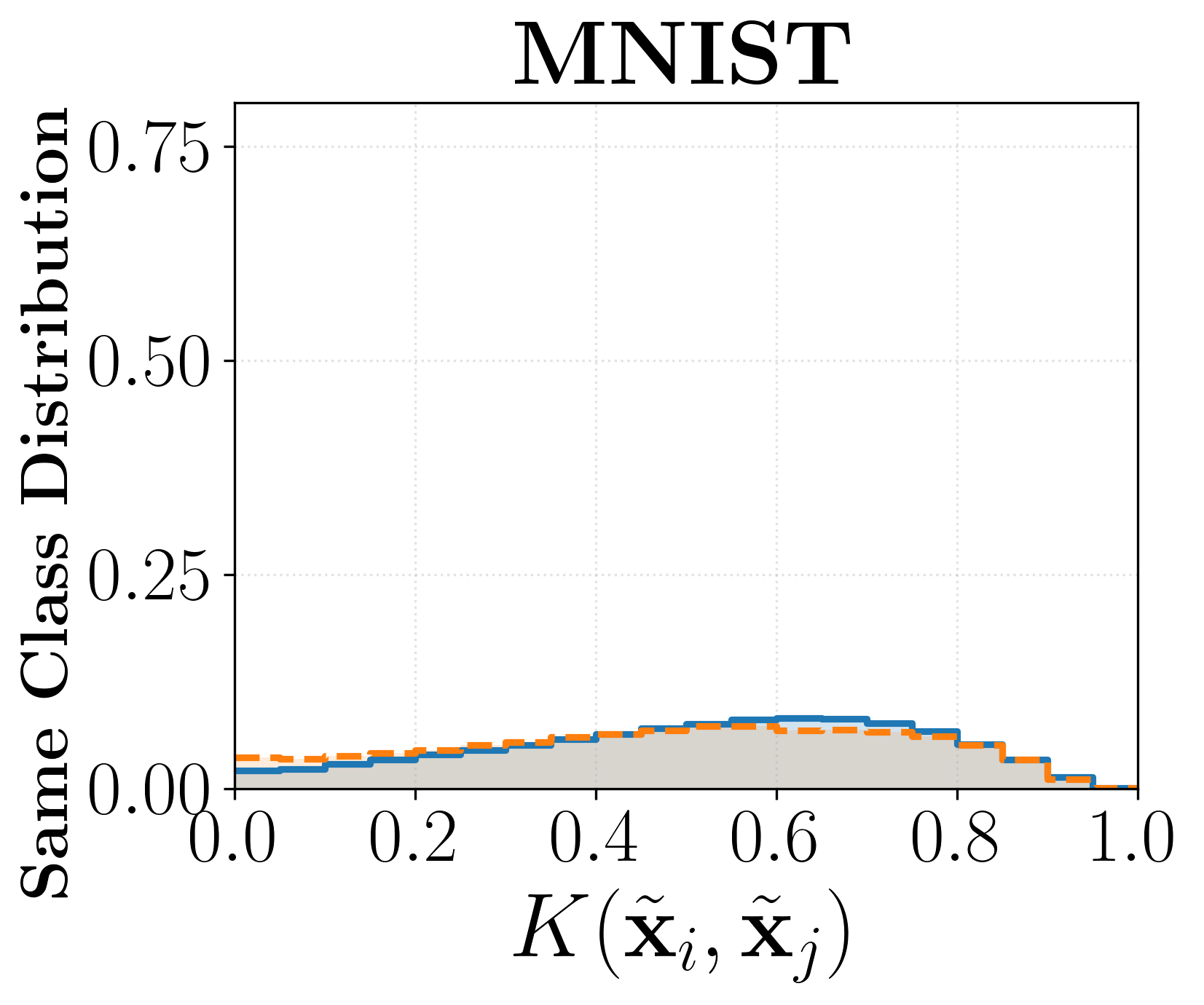}
					\put(4,90){\textbf{(e)}}
				\end{overpic}
				\label{fig:MNIST_hist_same}
			\end{subfigure}
			\begin{subfigure}{.23\textwidth}
				\begin{overpic}[width=\textwidth]{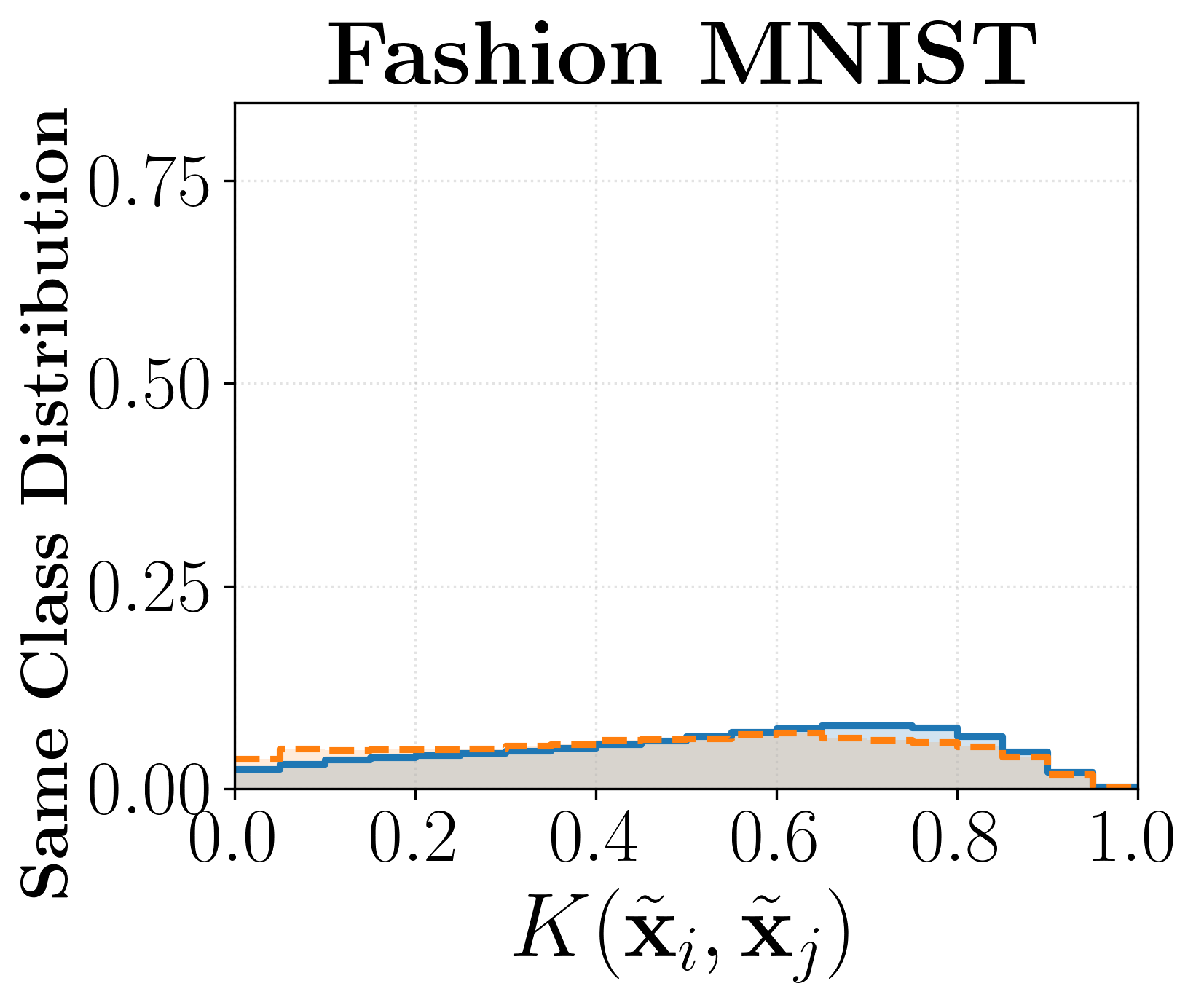}
					\put(4,90){\textbf{(g)}}
				\end{overpic}
				\label{fig:fashion_hist_same}
			\end{subfigure}
			
			\vspace{0.35cm}
			
			% =========================
			% Row 2: DIFFERENT-CLASS
			% =========================
			\begin{subfigure}{.23\textwidth}
				\begin{overpic}[width=\textwidth]{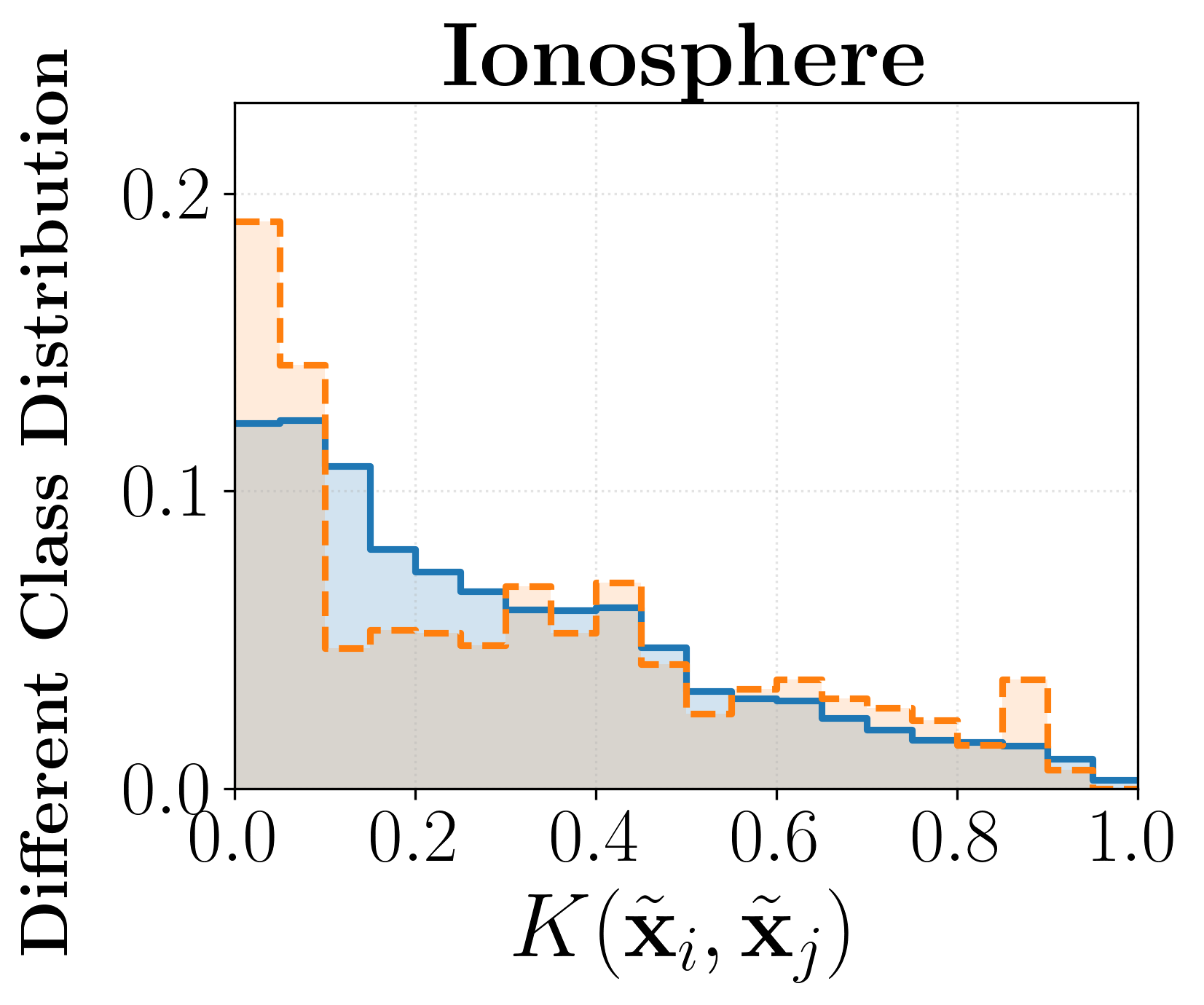}
					\put(4,90){\textbf{(b)}}
				\end{overpic}
				\label{fig:ion_hist_diff}
			\end{subfigure}
			\begin{subfigure}{.23\textwidth}
				\begin{overpic}[width=\textwidth]{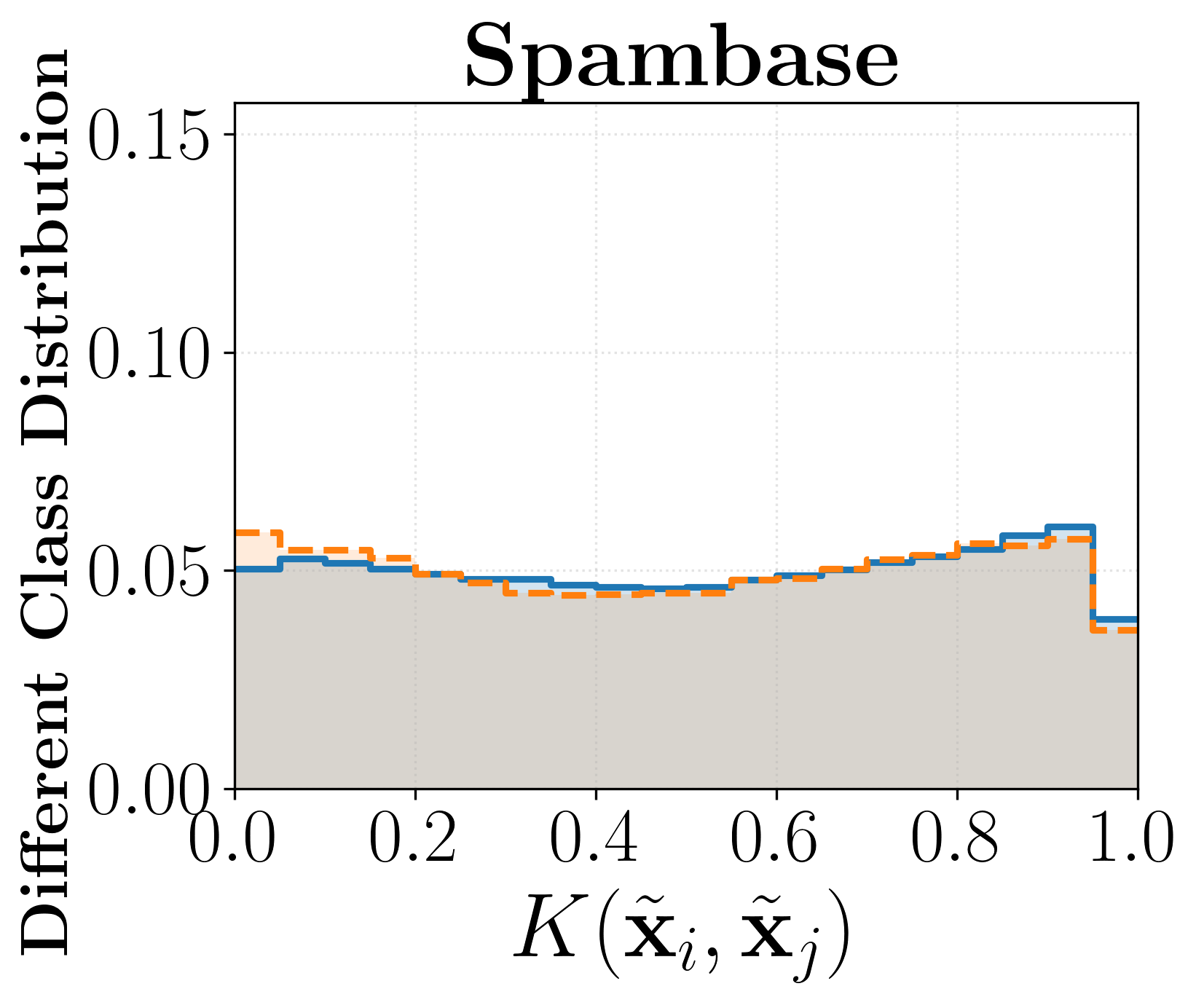}
					\put(4,90){\textbf{(d)}}
				\end{overpic}
				\label{fig:spam_hist_diff}
			\end{subfigure}
			\begin{subfigure}{.23\textwidth}
				\begin{overpic}[width=\textwidth]{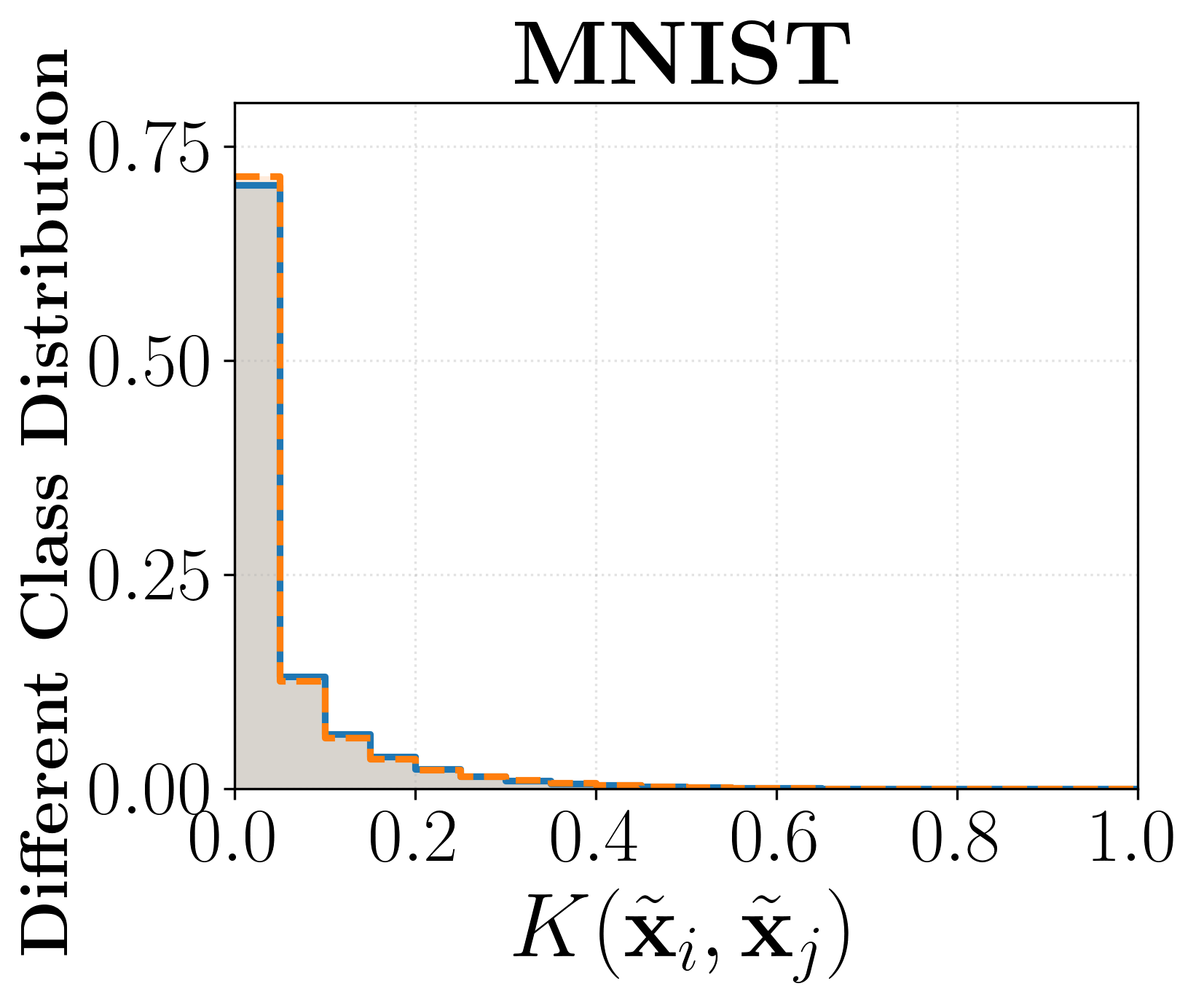}
					\put(4,90){\textbf{(f)}}
				\end{overpic}
				\label{fig:MNIST_hist_diff}
			\end{subfigure}
			\begin{subfigure}{.23\textwidth}
				\begin{overpic}[width=\textwidth]{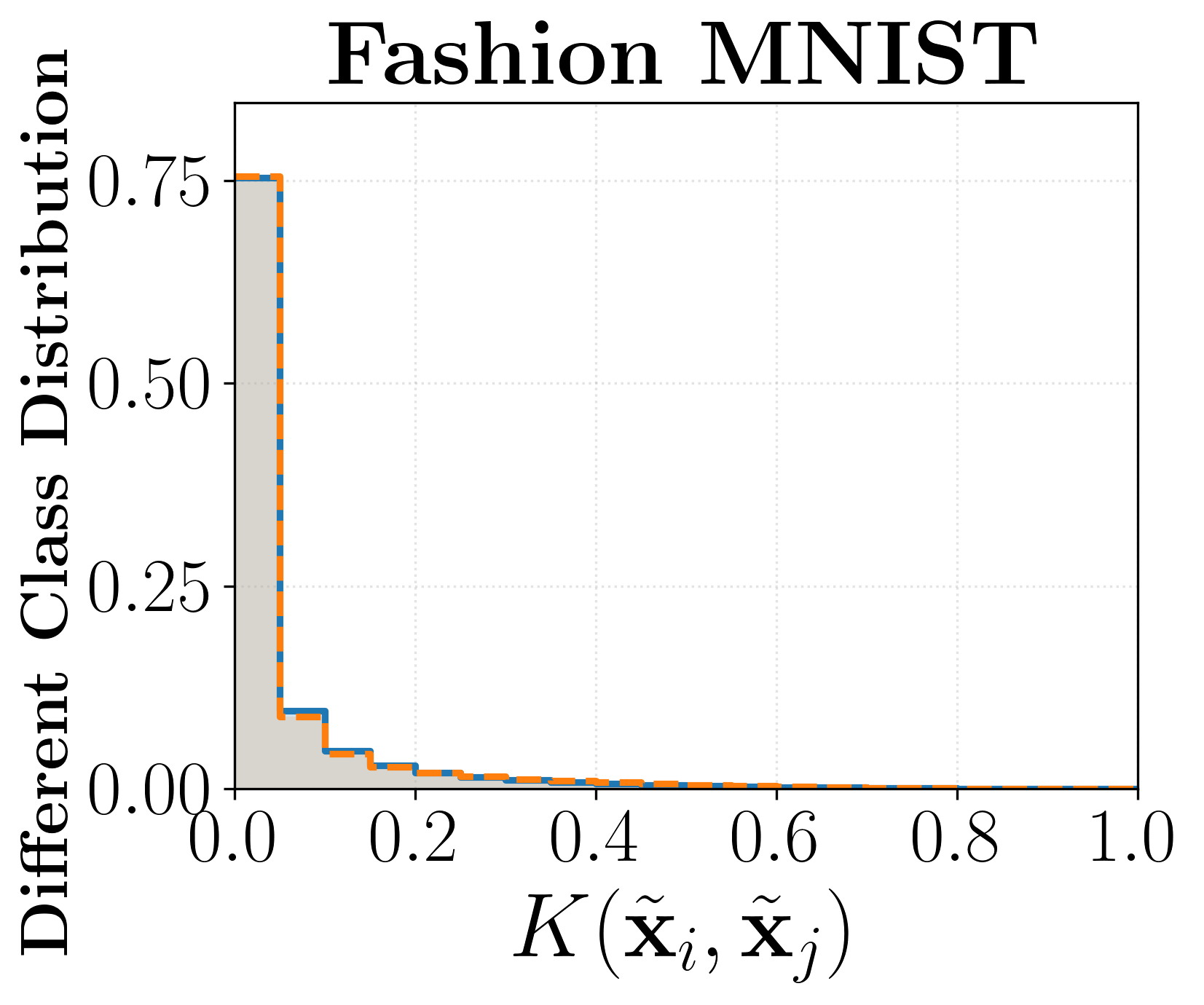}
					\put(4,90){\textbf{(h)}}
				\end{overpic}
				\label{fig:fashion_hist_diff}
			\end{subfigure}
			
			% ===== Caption =====
			\caption{Quantum-kernel value distributions. Panels show normalized occurrence distributions of quantum kernel values $K(\tilde{\mathbf{x}}_i,\tilde{\mathbf{x}}_j)$ for (top row) same-class pairs ($y_i{=}y_j$) and (bottom row) different-class pairs across four datasets: a,b) Ionosphere, c,d) Spambase, e,f) MNIST, and g,h) Fashion-MNIST. Each histogram uses bin width $0.05$ and is normalized by the respective number of counted pairs (train or test). Diagonal terms ($K(\tilde{\mathbf{x}}_i,\tilde{\mathbf{x}}_i){=}1$) are excluded and the results belong to one iteration of the model with three photons ($n{=}3$).}
			\label{fig:hist}
		\end{figure*}

		\noindent\textbf{Kernel quality and extractable information}
		
		\noindent A key issue in the accuracy of our protocol is the closeness of the kernel $K(\tilde{\mathbf{x}}_i,\tilde{\mathbf{x}}_j)$ with the ideal kernel in which the entry elements have to be $0$ when $\tilde{\mathbf{x}}_i$ and $\tilde{\mathbf{x}}_j$ belong to different classes and $1$ otherwise. Since the kernels are computed as the overlap between two quantum states, see Eq.~(\ref{eq:Qkernel}), such estimation requires several samplings in a boson sampling circuit. To illustrate the typical values of the kernel function, Figs.~\ref{fig:hist}(a)–(b) show the normalized occurrence distributions of the kernel values for the training and test data in a trained model on the Ionosphere dataset, corresponding to same-class and different-class data pairs, respectively. The same distributions are obtained for Spambase dataset, see Figs.~\ref{fig:hist}(c)-(d), MNIST dataset, see 
		Figs.~\ref{fig:hist}(e)-(f), and finally the Fashion-MNIST dataset, see Figs.~\ref{fig:hist}(g)-(h). The same-class and different-class kernel distributions are visually well separated across all datasets. To quantify this separation, we evaluate the Jensen–Shannon divergence (JSD) between the corresponding histograms of kernel values, denoted by $\mathcal{P}^{(1)}$ for same-class pairs and $\mathcal{P}^{(2)}$ for different-class pairs. Both histograms are constructed with a bin width of $0.05$ over the interval $[0,1]$. The JSD is defined as 
		\begin{equation} \mathrm{JSD}(\mathcal{P}^{(1)} \| \mathcal{P}^{(2)}) = \tfrac{1}{2} D_{\mathrm{KL}}(\mathcal{P}^{(1)} \| \mathcal{M}) + \tfrac{1}{2} D_{\mathrm{KL}}(\mathcal{P}^{(2)} \| \mathcal{M}), \end{equation}
		where $\mathcal{M} {=} \tfrac{1}{2}\big(\mathcal{P}^{(1)} {+} \mathcal{P}^{(2)}\big)$ and $D_{\mathrm{KL}}(P \| Q) {=} \sum_k P_k \ln(P_k/Q_k)$ is the Kullback–Leibler divergence, with the sum taken over histogram bins $k$. Identical distributions yield $\mathrm{JSD}{=}0$, whereas larger values indicate greater separability with a theoretical maximum of $\mathrm{JSD}{=}\ln 2 {\approx} 0.693$ for perfectly non-overlapping distributions. The obtained values for the datasets in Figs.~\ref{fig:hist}(a)–(h) are $\mathrm{JSD}{=}0.31$ for Ionosphere, $0.06$ for Spambase, $0.43$ for MNIST, and $0.40$ for Fashion-MNIST.

		To relate this separability to the expected number of readout samples, we compute the Chernoff information~\cite{Cover2012elements}, defined as $C {=} -\ln\!\big[\min_{0 \le s \le 1} \sum_k (\mathcal{P}^{(1)}_k)^{s} (\mathcal{P}^{(2)}_k)^{1-s}\big]$, where $s$ is a parameter in $[0,1]$ that balances the contributions of the two probability distributions. The Chernoff information quantifies the asymptotic rate at which the misclassification error decays with the number of samples. Using the asymptotic Chernoff error exponent, achieving a target misclassification probability $\varepsilon$ requires approximately $N_{\mathrm{readout}} {\simeq} \ln(1/\varepsilon)/C$. For a $90\%$ confidence level ($\varepsilon {=} 0.1$), we obtain $N_{\mathrm{readout}}{=}6$ for Ionosphere, $34$ for Spambase, and $4$ for both MNIST and Fashion-MNIST. These values provide heuristic estimates of the number of samples to distinguish the kernel distributions in the image datasets, consistent with the rapid increase in test accuracy shown in Fig.~\ref{fig:readout}. The results quantify, in an information-theoretic sense, the discriminative power encoded in the interconnected boson sampling and neural network learning model. \\

		\noindent\textbf{Effect of imperfections}
		
		\noindent In practice, we may face several imperfections for demonstration of our protocol on a realistic photonic chip, including: (i) finite sample shots; (ii) TBU loss; (iii) phase shifter noise; and (iv) photon indistinguishability. In this section, we investigate the resilience of our protocol against these imperfections.

\begin{figure*}[t]
			\centering
			
			% =========================
			% Left side: plots (two rows)
			% =========================
			\begin{minipage}[c]{.72\textwidth}
				\centering
				
				% ----- Row 1 -----
				\begin{subfigure}{.45\textwidth}
					\begin{overpic}[width=\textwidth]{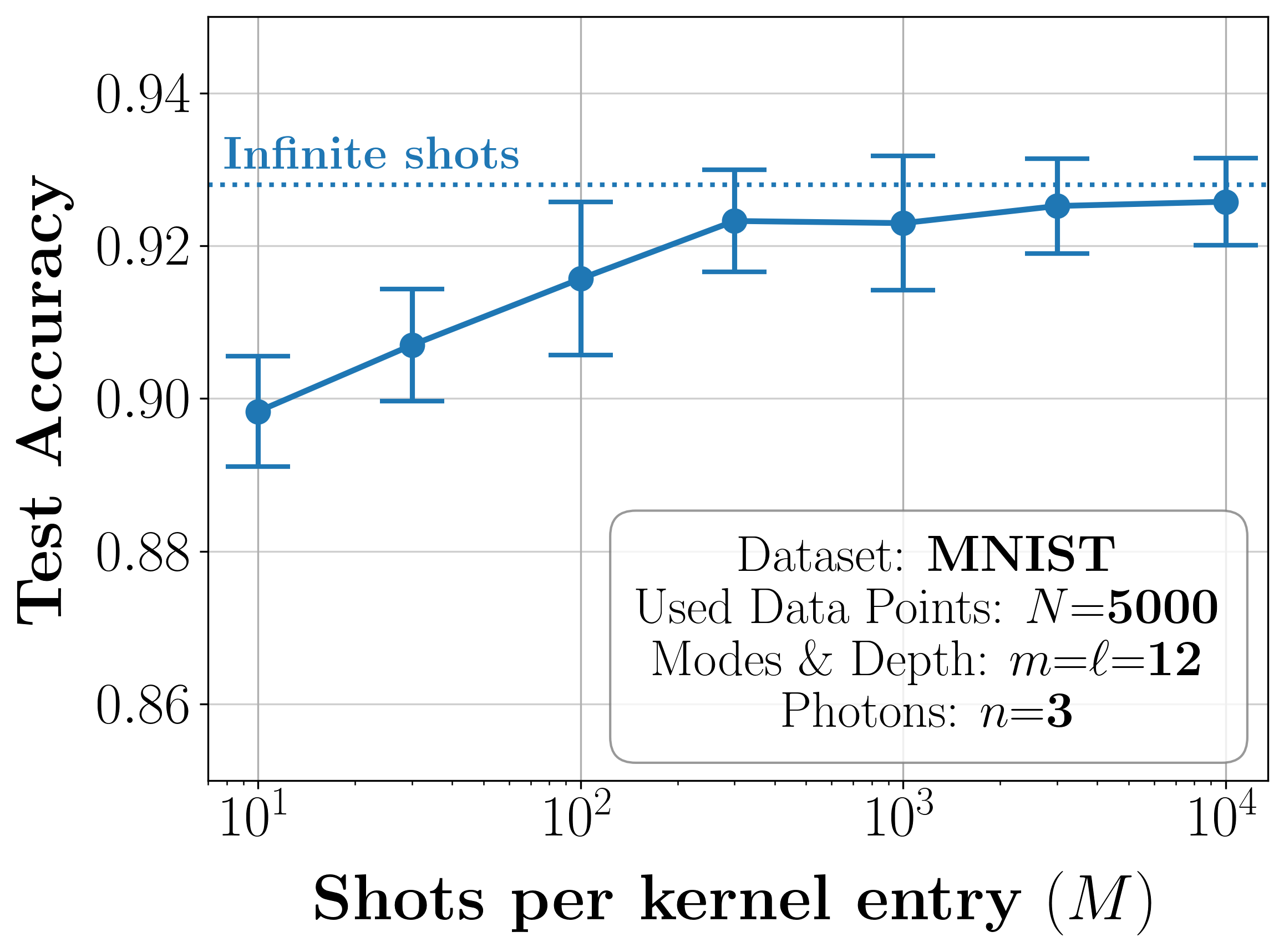}
						\put(-3,72){\textbf{(a)}}
					\end{overpic}
					%	\label{fig:ion_readout}
				\end{subfigure}
				\hspace{0.04\textwidth}
				\begin{subfigure}{.45\textwidth}
					\begin{overpic}[width=\textwidth]{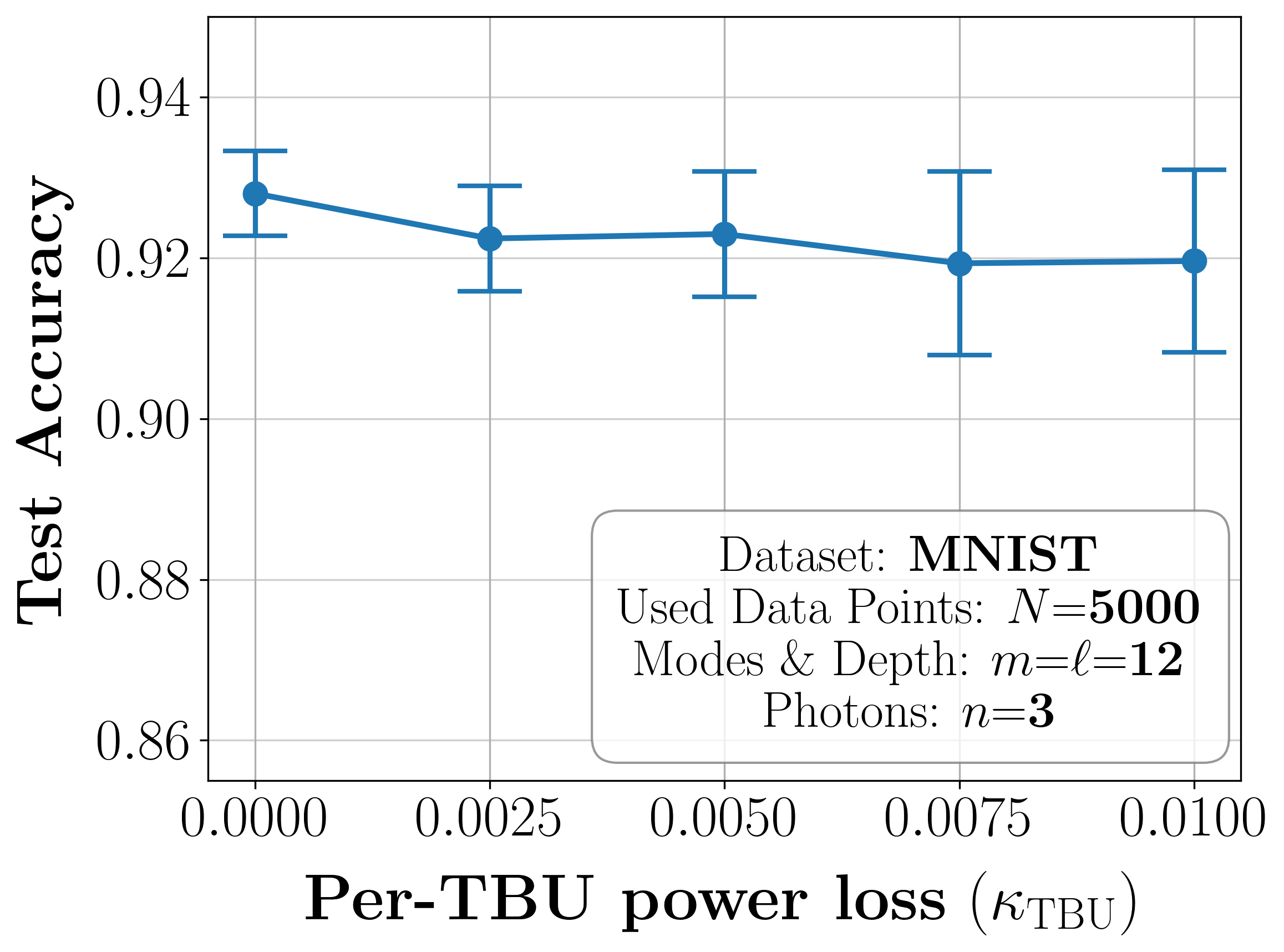}
						\put(-3,72){\textbf{(b)}}
					\end{overpic}
				\end{subfigure}
				
				\vspace{0.3cm}
				
				% ----- Row 2 -----
				\begin{subfigure}{.45\textwidth}
					\begin{overpic}[width=\textwidth]{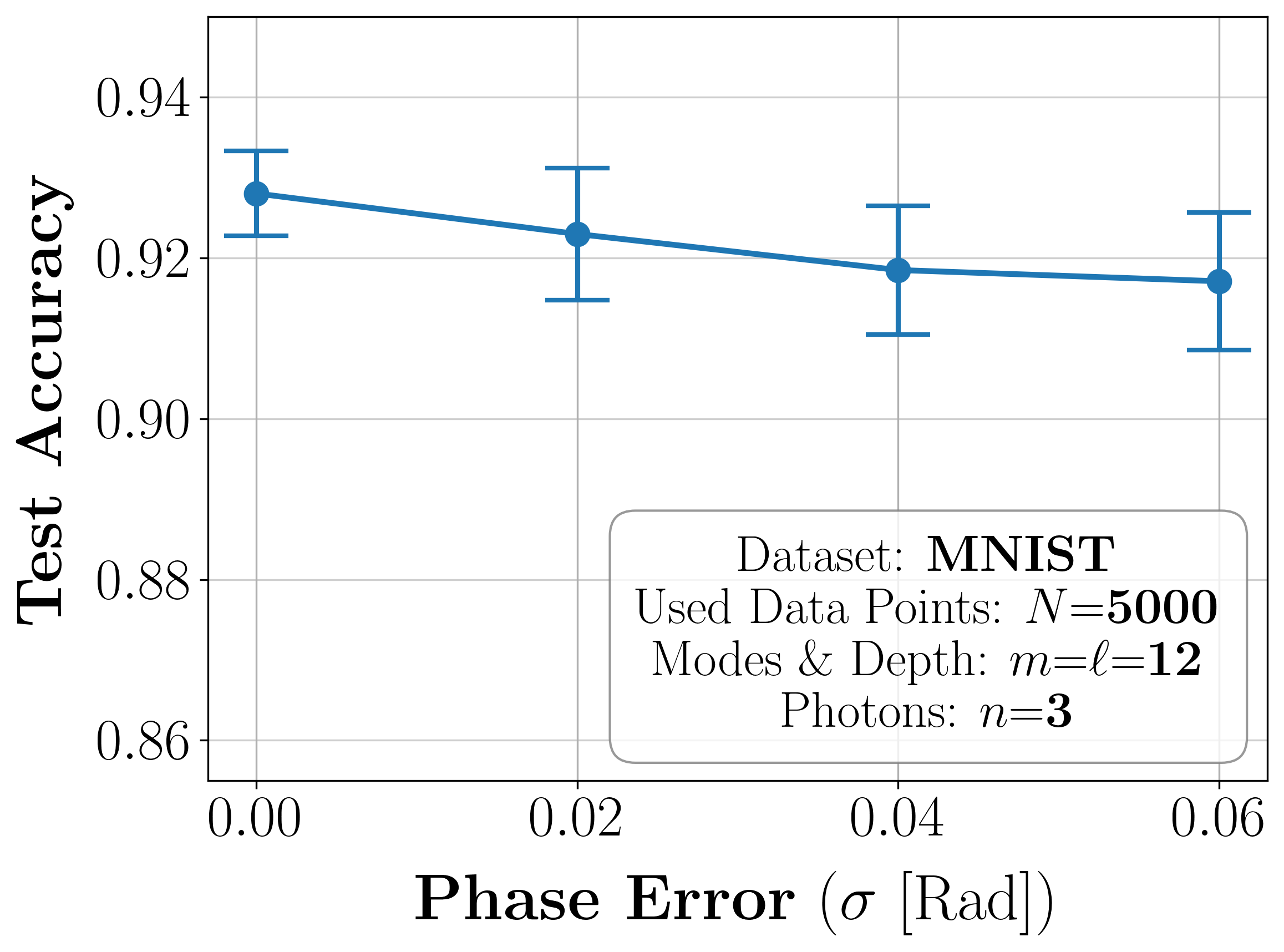}
						\put(-3,72){\textbf{(c)}}
					\end{overpic}
				\end{subfigure}
				\hspace{0.04\textwidth}
				\begin{subfigure}{.45\textwidth}
					\begin{overpic}[width=\textwidth]{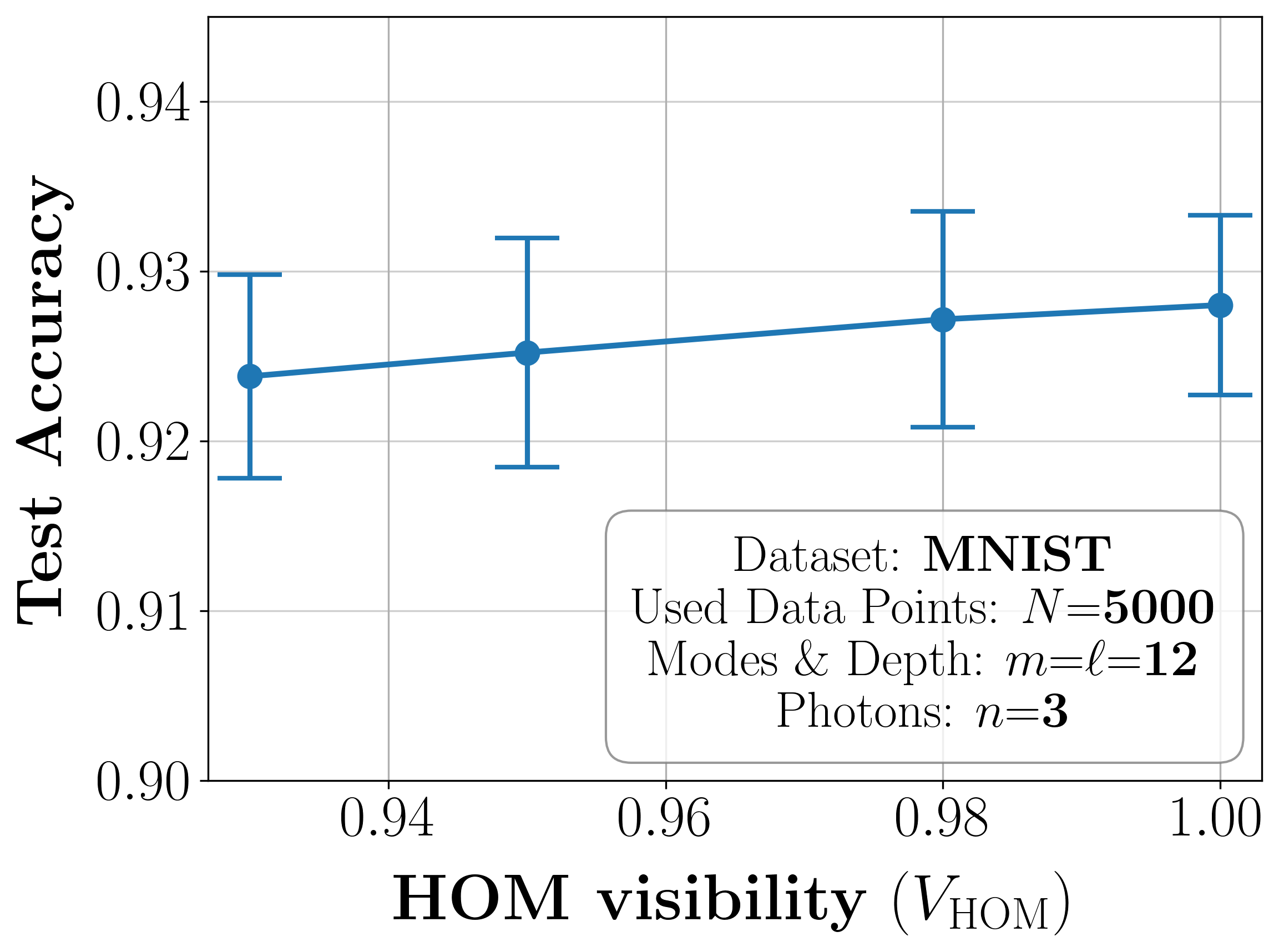}
						\put(-3,72){\textbf{(d)}}
					\end{overpic}
				\end{subfigure}
			\end{minipage}
			%	\hspace{0.03\textwidth}

			% ===== Caption =====
			\caption{Robustness of MNIST classification accuracy. Panels show the effect of a) finite sampling shots, b) loss in each TBU, c)  error in the phase shifters, and d) distinguishability of the photons, on the test accuracy of the model for the MNIST dataset. }
			
			\label{fig:imperfections}
		\end{figure*}

		The first imperfection to be considered is the finite sample shots. In our setup, each kernel entry $K(\tilde{\mathbf{x}}_i,\tilde{\mathbf{x}}_j){\in}[0,1]$ is the probability that, when the input Fock state $\ket{\phi}$ is injected into the interferometer, the same Fock state is detected at the output. In the ideal (infinite-shot) simulation we can compute the exact kernel value but in an experiment, $K(\tilde{\mathbf{x}}_i,\tilde{\mathbf{x}}_j)$ is not observed directly; instead one repeats the same setting for $M$ shots and only records how many times the outcome $\ket{\phi}$ occurs. 
		We therefore can treat each shot as an independent hit with probability $K(\tilde{\mathbf{x}}_i,\tilde{\mathbf{x}}_j)$, so that the total number of hits $C_{ij}{\in}\{0,1,\ldots,M\}$ follows a binomial distribution
		$C_{ij}{\sim} \mathrm{Binomial}(M,K(\tilde{\mathbf{x}}_i,\tilde{\mathbf{x}}_j))$.
		The shot-limited estimate of the kernel entry is then the observed hit fraction $\widehat K(\tilde{\mathbf{x}}_i,\tilde{\mathbf{x}}_j){=}\frac{C_{ij}}{M}$. Since we have a binomial distribution the variance of the number of hits is $\mathrm{Var}(C_{ij}){=}MK(\tilde{\mathbf{x}}_i,\tilde{\mathbf{x}}_j)(1-K(\tilde{\mathbf{x}}_i,\tilde{\mathbf{x}}_j))$ and therefore the variance of the kernel estimation is bounded by
		\begin{eqnarray}
			\mathrm{Var}(\widehat K(\tilde{\mathbf{x}}_i,\tilde{\mathbf{x}}_j))
			=\frac{K(\tilde{\mathbf{x}}_i,\tilde{\mathbf{x}}_j)(1-K(\tilde{\mathbf{x}}_i,\tilde{\mathbf{x}}_j))}{M}
			\le \frac{1}{4M},
		\end{eqnarray}
		because $0{\leq}K(\tilde{\mathbf{x}}_i,\tilde{\mathbf{x}}_j){\leq}1$. 
		Thus, the typical sampling error of $\widehat K(\tilde{\mathbf{x}}_i,\tilde{\mathbf{x}}_j)$ decreases by $1/\sqrt{M}$. To model a finite number of experimental measurements, we replace each kernel entry by the empirical estimator $\widehat K(\tilde{\mathbf{x}}_i,\tilde{\mathbf{x}}_j){=}\frac{C_{ij}}{M}$ with $C_{ij}{\sim} \mathrm{Binomial}(M,K(\tilde{\mathbf{x}}_i,\tilde{\mathbf{x}}_j))$. Then we sweep
		$M{\in}[10,10^{4}]$ shots per kernel and observe that the test accuracy gradually approaches
		the infinite-shot value. Fig.~\ref{fig:imperfections}(a) shows that $M{=} 100$
		shots per entry are sufficient to reach $91.6{\pm}0.01\%$ test accuracy for MNIST dataset with $n{=}3$ photons.

		The second imperfection to be considered is the TBU loss. One aspect of the model is that we operate in the collision-free regime (at most one photon per mode). While due to imperfections, photon decay can happen, its effect can always be detected, and experimental runs in which fewer photons are detected than injected can be identified as loss events and discarded. Indeed, in general, the impact of photon loss has to be treated through Lindbladian master equations in which the quantum state becomes a mixed state as the evolution takes place. However, the mentioned post-selection procedure allows for a simplified treatment of the photon loss through  a uniform-loss model. In this approach, the effect of photon loss is modeled by a reduced norm of the unitary operation. The reduced norm quantifies the probability of having no photon loss in the system. The uniform-loss model  is equivalent to monitoring the environment for detecting any decayed photon, see below for detailed analysis. Furthermore, the boson-sampling platforms can operate at sampling rates on the order of several kHz~\cite{Wang2018toward}, which allows one to partially compensate the reduced detection probability by collecting more samples. Therefore, one can use uniform-loss model for accounting for photon loss in the circuit. The impact of photon decay per TBU, is parameterized  by the per-TBU power loss $\kappa_{\mathrm{TBU}}{\in}[0,1]$, i.e., each TBU loses a fraction
		$\kappa_{\mathrm{TBU}}$ of optical power. The corresponding field-amplitude transmissivity is
		\begin{equation}
			\Gamma=\sqrt{1-\kappa_{\mathrm{TBU}}}.
		\end{equation}
		Accordingly, each TBU unitary is replaced by
		\begin{equation}
			U_{\mathrm{TBU}}\;\mapsto\;\Gamma\,U_{\mathrm{TBU}}=\sqrt{1-\kappa_{\mathrm{TBU}}}\,U_{\mathrm{TBU}}.
		\end{equation}
		
		We simulate propagation loss for the MNIST dataset with \(N{=}5000\), \(m{=}\ell{=}12\) modes and circuit depth, and \(n{=}3\) photons. As shown in Fig.~\ref{fig:imperfections}(b), the final test accuracy decreases gradually as the per-TBU power loss \(\kappa_{\mathrm{TBU}}\) increases but remains relatively stable for \(\kappa_{\mathrm{TBU}}{\lesssim}0.01\). This range is experimentally realistic and conservative. Indeed, based on the reported loss of the 12-mode universal photonic processor in Ref.~\cite{Taballione2020a} and a typical on-chip TBU length of about \(1~\mathrm{mm}\) (see Ref.~\cite{Marchisio2025comprehensive}), the estimated per-TBU loss is only \(\kappa_{\mathrm{TBU}} {\approx} 2.3 {\times} 10^{-3}\). Therefore, the simulated interval \(\kappa_{\mathrm{TBU}} {\in} [0,0.01]\) extends well beyond current experimental values, providing a conservative test of the model's robustness. These results confirm that the model is robust against realistic levels of TBU loss.
		
		The third imperfection to be considered is the phase shifter noise. We model imperfect control of each tunable element by adding independent Gaussian perturbations to the TBU parameters,
		\begin{equation}
			\tilde{\theta}=\theta +\delta\theta ,\quad 
			\tilde{\varphi} =\varphi +\delta\varphi ,\quad 
			\delta\theta \sim\mathcal{N}(0,\sigma^2),\quad \delta\varphi \sim\mathcal{N}(0,\sigma^2).
		\end{equation}
		For a given $\sigma$, the perturbations $\delta\theta$ and $\delta\varphi$ are sampled independently and applied across all TBUs during the circuit evaluation. Again, we simulate the effect of phase noise on the test accuracy for MNIST with $N{=}5000$, $m{=}\ell{=}12$ modes and depths, and $n{=}3$ photons.
		We sweep $\sigma {\in} [0,0.06]$~rad and the result is shown in Fig.~\ref{fig:imperfections}(c). 
		The error range is motivated by experimental programmable photonic circuits in Ref.~\cite{Zhong2021phase}. 
		Fig.~\ref{fig:imperfections}(c) shows that the accuracy of the model remains high for realistic phase-noise levels.

		The fourth nonideality in multi-photon linear optics is partial photon indistinguishability. To model it, we use the exact partially distinguishable multiphoton transition probability definition of Ref.~\cite{Tichy2015sampling}, in which distinguishability is encoded by the overlap matrix
		$S_{jk}{=}\langle \phi_j | \phi_k \rangle$,
		where $\phi_j$ and $\phi_k$ are the internal states of photons $j$ and $k$. Their overlap quantifies the degree of indistinguishability: $S_{jk}{=}1$ for perfectly indistinguishable photons and $S_{jk}{=}0$ for fully distinguishable photons.
		For a transition described by the effective submatrix
		$M_{ij}{=}\bigl[U^\dagger(\mathbf{x}_i)U(\mathbf{x}_j)\bigr]_{S,T},$
		the corresponding kernel entry is computed as
		\begin{equation}
			P_S(M_{ij})
			=
			\sum_{\tau\in S_n}
			\left(
			\prod_{a=1}^{n} S_{a,\tau(a)}
			\right)
			\mathrm{Per}\!\left(M_{ij}\odot M_{ij,\tau,:}^{*}\right),
		\end{equation}
		where $S_n$ is the symmetric group on \(n\) elements, \(a{\in}\{1,\dots,n\}\) labels the photons (equivalently, the rows of the effective submatrix), \(\tau(a)\) is the image of \(a\) under the permutation \(\tau\), \(M_{ij,\tau,:}\) denotes \(M_{ij}\) with its rows permuted according to \(\tau\), and \(\odot\) denotes entrywise multiplication. This reduces to the ideal bosonic limit
		$P_{\mathrm{id}}(M_{ij}){=}|\mathrm{Per}(M_{ij})|^2$
		for perfectly indistinguishable photons and to the fully distinguishable limit
		$P_{\mathrm{dist}}(M_{ij}){=}\mathrm{Per}(|M_{ij}|^2)$
		when the photons are completely distinguishable. A uniform-overlap approximation can be used for simulating the distinguishability
		\begin{equation}
			S_{jj}=1,\qquad |S_{jk}|^2 = V_{\mathrm{HOM}}\quad (j\neq k),
		\end{equation}
		so that the off-diagonal overlap amplitude is \(S_{jk}{=}\sqrt{V_{\mathrm{HOM}}}\). 
		The reported experimental distinguishability in Ref.~\cite{Taballione2020a} is about $V_\mathrm{HOM}{\sim}0.93$. We therefore sweep \(V_{\mathrm{HOM}}{\in}[0.93,1]\), where \(V_{\mathrm{HOM}}{=}1\) corresponds to ideal indistinguishable photons and smaller values model increasing distinguishability. Figure~\ref{fig:imperfections}(d) shows the final test accuracy for MNIST with \(N{=}5000\), \(m{=}\ell{=}12\), and \(n{=}3\) photons. The accuracy decreases as the HOM visibility is reduced but even when the visibility is reduced from the ideal value to practical range, the total decrease in test accuracy is low.  
		
		In our numerical simulations, the kernel is implemented via two sequential circuits of depth 
		$\ell{=}m$ for mapping both $\tilde{\mathbf{x}}_i$ and $\tilde{\mathbf{x}}_j$, giving an effective depth of $2m$. Experimentally, however, the circuit can be implemented using a universal $m$-mode linear optical interferometer where the circuit depth is equal to the number of modes. In this configuration, the general unitary transformation $U_G(\tilde{\mathbf{x}}_i,\tilde{\mathbf{x}}_j){=}U^\dagger(\tilde{\mathbf{x}}_i)U(\tilde{\mathbf{x}}_j)$ can be directly applied on an $m$-mode universal circuit~\cite{Yin2024experimental,Clements2016an,Reck1994experimental} eliminating the need to apply $U^\dagger(\tilde{\mathbf{x}}_i)$ and $U(\tilde{\mathbf{x}}_j)$ sequentially. This approach halves the circuit depth, improving experimental efficiency and reducing the imperfections effects. \\

		\noindent\textbf{Estimating wall-clock time}
		
		\noindent Here we try to calculate an approximate wall-clock time for two parts of the model, the training part and the testing part. We use the largest used datasets i.e. MNIST and Fashion-MNIST for this evaluation so the other datasets take less time. Regarding the training time, as mentioned above, only $N_\textrm{Nys}{=}20$ landmark training points and $M{=}100$ shots for an $n{=}3$ photons configuration are required to reach close to $91.6\%$ test accuracy. Therefore, the number of training kernel entries that must be estimated is
		\begin{eqnarray}
			N_{\mathrm{eval}} =N_{\mathrm{train}}N_\textrm{Nys}  = 85{,}720
		\end{eqnarray}
		evaluations, and the total shot budget is $M\,N_{\mathrm{eval}} {=} 8.6{\times}10^6$ shots. Thus, the wall-clock training time with a sampling rate of $r{=}10~\mathrm{kHz}$~\cite{Wang2018toward} is
		\begin{eqnarray}
			T = \frac{M\,N_{\mathrm{eval}}}{r}\approx14~\textrm{minutes}~.
		\end{eqnarray}
		
		Regarding the testing time, based on the readout evaluation (see Fig.~\ref{fig:readout}) for one new data point the trained SVM only requires evaluating the kernel between that test point and the $N_\textrm{readout}{=}100$ training data point. Therefore, the total shot budget is
		$
		M N_\textrm{readout} {=}10^4
		$
		shots. At the same sampling rate, a wall-clock testing time of
		$T_\textrm{test} {=}1~\textrm{s}$
		is enough for classifying one new datapoint.
	
	\section*{Discussion}

	We proposed a hybrid boson sampling–neural network framework that combines the quantum advantage of boson samplers together with the adaptability  of neural networks in handling large feature sets. The main task of the classical neural network, which for the ease of optimization is taken to be a minimal possible network, is dimensionality adaptation. The output of the neural network is fed as the parameters of the boson sampling chip to act on a given input Fock state. Consequently, any classical input data is mapped to a quantum state, generated by the boson sampling circuit. By employing classical optimization of a proper kernel function, measurable at the output of the boson sampling circuit, one can build an SVM classifier for the input data. The optimization trains the model to map the input data from the same class into high-fidelity quantum states while mapping the input data from different classes to orthogonal states. This improves the class separability in the Hilbert space allowing the accuracy of the SVM to be enhanced. 
	We demonstrate the performance of our protocol over four different datasets with different numbers of classes.  
	The achievable accuracy surpasses both the classical linear and nonlinear sigmoid kernels of the original data. Moreover, we show that the enhanced accuracy is truly coming from the quantum operation of the boson sampling circuit and not the classical neural network. We demonstrate this by benchmarking against a classical neural network of equivalent complexity to the one used in our protocol for feature mapping. These results demonstrate the capability of boson sampling–based kernels to capture complex, nonclassical correlations that remain inaccessible to conventional kernel methods. The accuracy can always be improved by increasing the number of photons and modes, effectively expanding the accessible Hilbert space. 
	From a practical point of view, boson sampling circuits are largely passive linear-optical interferometers that do not need cryogenic systems, suggesting to be very energy efficient for practical use. 
	
	Compared with recent photonic experiments for realizing a SVM algorithm, e.g. Ref.~\cite{Yin2024experimental}, which are only applicable for low-feature datasets, our hybrid protocol harnesses a classical neural network to adjust large dimensional data to be used in a NISQ-friendly quantum circuit. This allows our model to be flexible for the dimension of the input data and gives it the capability of handling large dimensional data. Furthermore, we have explored the effect of Hilbert space dimension on the accuracy of the protocol. Therefore, we benchmark multiple standard real-world datasets of varying input dimensionalities and examine how increasing photon number (and hence the accessible Hilbert-space dimension) influences classification performance.
	
	Beyond its current implementation, the framework establishes a general pathway for exploiting boson sampling devices in practical machine learning tasks. A natural next step is to implement the protocol on a programmable integrated photonic chip. Future extensions could explore more complex photonic architectures, such as three dimensional or time multiplexed integrated circuits, to further enlarge the Hilbert space and increase circuit reconfigurability~\cite{Hoch2022reconfigurable,Tan2023scalable}. Additionally, incorporating alternative Fock state encodings, e.g. multi-photon injections per mode, may enhance the expressive capacity of the model and enable richer quantum feature representations. 
	\\
	
	\section*{Methods}

	\noindent\textbf{Numerical simulation framework}
	
	\noindent For numerical computation of the SVM classification algorithm we rely on  \verb*|Scikit-Learn|~\cite{Pedregosa2011scikit} in Python. The classical neural network is simulated using \verb*|PyTorch| library~\cite{Paszke2019pytorch}. The neural network parameters are initialized from a uniform distribution and optimized using the Adam optimizer with learning rate $10^{-3}$. 
	All simulations are performed with random seeds to ensure reproducibility. For classification, the SVM is trained on the precomputed training kernel matrix and tested on the corresponding test-versus-train kernel matrix. The global interferometer unitary is initialized as the identity matrix,
	\(U^{(0)} {=} I_m,\)
	where $m$ is the number of modes. The circuit is constructed layer by layer, with each layer corresponding to the two-mode transformations of TBUs corresponding to that layer. At each layer \(r\), a unitary matrix \(U_{\mathrm{layer}}^{(r)}\) is constructed by embedding the corresponding two-mode TBU operations into the full \(m{\times }m\) space. The total unitary is then built recursively as
	\begin{eqnarray}
		U^{(r)} = U_{\mathrm{layer}}^{(r)} U^{(r-1)}, 
		\qquad r = 1,\dots,\ell,
	\end{eqnarray}
	where \(\ell\) is the circuit depth. The final interferometer corresponding to the input \(\tilde{\mathbf{x}}_i\) is therefore given by
	\begin{eqnarray}
		U(\tilde{\mathbf{x}}_i) = U^{(\ell)} = U_{\mathrm{layer}}^{(\ell)} \cdots U_{\mathrm{layer}}^{(1)}~.
	\end{eqnarray}
	
	These unitary transformations are then used to compute the corresponding kernel matrices, which serve as inputs to the SVM classifier for accuracy evaluation.\\

		\begin{table}[t]
		\centering
		\begin{tabular}{|l|c|c|c|}
			\hline
			\multirow{2}{*}{\textbf{Metric}} &
			\multicolumn{3}{c|}{\textbf{Dataset}} \\
			\cline{2-4}
			& \textbf{Spambase} & \textbf{MNIST} & \textbf{Fashion-MNIST} \\
			\hline
			$\mathrm{MI}_{\mathrm{mean}}(\mathbf{x}_i)$
			& $0.06{\pm}0.0008$ & $0.07{\pm}0.0009$ & $0.26{\pm}0.004$ \\
			\cline{1-4}
			$\mathrm{MI}_{\mathrm{mean}}(\tilde{\mathbf{x}}_i)$
			& $0.15{\pm}0.007$ & $0.30{\pm}0.004$ & $0.43{\pm}0.01$ \\
			\hline
			$\mathrm{MSE}_{\mathrm{rec}}(\tilde{\mathbf{x}}_i)$
			& $0.001{\pm}0.00008$ & $0.01{\pm}0.0002$ & $0.01{\pm}0.0005$ \\
			\hline
			SVM $(\mathbf{x}_i)$
			& $0.89{\pm}0.006$ & $0.92{\pm}0.004$ & $0.80{\pm}0.008$ \\
			\cline{1-4}
			SVM $(\tilde{\mathbf{x}}_i)$
			& $0.86{\pm}0.005$ & $0.91{\pm}0.01$ & $0.81{\pm}0.009$ \\
			\hline
		\end{tabular}
		\caption{Summary of class-discrimination diagnostics comparing the raw input $\mathbf{x}_i$ and the trained output $\tilde{\mathbf{x}}_i$.}
		\label{tab:nn_preservation_all}
	\end{table}
	
	\noindent\textbf{Quantification of feature-space discrimination}
	
	\noindent We next explain some methods about how effectively the learned feature adaptation retains the class-discriminative structure of the original data. Here, we evaluate three quantitative measures, i.e. mutual information, reconstruction error, and linear separability scores for three high-dimensional datasets (Spambase, MNIST, and Fashion-MNIST) to evaluate the discrimination structure. This evaluation is done before ($\mathbf{x}_i{\in}\mathbb{R}^{d_\textrm{raw}}$) and after ($\tilde{\mathbf{x}}_i{\in}\mathbb{R}^{d}$) the neural network. We use a three-photon model with the same number of modes as in Fig.~\ref{fig:acc_photons} ($m{=}11$ for Spambase and $m{=}12$ for MNIST and Fashion-MNIST).
	
	First we obtain the mutual information between each feature dimension and the class label for the raw features $\mathbf{x}_i$
	\begin{equation}
		I\!\left(X^{(j)};Y\right)=\mathbb{E}_{(X^{(j)},Y)}\!\left[\log\frac{p(X^{(j)},Y)}{p(X^{(j)})\,p(Y)}\right],
	\end{equation}
	where $X^{(j)}$ denotes the random variable corresponding to the $j$-th feature (and $x_i^{(j)}$ is its observed value for sample $i$), and $Y$ is the class-label random variable (with observations $y_i$). We then get the mean value of the information across the dimensions
	\begin{equation}
		\mathrm{MI}_{\mathrm{mean}}(\mathbf{x})
		= \frac{1}{d_\mathrm{raw}}\sum_{j=1}^{d_\mathrm{raw}}I(X^{(j)};Y).
	\end{equation}
	The same procedure is applied to $\tilde{\mathbf{x}}_i$ by replacing $\mathbf{x}_i$ with $\tilde{\mathbf{x}}_i$ (and correspondingly replacing $d_\mathrm{raw}$ by $d$). The results in Table.~\ref{tab:nn_preservation_all} illustrate that the mutual information for all high-dimensional datasets are improved and the feature mapping does not destroy this class-discrimination metric.

	The second metric that we evaluate is the reconstruction error. We fit a linear ridge decoder that reconstructs the raw input $\mathbf{x}_i$ from the transformed feature data $\tilde{\mathbf{x}}_i$ by learning a weight matrix $W{\in}\mathbb{R}^{d_{\mathrm{raw}}\times d}$ on the training set of size $N_{\mathrm{train}}$
	\begin{equation}
		W^\star
		=
		\arg\min_W
		\sum_{i=1}^{N_{\mathrm{train}}}
		\left\|
		\mathbf{x}_i - W\,\tilde{\mathbf{x}}_i
		\right\|_2^2
		+
		\alpha \|W\|_F^2,
	\end{equation}
	where $\|\cdot\|_2$ is the Euclidean norm, $\|\cdot\|_F$ is the Frobenius norm, and $\alpha{\ge} 0$ controls the ridge regularization strength (we set $\alpha{=}1$).
	We then evaluate on the test set of size $N_{\mathrm{test}}$ via
	\begin{equation}\label{eq:reconstruction_err}
		\mathrm{MSE}_{\mathrm{rec}}(\tilde{\mathbf{x}})
		=
		\frac{1}{N_\mathrm{test}\,d_\mathrm{raw}}
		\sum_{i=1}^{N_{\mathrm{test}}}
		\left\|
		\mathbf{x}_i - W^\star\,\tilde{\mathbf{x}}_i
		\right\|_2^2.
	\end{equation}
	Table~\ref{tab:nn_preservation_all} shows that the ridge reconstruction error remains consistently low across three high-dimensional datasets.

	As a third metric we use the linear separability. Comparable linear-SVM accuracy for $\mathbf{x}_i$ and $\tilde{\mathbf{x}}_i$ indicates that the transformation preserves the class-discriminative geometry. We already have a linear SVM accuracy on $\mathbf{x}_i$ that is reported in Table~\ref{tab:merged_datasets}. Here we perform another linear SVM on  $\tilde{\mathbf{x}}_i$ and compare its test accuracy with the one on $\mathbf{x}_i$. 
	Table~\ref{tab:nn_preservation_all} shows that the linear-separability accuracy on the boson sampling-neural network output $\tilde{\mathbf{x}}_i$ is comparable to (and in some cases slightly higher than) that of the raw features $\mathbf{x}_i$, indicating that the dimensionality adaptation largely preserves the class-discriminative structure of the datasets.\\

	\begin{figure}[t]
		\centering
		
		\begin{overpic}[width=0.48\textwidth]{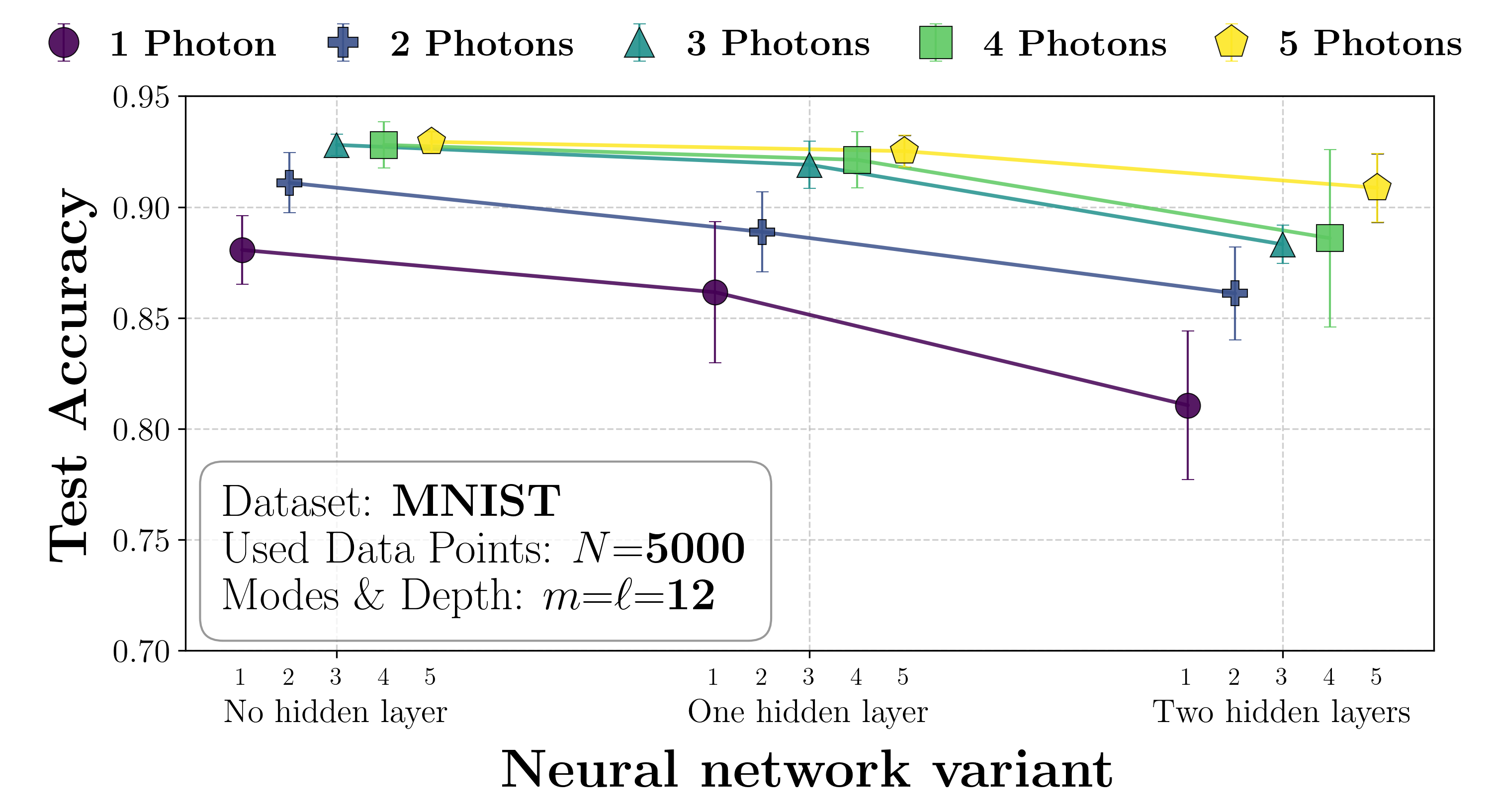}
			% \put(-5,50){\textbf{(a)}}
		\end{overpic}
		
		% ===== Caption =====
		\caption{Effect of the neural network complexity on the test accuracy of the model for the MNIST dataset.}
		\label{fig:layered}
	\end{figure}

	\noindent\textbf{Neural network complexity}\\
	In this model we use a shallow fully connected network without any hidden layer to match the feature dimension into the circuit parameters without classically smearing the quantum effects. Here, we examine whether increasing the depth of the neural network improves performance or degrades stability. To demonstrate the effect we use MNIST with \textbf{$N{=}5000$} samples and a boson-sampling circuit with $m{=}\ell{=}12$ modes and depths (same parameters as Fig.~\ref{fig:acc_photons}).

	We compare three neural network variants: the shallow one without any hidden layer that is used in the model, a deeper neural network with a 256-nodes hidden layer, and another one with two 256-nodes hidden layers.  The results show that the shallow encoder achieves the highest test accuracy and the best stability, while deeper encoders lead to reduced accuracy. Increasing the neural network depth  may induce information attenuation whose consequence is  discrimination reduction of the quantum states. The uncertainty (standard deviation of the test accuracy) can be used as a proxy for stability, which increases with encoder depth as shown in Fig.~\ref{fig:layered}, indicating degraded stability for deeper encoders. 
	
	Also, we evaluate the reconstruction error (Eq.~\eqref{eq:reconstruction_err}) for the three neural networks. The reconstruction error increases with implementing deeper neural networks, yielding $\mathrm{MSE}_{\mathrm{rec}}{=}0.013{\pm}0.0002$ for the model with no hidden layer, $0.020{\pm}0.0004$ for one hidden layer, and $0.028{\pm}0.0004$ for two hidden layers. Therefore, increasing encoder depth does not improve accuracy for high-dimensional datasets and tends to reduce both accuracy and stability.\\

%	\section*{Data Availability}
%	
%	The datasets analyzed during the current study are publicly available as follows:
%	
%	The Ionosphere dataset is available from the UCI Machine Learning Repository 
%	(https://doi.org/10.24432/C5W01B).
%	
%	The Spambase dataset is available from the UCI Machine Learning Repository 
%	(https://doi.org/10.24432/C53G6X).
%	
%	The MNIST dataset is available from Kaggle at 
%	https://www.kaggle.com/datasets/avnishnish/mnist-original.
%	
%	The Fashion-MNIST dataset is available from Kaggle at 
%	https://www.kaggle.com/datasets/zalando-research/fashionmnist.\\
	
	\section*{Code Availability}
The code scripts used for analyzing the present study are publicly available on the project page at
	\href{https://github.com/mohammadshariifian/boson-sampling-neural-network}{https://github.com/mohammadshariifian/boson-sampling-neural-network}.
	\\
	\section*{Acknowledgments}
	AB acknowledges support
	from the National Natural Science Foundation of China
	(grants No. W2541020, No. 12274059, No. 12574528, and
	No. 1251101297). The funder played no role in study design,
	data collection, analysis and interpretation of data, or the
	writing of this manuscript. The authors would like to thank S. Sarkar and C. Mukhopadhyay for useful discussions.

	%==============================
	
\nocite{apsrev42Control}
\bibliographystyle{apsrev4-2}
\bibliography{refs}
	%==============================
	%	\pagebreak

\end{document}